\documentclass[]{aa} 
\usepackage{graphicx}
\usepackage{txfonts}
\usepackage{comment}
\usepackage{float}
\usepackage[dvipsnames]{xcolor}
\usepackage{subfig}
\usepackage[utf8]{inputenc}
\usepackage[T1]{fontenc}
\usepackage{amsfonts}
\usepackage{pdflscape}
\usepackage{longtable}
\usepackage{xcolor}
\usepackage{lscape}
\usepackage{nicefrac}
\usepackage[colorlinks=true,allcolors=blue]{hyperref}

\newcommand{\Trot}{$T\rm{_{rot}}$}

\graphicspath{{./}{figures/}}

\begin{document} 

\title{Far-infrared line emission from the outer Galaxy cluster Gy~3--7 with SOFIA/FIFI-LS: Physical
conditions and UV fields\thanks{Tables \ref{table:rot}, \ref{t:physpar}, 
\ref{tab:maser}, \ref{tab:flux_2cores}, \ref{tab:phot}, \ref{tab:Trot_compared},
and \ref{tab:summary_distance_Luminosity_YSOs_MW_MCs} are available in electronic form
at the CDS via anonymous ftp to \url{cdsarc.cds.unistra.fr} (130.79.128.5)
or via \url{https://cdsarc.cds.unistra.fr/cgi-bin/qcat?J/A+A/}
}}

\author{Ngân Lê\inst{1}, A.~Karska\inst{1,2}\thanks{Corresponding author: Agata Karska \newline
\email{agata.karska@umk.pl}}, M.~Figueira\inst{3,1}, M.~Sewiło\inst{4,5,6}, A.~Mirocha\inst{7},
Ch.~Fischer\inst{8}, M.~Kaźmierczak-Barthel\inst{8}, R.~ Klein\inst{9}, M.~Gawroński\inst{1},
M.~Koprowski\inst{1}, K.~Kowalczyk\inst{1}, W.~J.~Fischer\inst{10}, K.~M.~Menten\inst{2},
F.~Wyrowski\inst{2}, C.~K\"{o}nig\inst{2}, L.~E.~Kristensen\inst{11}}

\institute{$^{1}$ Institute of Astronomy, Faculty of Physics, Astronomy and Informatics, Nicolaus
Copernicus University, Grudzi\k{a}dzka 5, 87-100 Toruń, Poland\\
$^{2}$ Max-Planck-Institut für Radioastronomie, Auf dem Hügel 69, 53121, Bonn, Germany \\
$^{3}$ National Centre for Nuclear Research, ul. Pasteura 7, 02-093, Warszawa, Poland \\
$^{4}$ Exoplanets and Stellar Astrophysics Laboratory, NASA Goddard Space Flight Center, Greenbelt,
MD 20771, USA \\
$^{5}$ Center for Research and Exploration in Space Science and Technology, NASA Goddard Space
Flight Center, Greenbelt, MD 20771 \\
$^{6}$ Department of Astronomy, University of Maryland, College Park, MD 20742, USA \\
$^{7}$ Astronomical Observatory of the Jagiellonian University, Orla 171, 30-244, Kraków, Poland\\
$^{8}$ Deutsches SOFIA Institut, University of Stuttgart, Pfaffenwaldring 29, 70569, Stuttgart\\
$^{9}$ SOFIA/USRA, NASA Ames Research Center, P.O. Box 1, MS 232-12, Moffett Field, CA 94035, USA\\
$^{10}$ Space Telescope Science Institute, 3700 San Martin Dr., Baltimore, MD 21218, USA\\
$^{11}$ Niels Bohr Institute, Centre for Star and Planet Formation, University of Copenhagen, \O
ster Voldgade 5-7, 1350 Copenhagen, Denmark} 

\date{Received February 14, 2023; accepted April 10, 2023}
\titlerunning{FIFI-LS spectroscopy of Gy~3--7 in the outer Galaxy}
\authorrunning{N.~Lê et al., 2023}

\abstract
{Far-infrared (FIR) line emission provides key information about the gas cooling and heating due to shocks
and UV radiation associated with the early stages of star formation. Gas cooling via FIR
lines might, however, depend on metallicity.}
{We aim to quantify the FIR line emission and determine the spatial distribution of the CO
rotational temperature, ultraviolet (UV) radiation field, and H$_2$ number density toward the
embedded cluster Gy~3--7 in the CMa--$l224$ star-forming region, whose metallicity is expected to be
intermediate between that of the Large Magellanic Cloud and the Solar neighborhood. By comparing the total
luminosities of CO and [\ion{O}{i}] toward Gy~3--7 with values found for low- and high-mass
protostars extending over a broad range of metallicities, we also aim to identify the possible
effects of metallicity on the FIR line cooling within our Galaxy.}
{We studied SOFIA/FIFI-LS spectra of Gy~3--7, covering several CO transitions from $J=14-13$ to
$31-30$, the OH doublet at 79~$\mu$m, the [\ion{O}{i}] 63.2 and 145.5~$\mu$m, and the  [\ion{C}{ii}]
158~$\mu$m lines. The field of view covers a $2'\times1'$ region with a resolution of
$\sim$7$\arcsec$-18$\arcsec$. }
{The spatial extent of CO high$-J$ ($J_{\mathrm{up}}\geq$14) emission resembles that of the
elongated 160~$\mu$m continuum emission detected with \textit{Herschel}, 
but its peaks are offset from the positions of the dense cores. 
The [\ion{O}{i}] lines at 63.2~$\mu$m and 145.5~$\mu$m follow a similar pattern, but  their peaks
are found closer  
to the positions of the cores. The CO transitions from $J=14-13$ to $J=16-15$ are detected throughout
the cluster and show a median rotational temperature of 170$\pm$30~K on Boltzmann diagrams. 
Comparisons to other protostars observed with \textit{Herschel} show a good agreement with
intermediate-mass sources in the inner Galaxy. 
Assuming an origin of the [\ion{O}{i}] and high$-J$ CO emission in UV-irradiated $C-$shocks, we
obtained pre-shock H$_2$ number densities of 10$^4$-10$^{5}$~cm$^{-3}$ and UV radiation field
strengths of 0.1-10~Habing fields ($G_\mathrm{0}$). 
}
{Far-IR line observations reveal ongoing star formation in Gy~3--7, dominated by intermediate-mass
Class 0/I young stellar objects. The ratio of molecular-to-atomic far-IR line emission shows a
decreasing trend with bolometric luminosities of the protostars. However, it does not indicate that
the low-metallicity has an impact on the line cooling in Gy~3--7.}
\maketitle

\section{Introduction}

During the earliest stages of star formation, the gravitational collapse of dense cores is
accompanied by an ejection of bipolar jets originating from the resulting protostars, which may
alter the physical conditions and chemistry of their environment, even on clump scales
\citep{arce07,frank14}. Non-dissociative shock waves develop as the jets (and winds) interact with
the surrounding medium \citep{kn96,fl12} and heat up the gas up to typically $\sim$300~K
\citep{karska18,yang18}. Additionally, ultraviolet (UV) photons contribute to the gas heating and
influence the chemical composition of the low- to high-mass protostars' envelopes
\citep{bru09,vi12}. Similarly to some pc-scale outflows, UV photons may operate over a significant
fraction of low-mass star-forming clumps and clusters \citep{mirocha21}. The cooling of the gas,
which in the case of embedded objects is dominated by line emission in the far-infrared (FIR) and
(sub)millimeter domains, provides important constraints on the heating mechanisms and observations
of these cooling lines allow us to constrain gas temperatures, densities, and UV fields
\citep{GL78,Hol89}. 

Recent observations with the \textit{Herschel} Space Observatory
\citep{Pi10}\footnote{\textit{Herschel} 
was an ESA space observatory with science instruments provided by European-led Principal
Investigator consortia and with important participation from NASA.} targeted the main gas cooling
lines toward a significant sample of protostars spanning a broad range of masses \citep{vD21}. In
particular, the Photodetector Array Camera and Spectrometer \citep[PACS;][]{pacs} provided
detections of high$-J$ CO ($J_\mathrm{up}\geq$14), H$_2$O, and OH lines, as well as forbidden transitions
of [\ion{O}{i}] and [\ion{C}{ii}], all of them being important diagnostic tools in molecular clouds.
Among the key findings with PACS are the following: (i) the presence of ubiquitous "warm" gas
($\sim$300~K) associated with low- to high-mass protostars
\citep{green2013,manoj2013,karska13,karska14,matuszak2015}; (ii) the detection of a plethora of
high$-J$ CO (up to $49-48$; $E_\mathrm{u}$ $\sim$6700~K) and H$_2$O (up to $E_\mathrm{u}$
$\sim$1500~K) lines tracing the \lq" hot" gas component \citep{her12,goi12}, with a median
temperature of $\sim$720~K \citep{karska18}; (iii) the identification of the origin of the FIR
line emission in UV-irradiated non-dissociative shocks extending along the outflows
\citep{karska14b,Ben16, kri17co,karska18}; and (iv) the recognition of the dominant role of CO and
H$_2$O in the gas cooling budget of low-mass (LM) protostars \citep{karska13,karska18} as well as CO and
[\ion{O}{i}] as coolants of high-mass (HM) protostars \citep{karska14}. The above observations
have provided a surprisingly uniform picture of the FIR line emission from deeply-embedded protostars,
but targeted only on relatively nearby regions \citep[$d$<450~pc in case of LM objects][]{vD21}.

\begin{figure*}
\centering
\includegraphics[width=\linewidth]{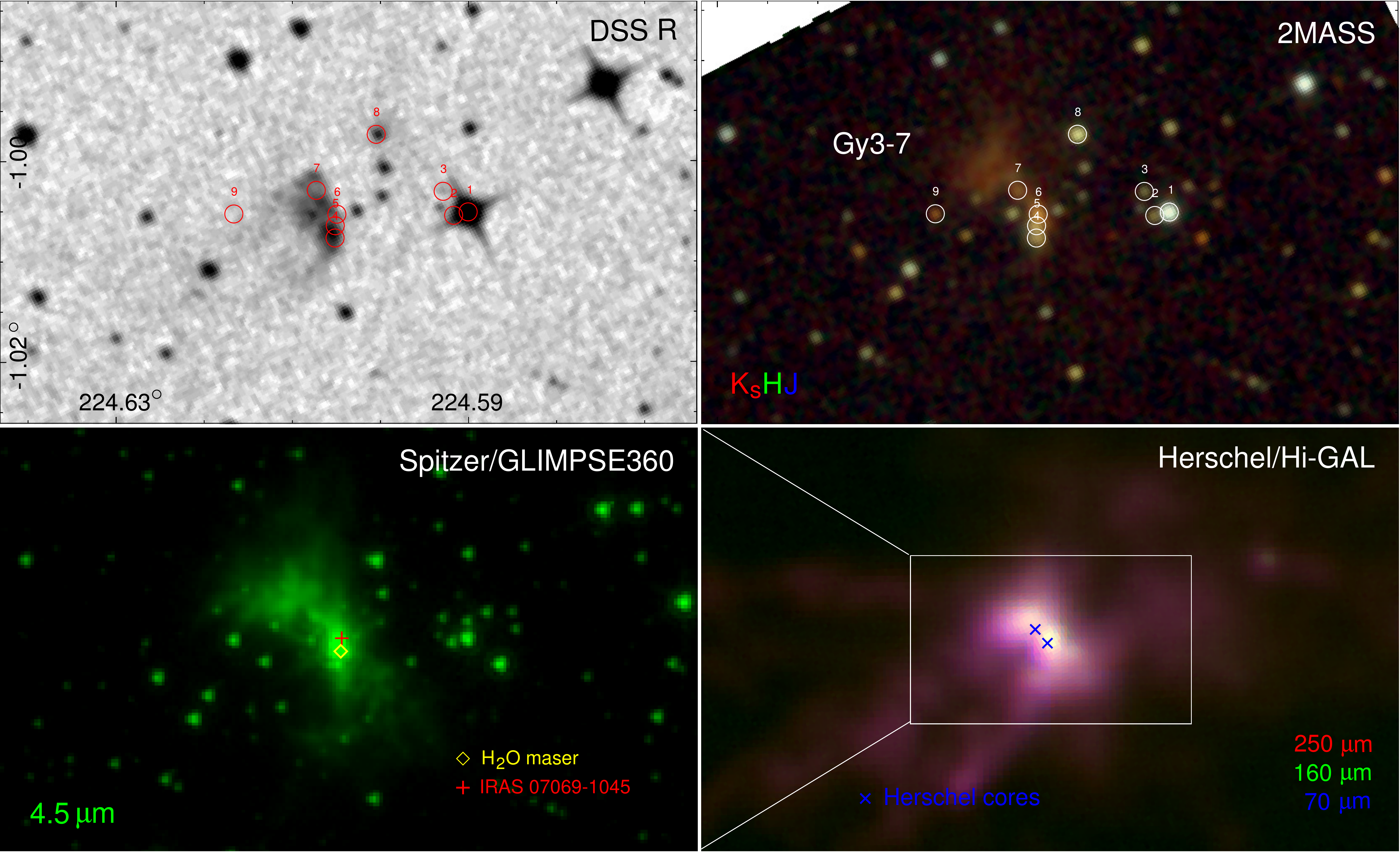}
\caption{Digital Sky Survey (DSS) R (top left), 2MASS composite image using the J, H, and
K$_\mathrm{s}$ filters (top right), \textit{Spitzer}/GLIMPSE360 4.5~$\mu$m 
(\citealt{sewilo19}; bottom left), and \textit{Herschel}/Hi-GAL composite image at 70, 160, and
250~$\mu$m (\citealt{Elia13}; bottom right) of the Gy~3--7 cluster. Circles in the top panels show
the positions of YSO candidates from \citet{tapia97}. Yellow diamond and red cross in the bottom-left panel show the position of the H$_2$O maser \citep{Urquhart11} and the IRAS source
at the south-west side (IRAS 07069--1045), respectively. The blue $"\times"$ symbols in the bottom
right panel show the positions of dense cores as traced by the H$_2$ column density \citep{Elia13}. Dense core in the west corresponds to IRAS 07069--1045.}
\label{fig:multi}
\end{figure*}

Observations of the Large and Small Magellanic Clouds (LMC and SMC) with \textit{Herschel} have shed some
light on the FIR line cooling from protostars in a significantly different, low-metallicity
environment ($Z$ of 0.2--0.5~$Z_{\odot}$; \citealt{rd92}). The Spectral and Photometric Imaging
Receiver \citep[SPIRE, ][]{griffin10} and PACS provided detections of CO (up to $J$=14-13), H$_2$O,
OH, and [\ion{O}{i}], and [\ion{C}{ii}] lines, facilitating comparisons with Galactic sources
\citep{Oliveira19}. The CO emission toward 15 sources in the LMC showed two relatively cool gas
components, with temperatures of $\sim$40 and $\sim$120~K, consistent with Galactic measurements
using SPIRE \citep{whi10,jim17,yang18}. The line cooling budget of
protostars in the Magellanic Clouds is dominated by [\ion{O}{i}] and [\ion{C}{ii}] emission, with an
increasing contribution of CO as the metallicity increases from the SMC to LMC and to the Galactic
young stellar objects \citep[YSOs,][]{karska14,Oliveira19}. The low fraction of CO line cooling is
interpreted as a metallicity effect: the combined result of a reduced carbon abundance and higher
grain temperatures due to a lower shielding from UV photons \citep{Oliveira19}. 

The outer parts of our Galaxy provide an alternative site for testing the impact of metallicity on
the FIR gas properties of protostars. Due to the negative-metallicity gradient, the abundances of
dust and molecules decrease in the outer Galaxy \citep{sodroski1997}. The metallicity affects the
gas and dust cooling budget of molecular clouds and results in lower CO rotational temperatures,
\Trot~\citep{roman2010}. Despite the overall decreasing trend of the mass surface density of
molecular clouds in the outer Galaxy \citep[for a review, see][]{hd15}, some star-forming regions
show a significant star-formation activity. For example, the CMa--$l224$ star-forming region at
a Galactocentric radius, $R_\mathrm{GC}$, of 9.1~kpc consists of $\sim$290 Class I/II YSOs, as
identified by \cite{sewilo19} using data from  GLIMPSE360: Completing the Spitzer Galactic Plane
Survey (PI: B. Whitney) and the  Herschel infrared Galactic Plane Survey (Hi-GAL;
\citealt{molinari2010}, see also \citealt{Elia13}). The expected metallicity of this region is
$\sim$0.55-0.73~$Z_\odot$, depending on the adopted O/H Galactocentric radial gradient
\citep{balser11,fernandez2017,esteban18}. 
\begin{table*}
\caption{Catalog of lines observed with FIFI-LS toward Gy~3--7 \label{table:sofialines}} 
\centering 
\begin{tabular}{c c r r r r r r}
\hline \hline 
Line & Transition & \multicolumn{1}{c}{$E_\mathrm{u}/k_\mathrm{B}$} &
\multicolumn{1}{c}{$A_{\rm{u}}$} & $g_{\rm{u}}$ & \multicolumn{1}{c}{$\lambda_\mathrm{lab}$} &
\multicolumn{1}{c}{Map size} & Beam \\
~ & & \multicolumn{1}{c}{(K)} & \multicolumn{1}{c}{(s$^{-1}$)} & & \multicolumn{1}{c}{($\mu$m)} &
($\arcsec\times\arcsec$)& \multicolumn{1}{c}{($\arcsec$)} \\ 
\hline
~CO & $14-13$ & 580.5 & 2.7(-4) & 29 & 185.99 & 60$\times$60& 18.3 \\
~CO& $16-15$ & 751.7 & 4.1(-4) & 33 & 162.81 & 60$\times$60& 16.1 \\
~CO & $17-16$ & 845.6 & 4.8(-4) & 35 & 153.27 & 60$\times$60& 15.2\\
~CO & $22-21$ & 1397.4 & 1.0(-3) & 45 & 118.58 & 30$\times$30& 11.9\\
~CO & $30-29$ & 2564.9 & 2.3(-3) & 61 & 87.19 & 30$\times$30& 8.6\\
~CO\tablefootmark{a} & 31-30 & 2735.3 & 2.5(-3) & 63 & 84.41 & 30$\times$30& 8.3\\
~OH\tablefootmark{b} & $\frac{1}{2}$,$\frac{1}{2}-\frac{3}{2}$,$\frac{1}{2}$ & 270.1 & 6.5(-2) & 4 &
163.12 & 60$\times$60 & 16.1\\
~OH\tablefootmark{c} & $\frac{1}{2}$,$\frac{1}{2}-\frac{3}{2}$,$\frac{3}{2}$ & 181.7 & 2.9(-2) & 3 &
79.18 & 30$\times$30 & 7.9\\
~OH\tablefootmark{a} & $\frac{7}{2}$,$\frac{3}{2}-\frac{5}{2}$,$\frac{3}{2}$ & 290.5 & 5.1(-1) & 9 &
84.42 & 30$\times$30& 8.3\\
~[\ion{O}{i}] & $^3$P$_0$--$^3$P$_{2}$ & 227.7 & 8.9(-5) & 3 & 63.18 & 60$\times$60 & 7.1 \\
~[\ion{O}{i}] & $^3$P$_1$--$^3$P$_{2}$ & 326.6 & 1.8(-5) & 1 & 145.52 &60$\times$60 & 14.4\\
~\ion{[C}{ii}] & $^2$P$_{3/2}$--$^2$P$_{1/2}$ & 91.2 & 2.4(-6) & 5 & 157.74 &90$\times$90 & 15.6\\
\hline 
\hline 
\end{tabular}
\begin{flushleft}
\tablefoot{Molecular data adopted from the Leiden Atomic and Molecular Database (LAMDA,
\citealt{schoier05}) and the JPL database \citep{Pic98}.  
\tablefoottext{a}{CO 31-30 and OH line at 84.42 $\mu$m are blended.}
\tablefoottext{b}{OH line at 163.12 $\mu$m is not detected.}
\tablefoottext{c}{The 79.18 $\mu$m is detected at the band edge, and its line fluxes cannot be
securely measured due to the lack of a baseline.}
} 
\end{flushleft}
\end{table*}
\begin{figure*}[tb]\centering
\includegraphics[height=0.34\linewidth]{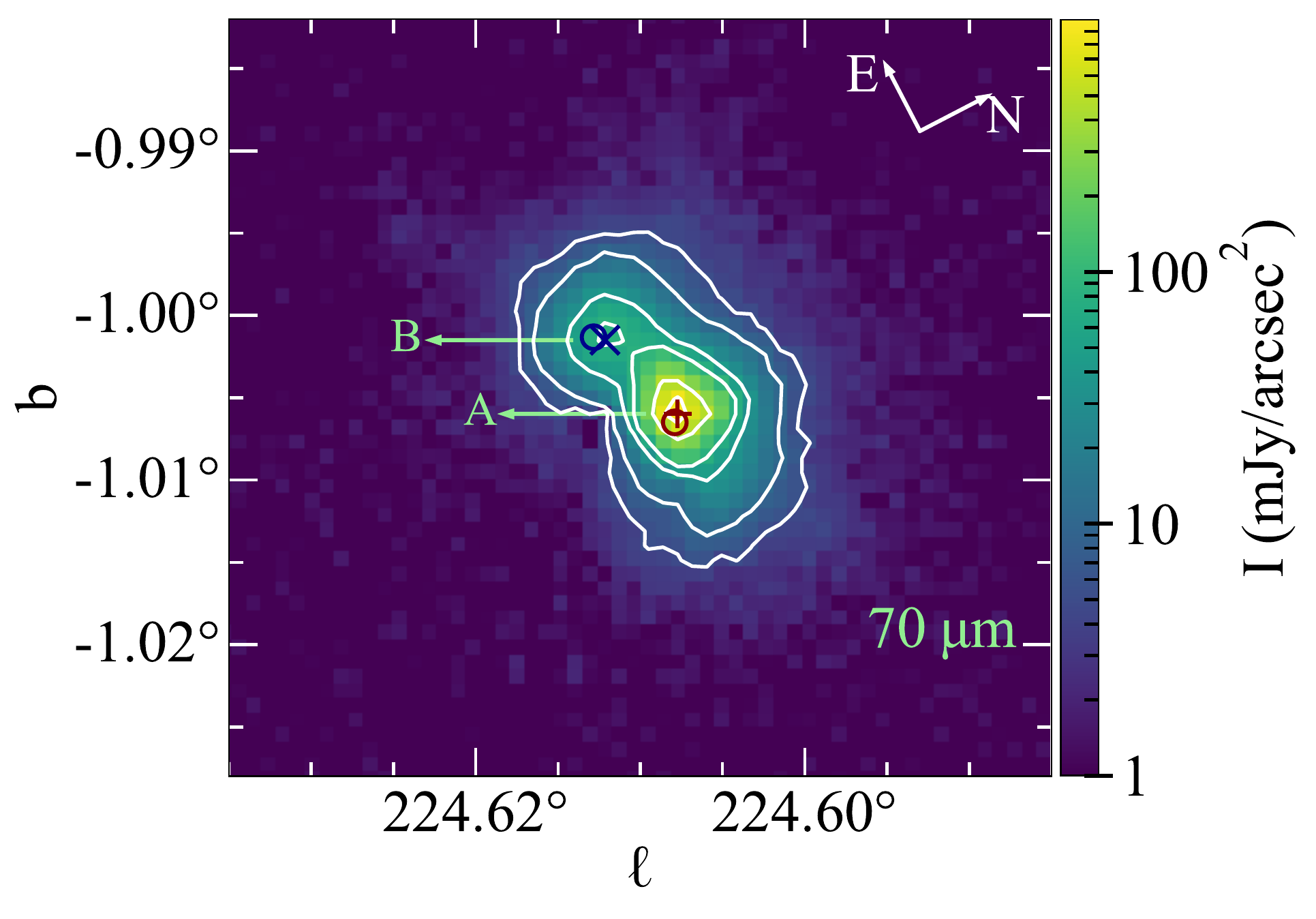}
\includegraphics[height=0.34\linewidth]{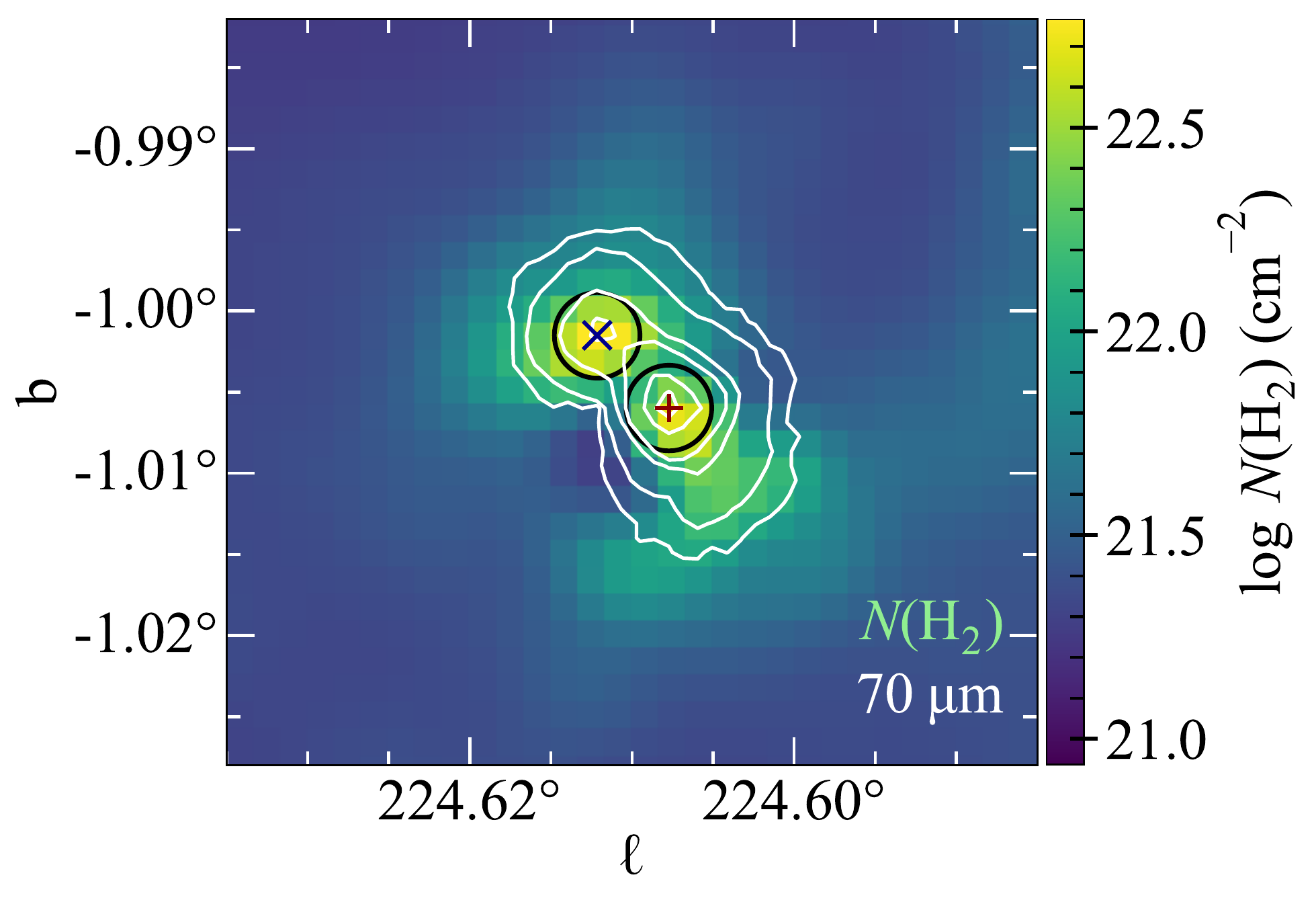}
\caption{Distribution of dust continuum emission toward Gy~3--7. Left: Continuum map of Gy~3--7
at 70 $\mu$m observed with \textit{Herschel}/PACS. Red $"+"$ and blue $"\times"$ symbols refer to
the 70 $\mu$m continuum peaks, adopted as the positions for the dense cores A and B in the
subsequent analysis. Circles show the positions of HIGALBM224.6079--1.0065 and
HIGALBM224.6128--1.0013 cores from the \textit{Herschel}/Hi-GAL catalog \citep{elia21}. Right: Map of the H$_2$ column density ($N_\mathrm{H_2}$). Black circles with the beam size of
20$\arcsec$ indicate the extract regions of the SOFIA FIFI-LS spectra toward the two dense cores A
and B (see more in Section~\ref{ssec:det}). White contours in each map show the continuum at 70
$\mu$m, with contour levels at 5, 10, 40, 80, 400, and 800~mJy/arcsec$^{2}$. 
}
\label{fig:tdust_nh2_ppmap}
\end{figure*}

Gy~3--7 is a deeply-embedded cluster with exceptionally bright FIR continuum emission, located at
the second-most massive filament in the CMa--$l224$ region \citep[][]{sewilo19} at a distance of
$\sim$1~kpc \citep[e.g.,][]{Lombardi11}. It is associated with IRAS~07069--1045, which was recognized early on
 as a star-forming region driving a CO outflow, with a bolometric luminosity
($L_{\rm{bol}}$) of 980~$L_{\rm{\odot}}$ \citep[assuming a distance of 1.4~kpc;][]{Wouterloot89}. 
The source was considered as a candidate HM star-forming region based on its position in the IRAS
color-color diagrams and the presence of dense gas traced by CS~$2-1$ \citep{WC89,Bronfman96}.
However, several attempts have failed to detect the CH$_3$OH maser emission, which is a common signature
of HM protostars \citep{menten92,szymczak}. Instead, H$_2$O maser and thermal NH$_3$ emission was
detected as part of the Red MSX Survey \citep[RMS; ][see Figure~\ref{fig:multi}]{Urquhart11}.
\begin{figure*}[htp]
\includegraphics[height = 0.34\textwidth]{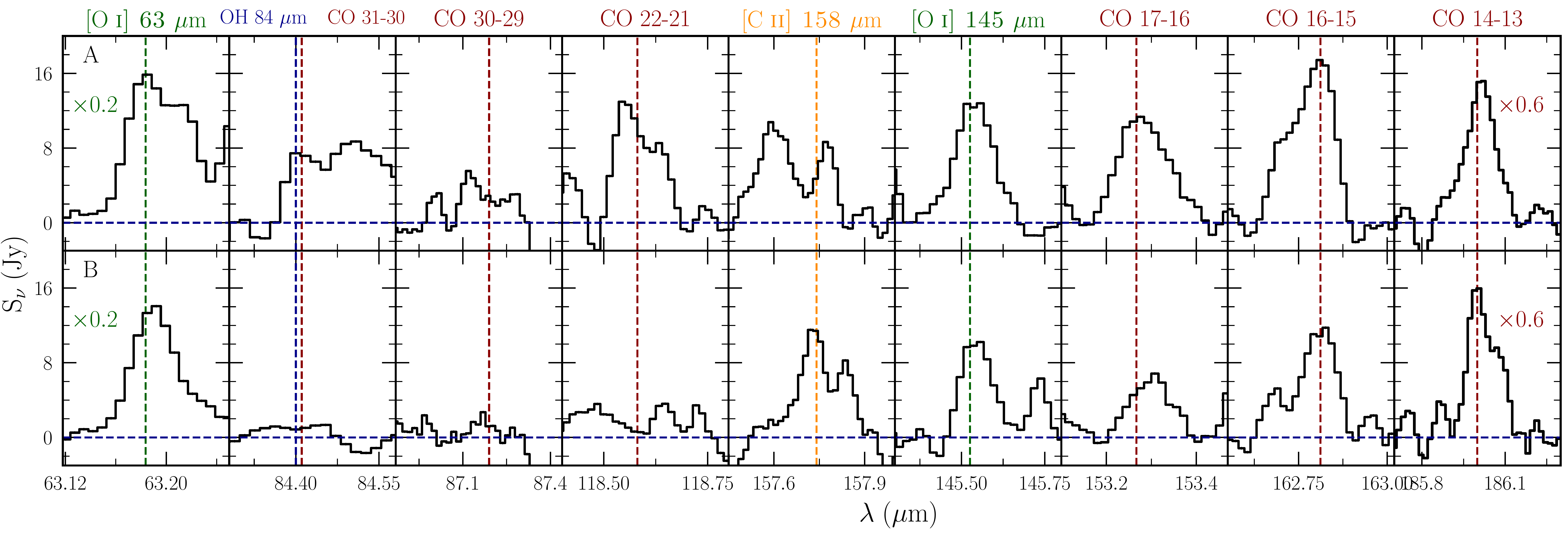}
\caption{SOFIA FIFI-LS continuum-subtracted spectra toward the two dense cores in Gy~3--7:
HIGALBM224.6079--1.0065 (source \lq \lq A'') and HIGALBM224.6128--1.0013 (source \lq \lq B''). The
emission is extracted within a beam size of 20$\arcsec$ indicated by black circles in the right
panel of Figure~\ref{fig:tdust_nh2_ppmap}. Vertical lines show the laboratory wavelengths of the
detected lines. Spectra of the [\ion{O}{i}] line at 63 $\mu$m are multiplied by a factor of 0.2 and
those of the CO~$14-13$ line by a factor of 0.6 to better illustrate the line detections.
}\label{fig:fifi_spectra}
\end{figure*}

Recent FIR observations with \textit{Herschel}/Hi-GAL spatially resolved two dense cores in
Gy~3--7 with $L_\mathrm{bol}$ of 75.9 and 324.2~$L_{\rm{\odot}}$, the latter corresponding to
IRAS~07069--1045 \citep[][see Figure~\ref{fig:multi}]{Elia13,elia21}. Near-IR observations revealed
an extended H$_2$ emission, which may arise from the jets from protostars \citep{Navarete15}.
Gy~3--7 contains several more evolved YSOs, with spectral types ranging from B1 to B5
\citep{gy12,tapia97} and it is cataloged as a young stellar cluster \citep{Soares02,Soares03,bica03}. 

In this paper, we investigate the FIR line emission toward Gy~3--7 obtained using the
Stratospheric Observatory for Infrared Astronomy (SOFIA) observations with the Field-Imaging
Far-Infrared Line Spectrometer \citep[FIFI-LS; ][]{fifi,fischer18}. We also consider the gas
densities and UV radiation fields in Gy~3--7, as well as the stellar content of the cluster. We examine
whether the rotational temperatures and the ratio of CO and [\ion{O}{i}] line luminosities in
Gy~3--7 are consistent with the picture of star formation in the inner Galaxy and/or the
low-metallicity environment of the Magellanic Clouds. Finally, we also present the results of our
search for water masers with the Toruń 32-m radio telescope (RT4). 

The paper is organized as follows. Section~\ref{sec:obs} describes the observations with SOFIA and
RT4. In Section~\ref{sec:results}, we present line and continuum maps, as well as the line profiles at
selected positions of interest. Section~\ref{sec:Analysis} shows the analysis of the results and
Section~\ref{sec:dis} provides the discussion of the results in the context of previous studies. We
provide our summary and conclusions in Section~\ref{conclusions}.

\section{Observations}\label{sec:obs}
\subsection{SOFIA FIFI-LS}\label{obs:fifi}

SOFIA/FIFI-LS observations were collected in November 2019 as part of the SOFIA Cycle 7 (Project ID
$07_{-}0157$, PI: M. Kaźmierczak-Barthel). FIFI-LS is an integral field unit consisting of two
grating spectrometers with a spectral coverage ranging from 51 to 120~$\mu$m (blue) and from 115 to
200~$\mu$m (red), facilitating simultaneous observations of selected wavelength intervals
(0.3-0.9~$\mu$m) in both channels (SOFIA Observer's Handbook for Cycle
10\footnote{\url{https://www-sofia.atlassian.net/wiki/spaces/OHFC1/overview}}). The spectral
resolution, $R$, ranges from $\sim$500 to 2000 and increases with wavelength for a given grating
order. The corresponding velocity resolution of $\sim$150 to 600~km~s$^{-1}$ provides unresolved
spectral profiles of all the FIR lines, including H$_2$O \citep{kri12,mot17}, CO \citep{kri17co}, and
[\ion{O}{i}] \citep{kri17oxy,yang22}. 
\begin{figure*}[ht!]
\centering
\includegraphics[width = 0.47\textwidth]{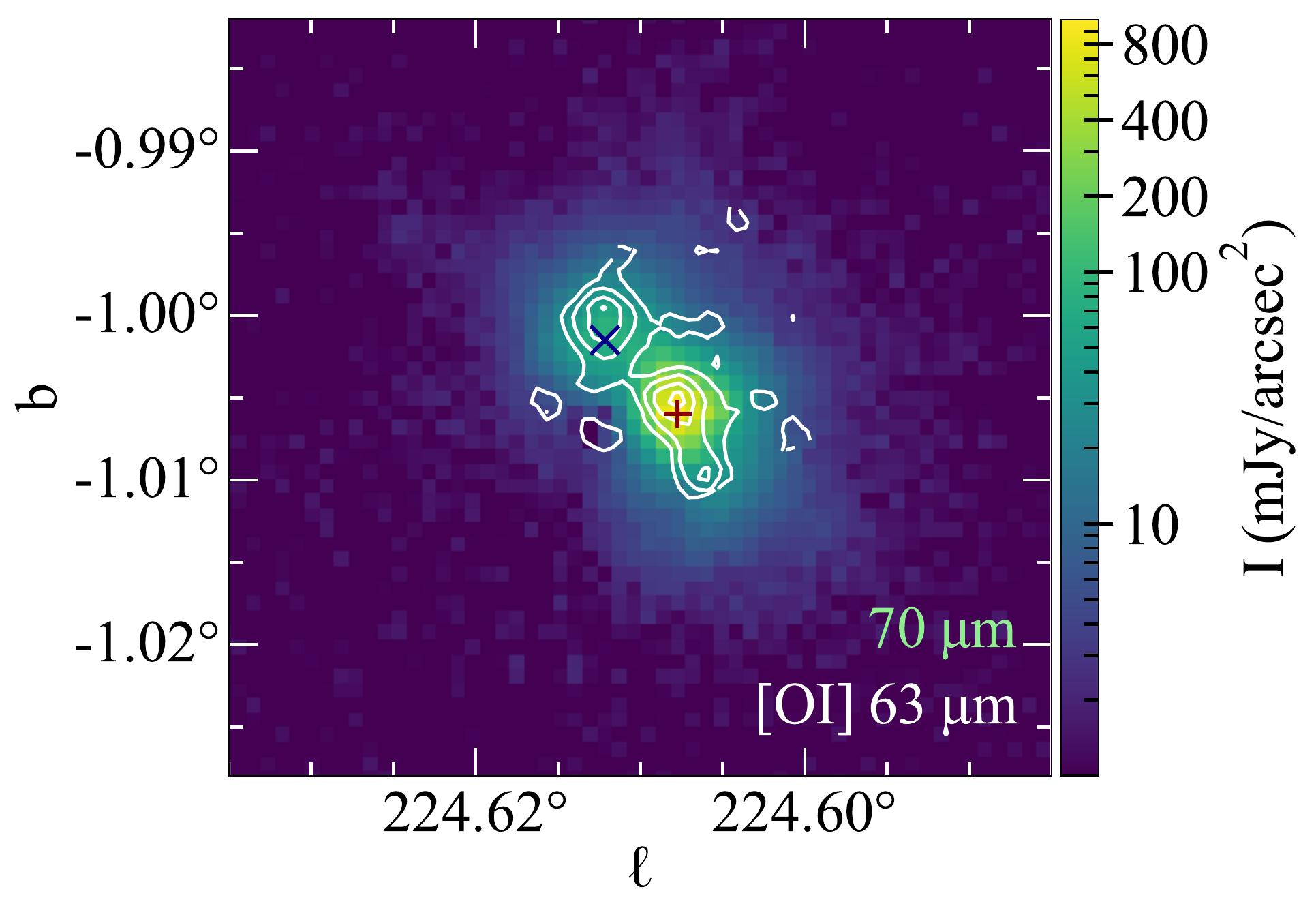}
\includegraphics[width = 0.47\textwidth]{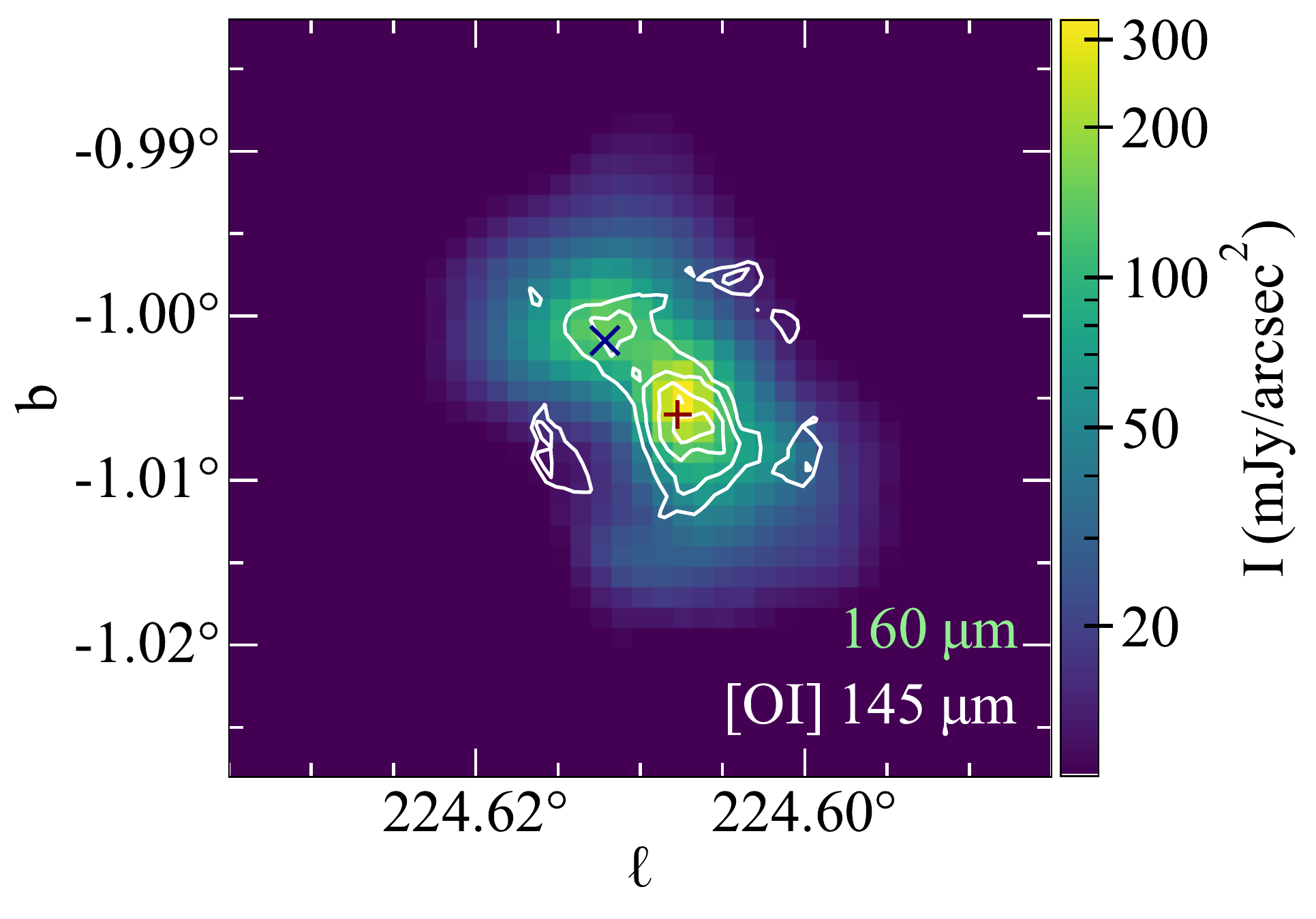}
\includegraphics[width = 0.47\textwidth]{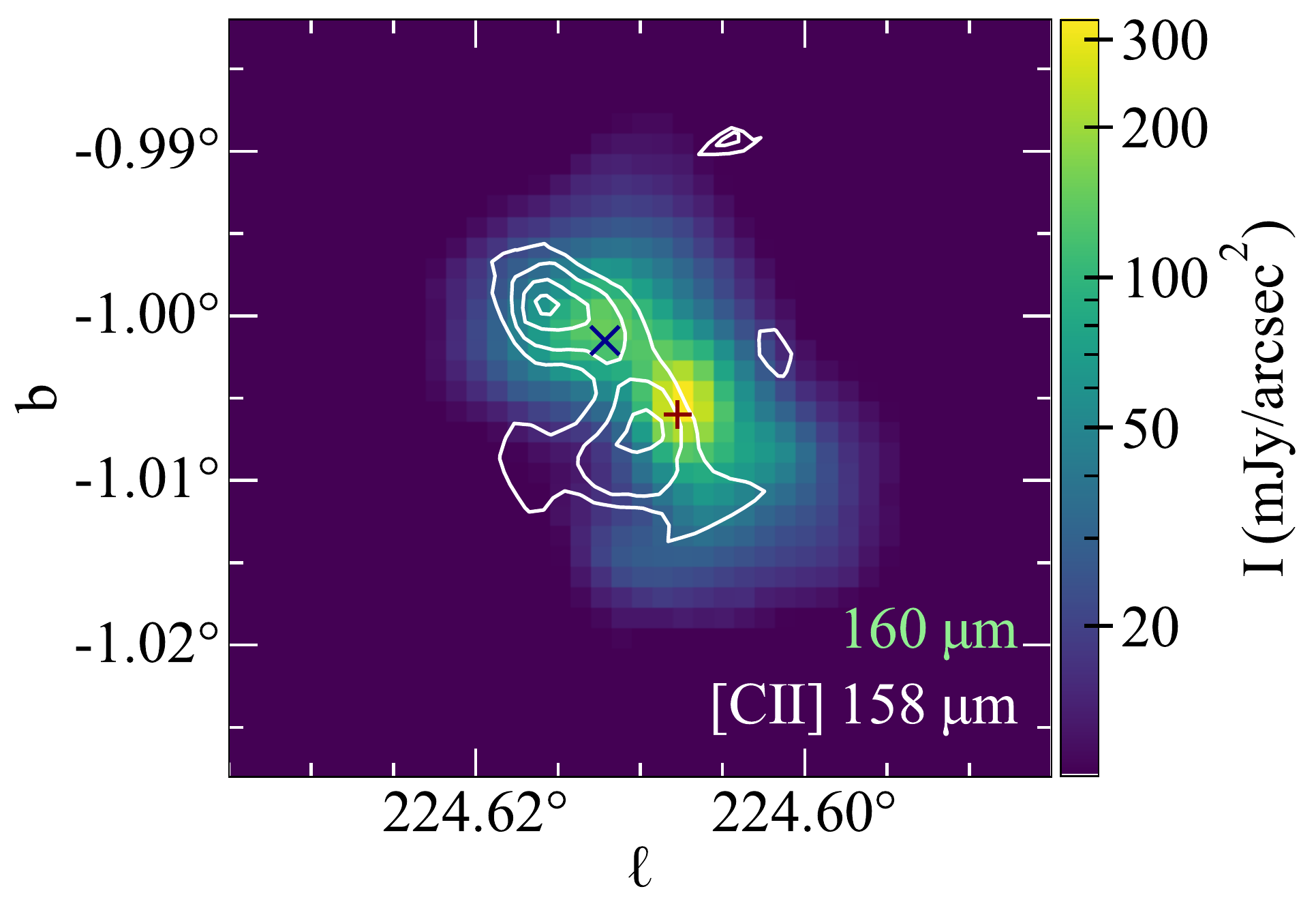}
\includegraphics[width = 0.47\textwidth]{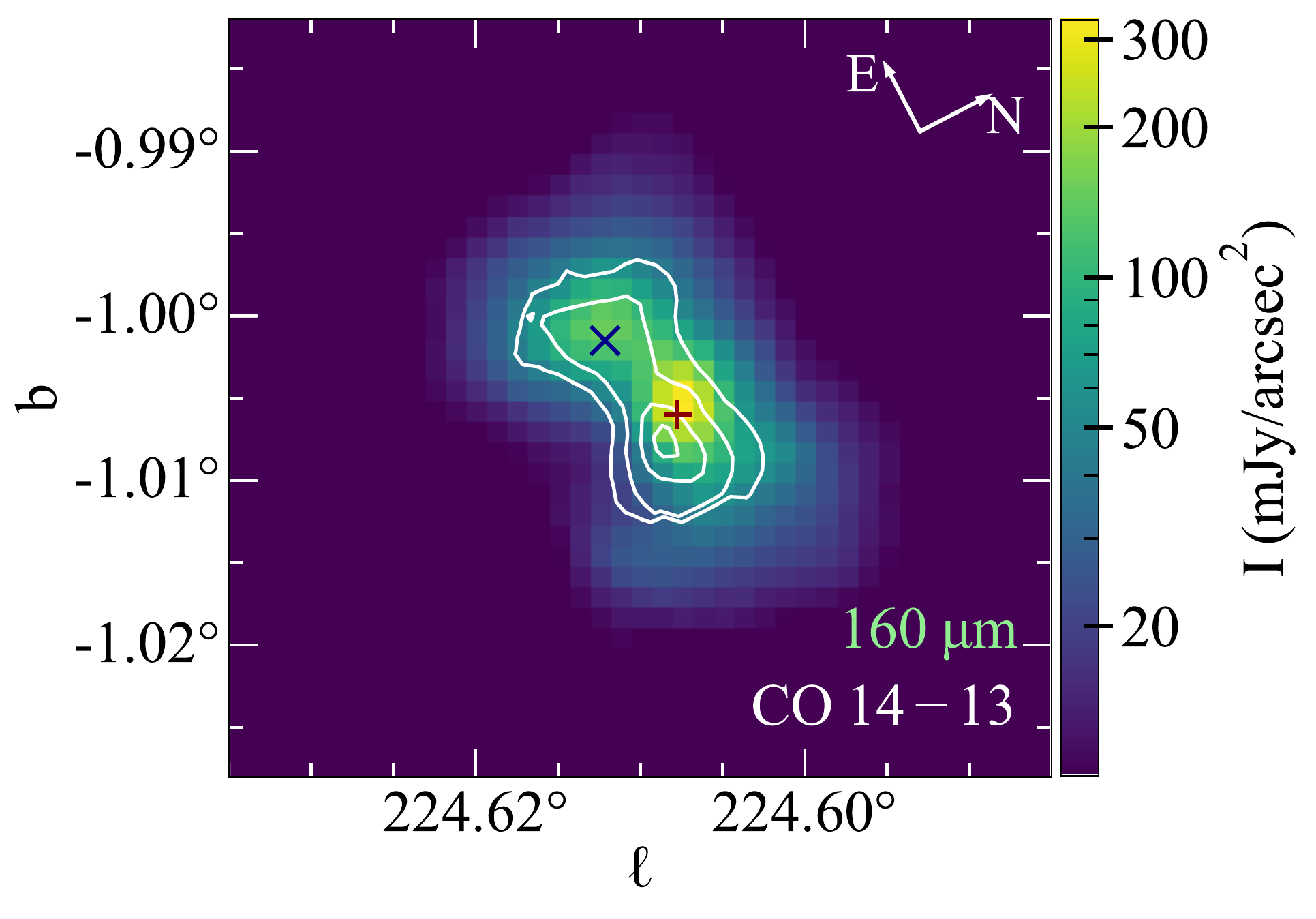}
\includegraphics[width = 0.47\textwidth]{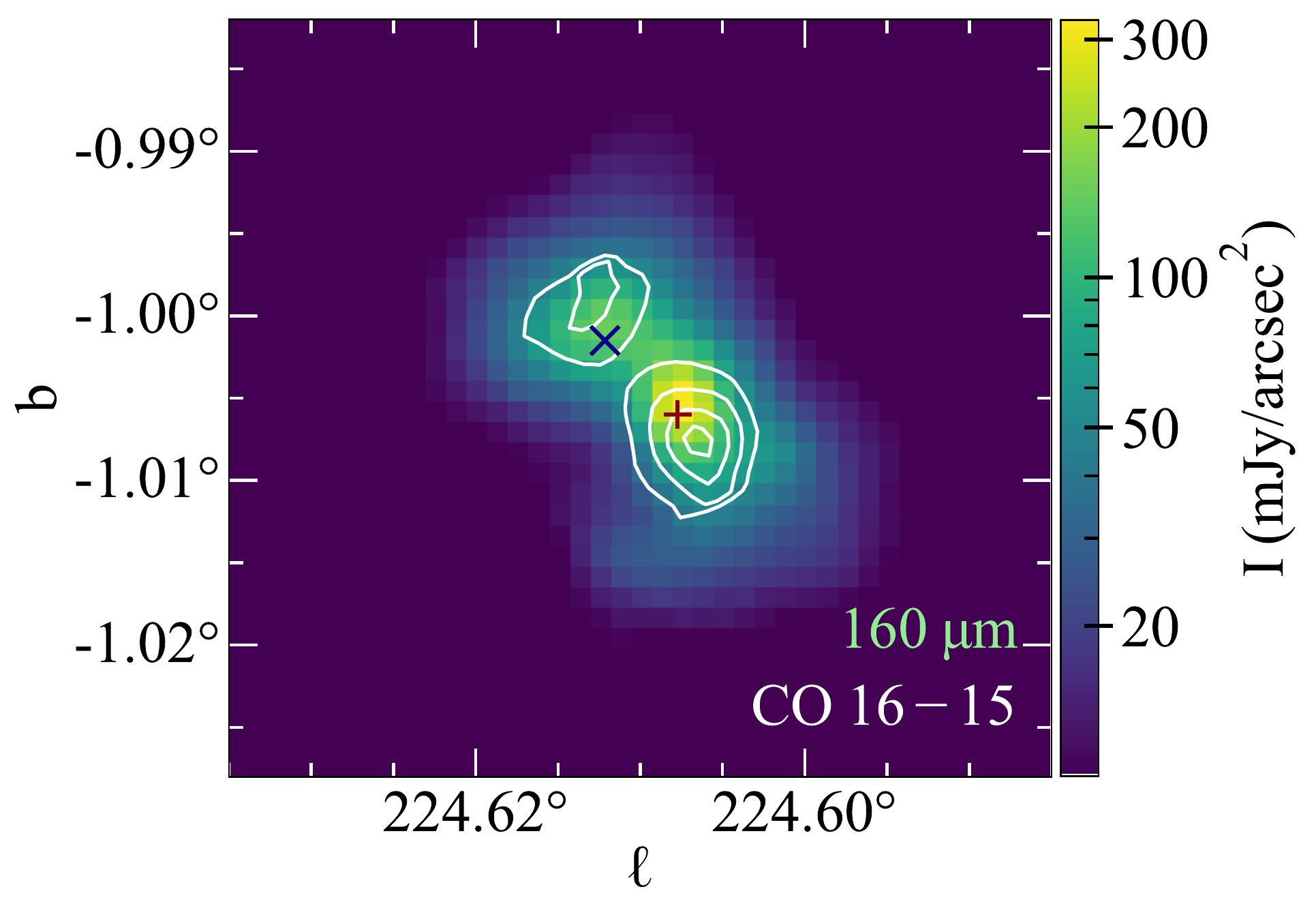}
\includegraphics[width = 0.47\textwidth]{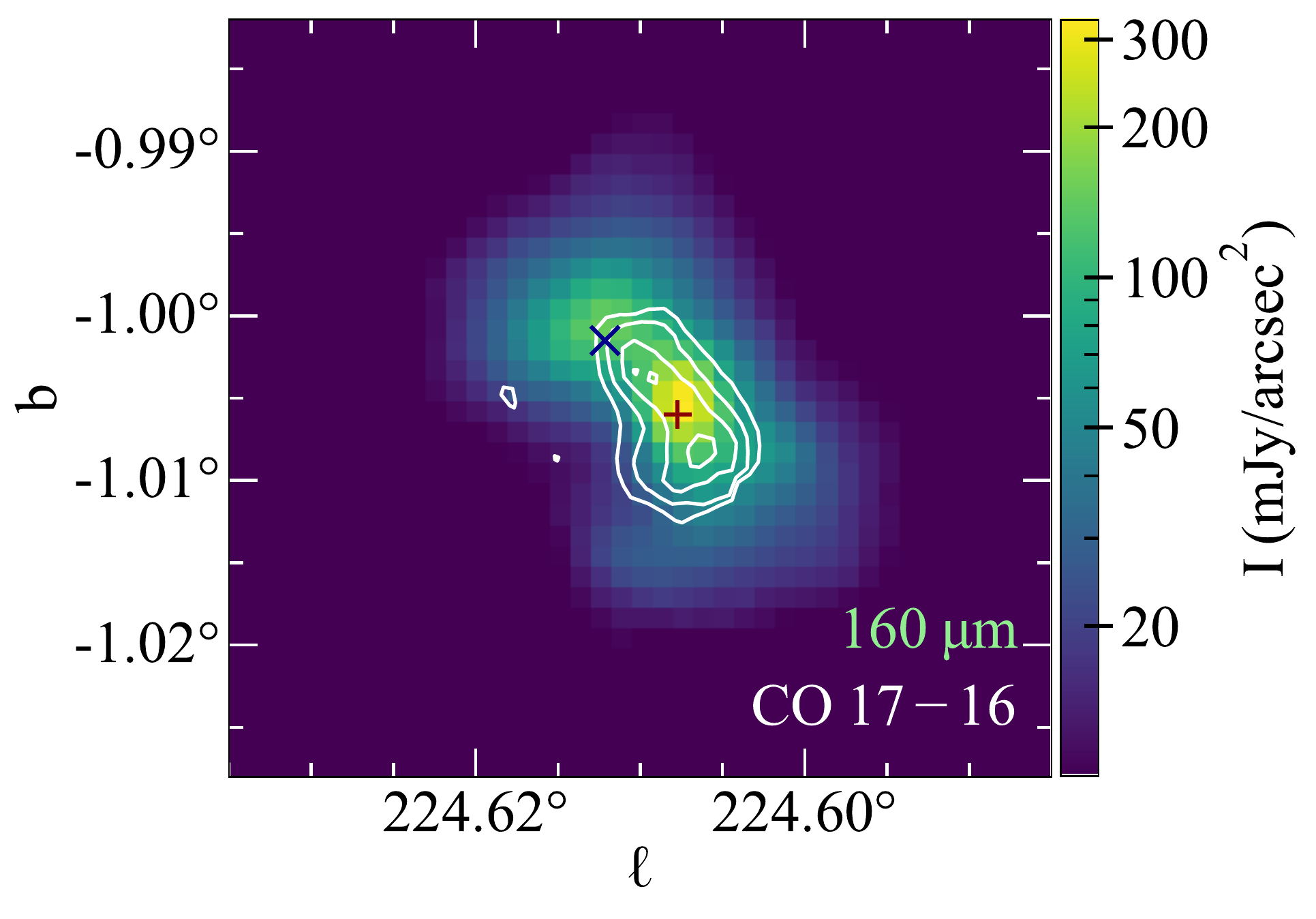}
\caption{FIFI-LS contour maps of the [\ion{O}{i}] lines at 63.2 and 145.5~$\mu$m, the [\ion{C}{ii}]
line at 157.7~$\mu$m, the CO lines with $J_\mathrm{up}$ = 14, 16, 17 at 186, 163, and 153~$\mu$m
respectively (white contours) on top of the continuum emission at 160~$\mu$m (at 70~$\mu$m for the
[\ion{O}{i}] line at 63.2~$\mu$m) from \textit{Herschel}/PACS. The white contours show line emission
at 25\%, 50\%, 75\%, and 95\% of the corresponding line emission peak.
The $"+"$ and $"\times"$ signs show the positions of the dense cores A and B. }
\label{fig:far-IRoverlaid_maps}
\end{figure*}
\begin{figure*}[ht!]
\centering
\includegraphics[width = 0.47\textwidth]{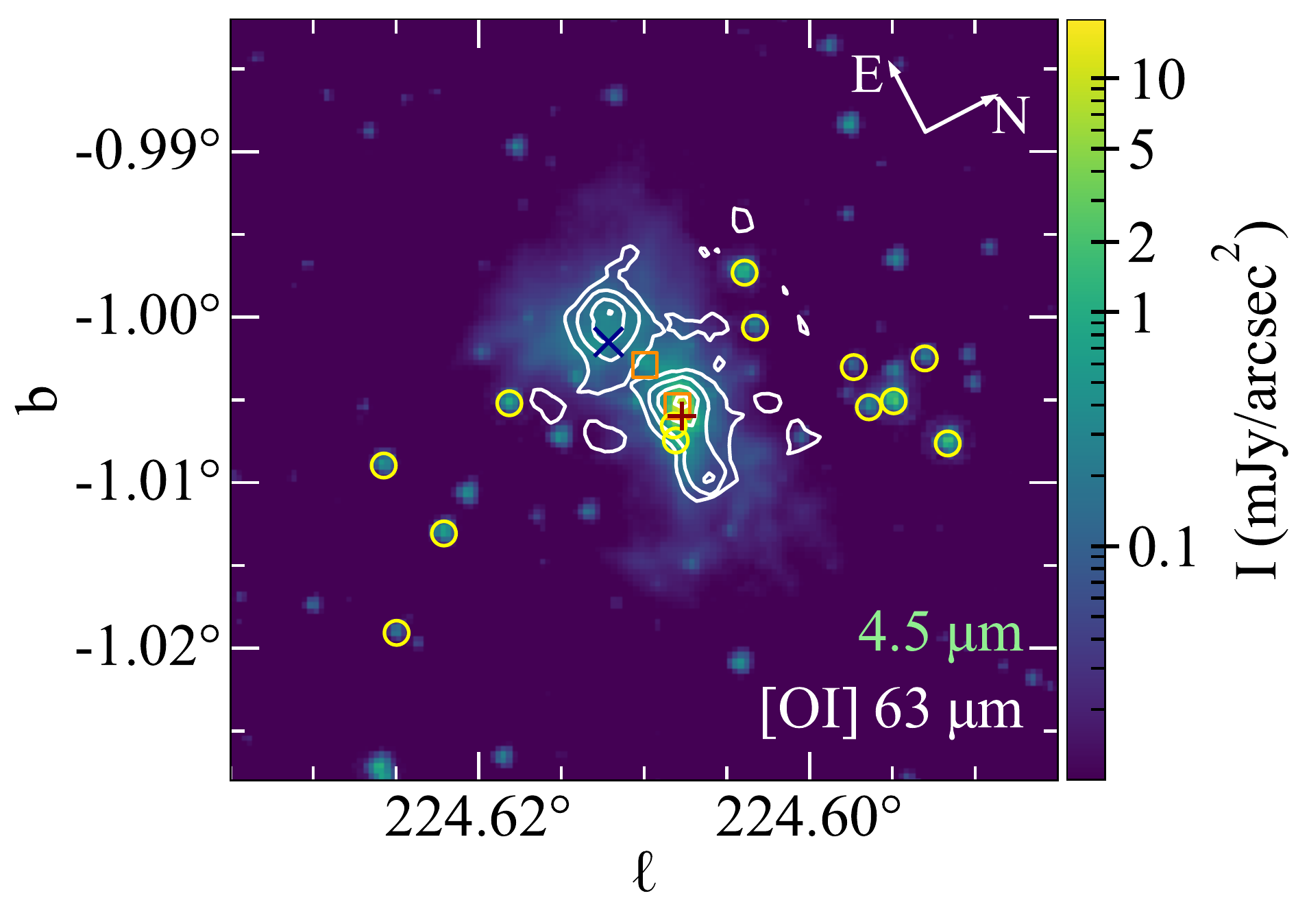}
\includegraphics[width = 0.47\textwidth]{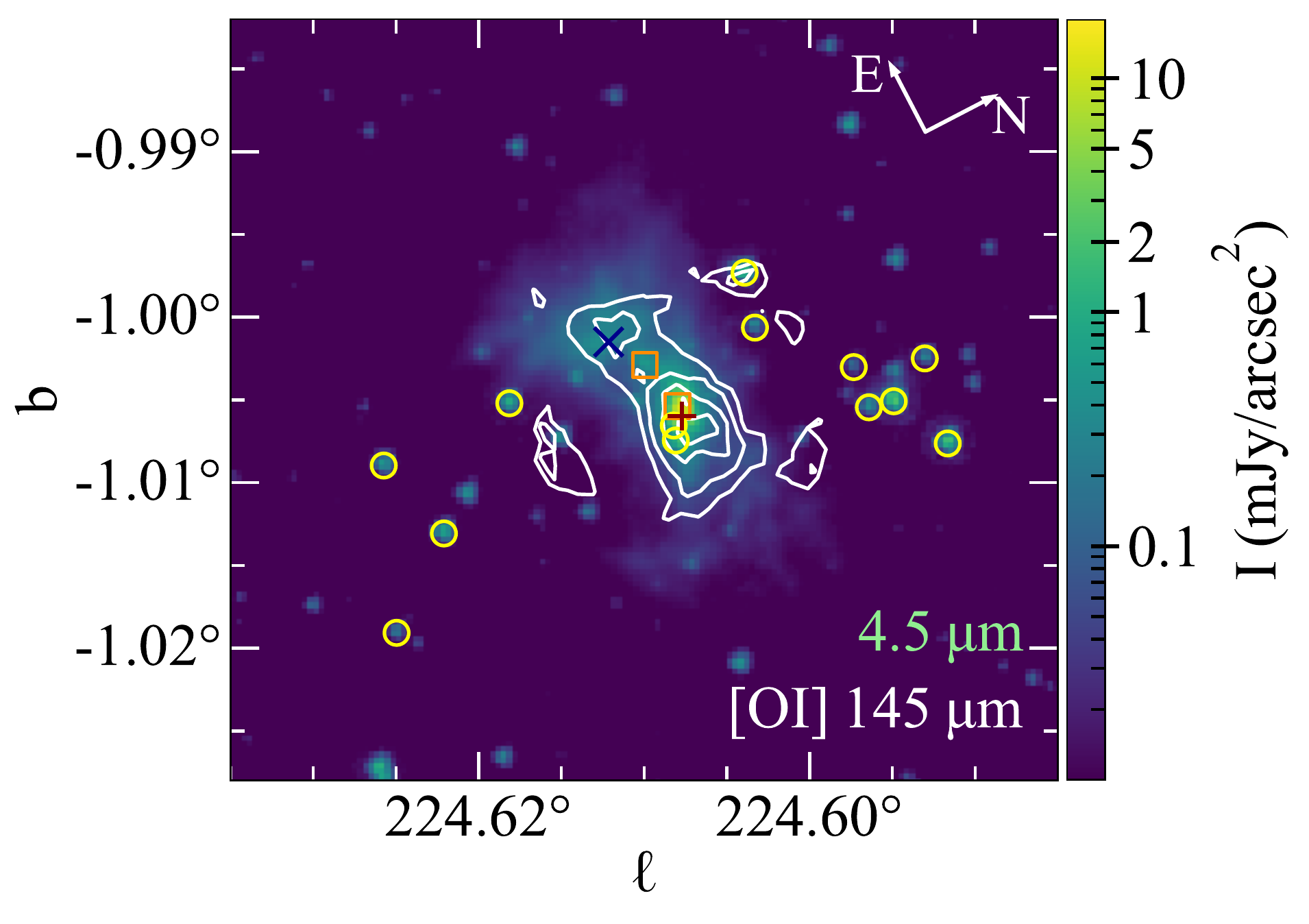}
\includegraphics[width = 0.47\textwidth]{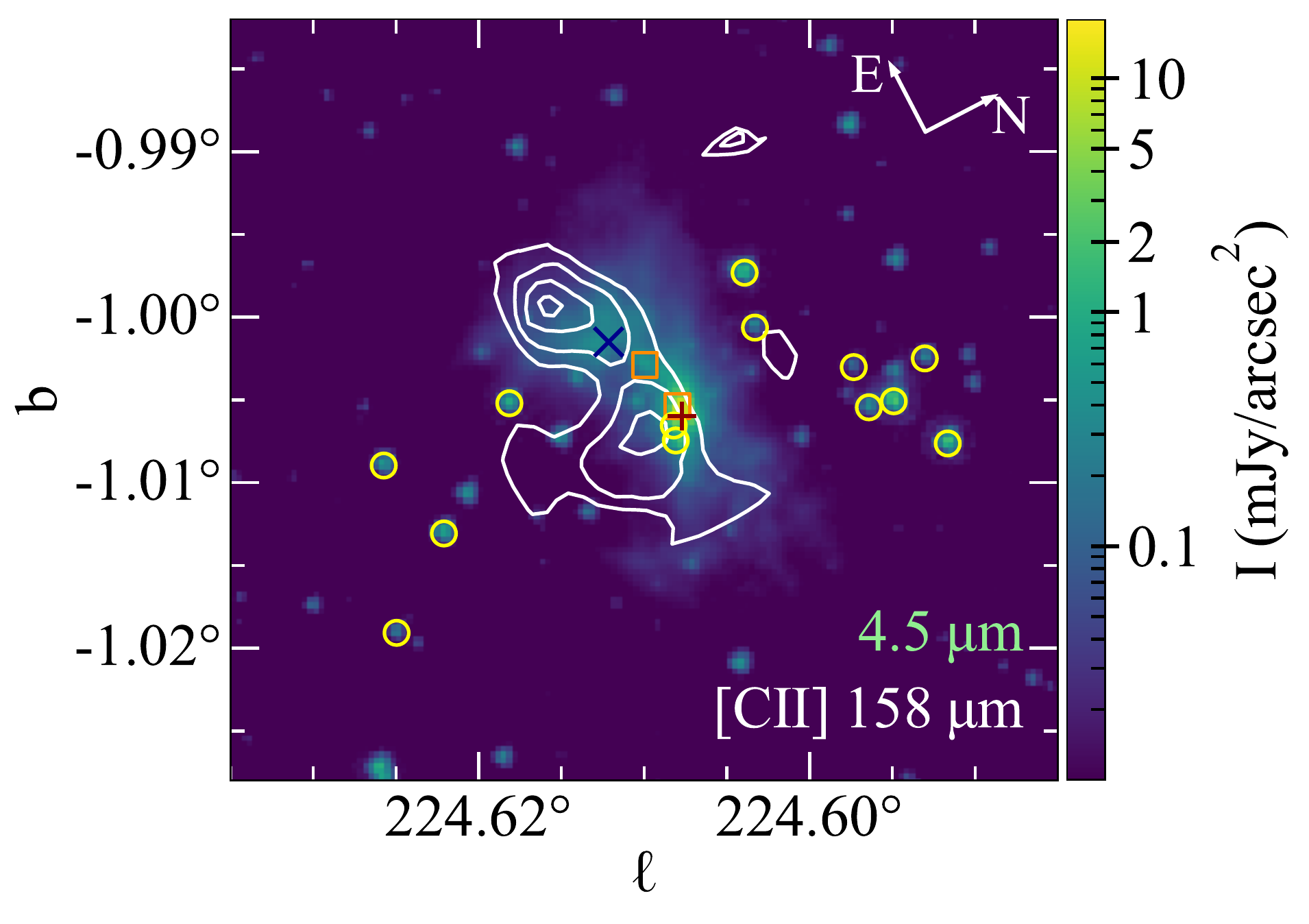}
\includegraphics[width = 0.47\textwidth]{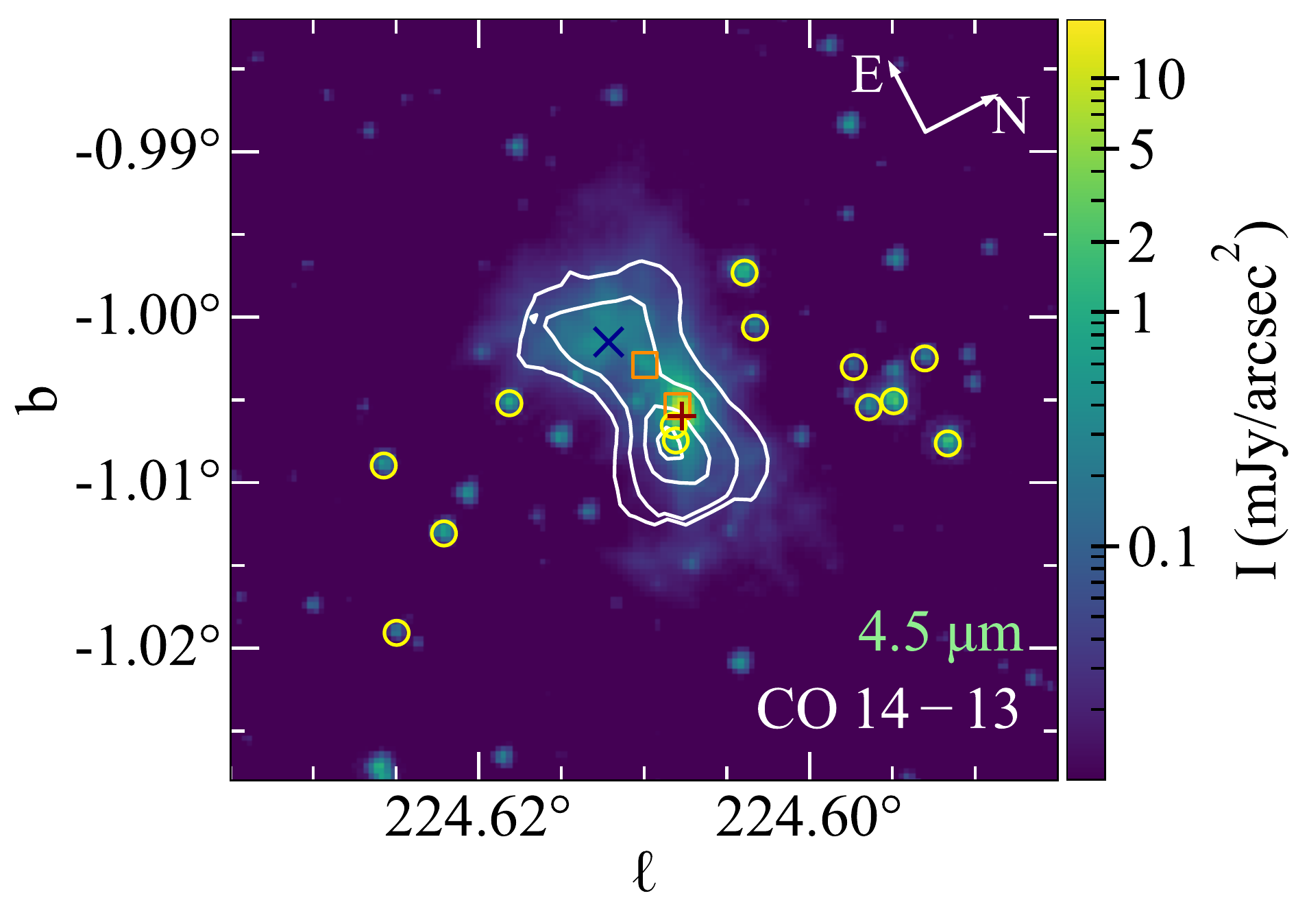}
\caption{FIFI-LS contour maps of the [\ion{O}{i}] lines at 63.2 and 145.5~$\mu$m, the [\ion{C}{ii}]
line at 157.7~$\mu$m, and the CO~$14-13$ line at 186~$\mu$m (white contours) on top of the continuum
emission at 4.5~$\mu$m from \textit{Spitzer}/IRAC. The contours show line emission at 25\%, 50\%,
75\%, and 95\% of the corresponding line emission peak. The "$+$" and "$\times$" signs show the
positions of the dense cores A and B, respectively. The orange squares show the positions 
of two YSO candidates with envelopes, and the yellow circles show 
the positions of the remaining YSOs (Section~\ref{ssec:sed_fit}).}
\label{fig:near-IRoverlaid_maps}
\end{figure*}
\begin{figure*}[htp]
\centering
\includegraphics[width = 0.49\textwidth]{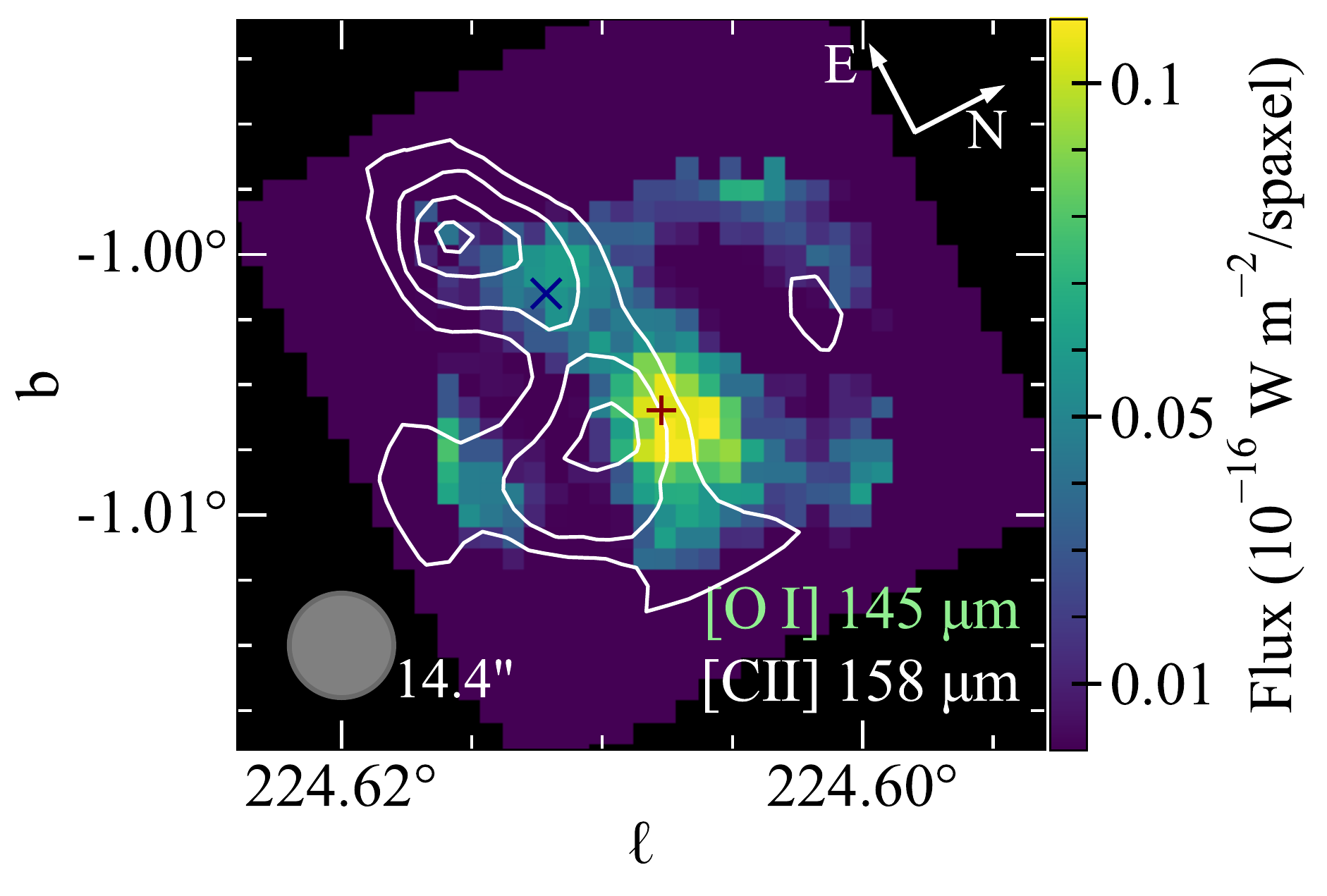}
\includegraphics[width = 0.49\textwidth]{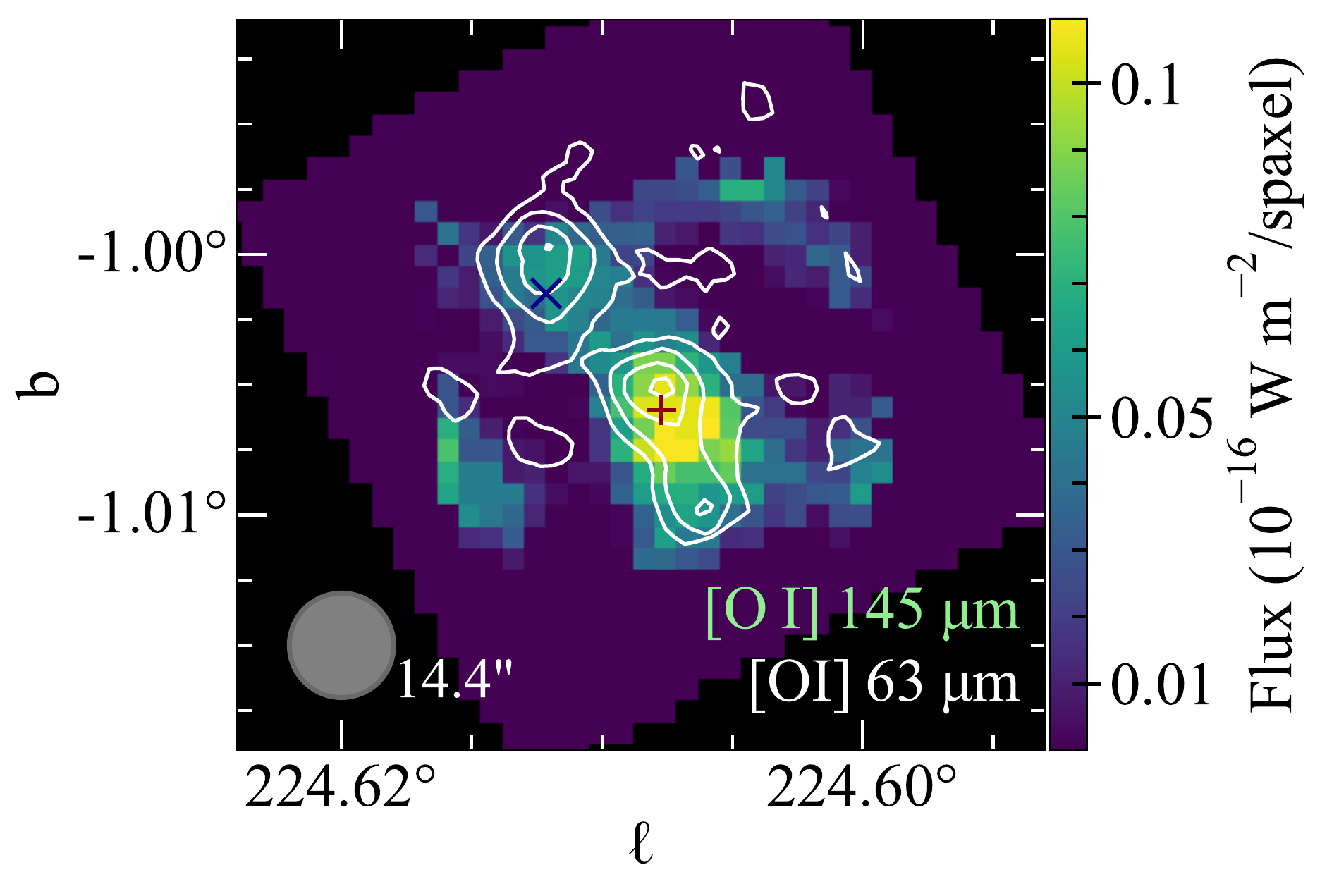}
\includegraphics[width = 0.49\textwidth]{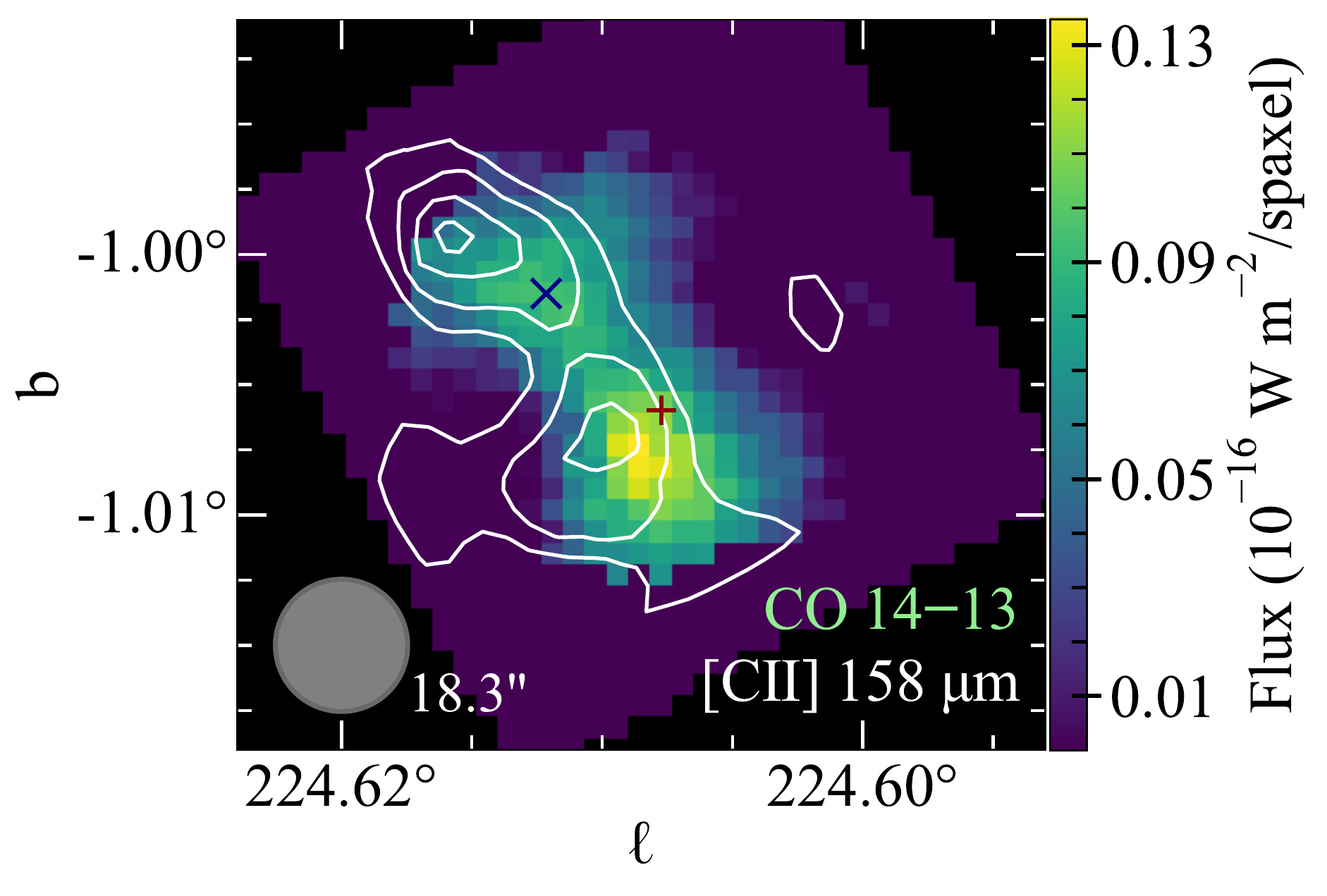}
\includegraphics[width = 0.49\textwidth]{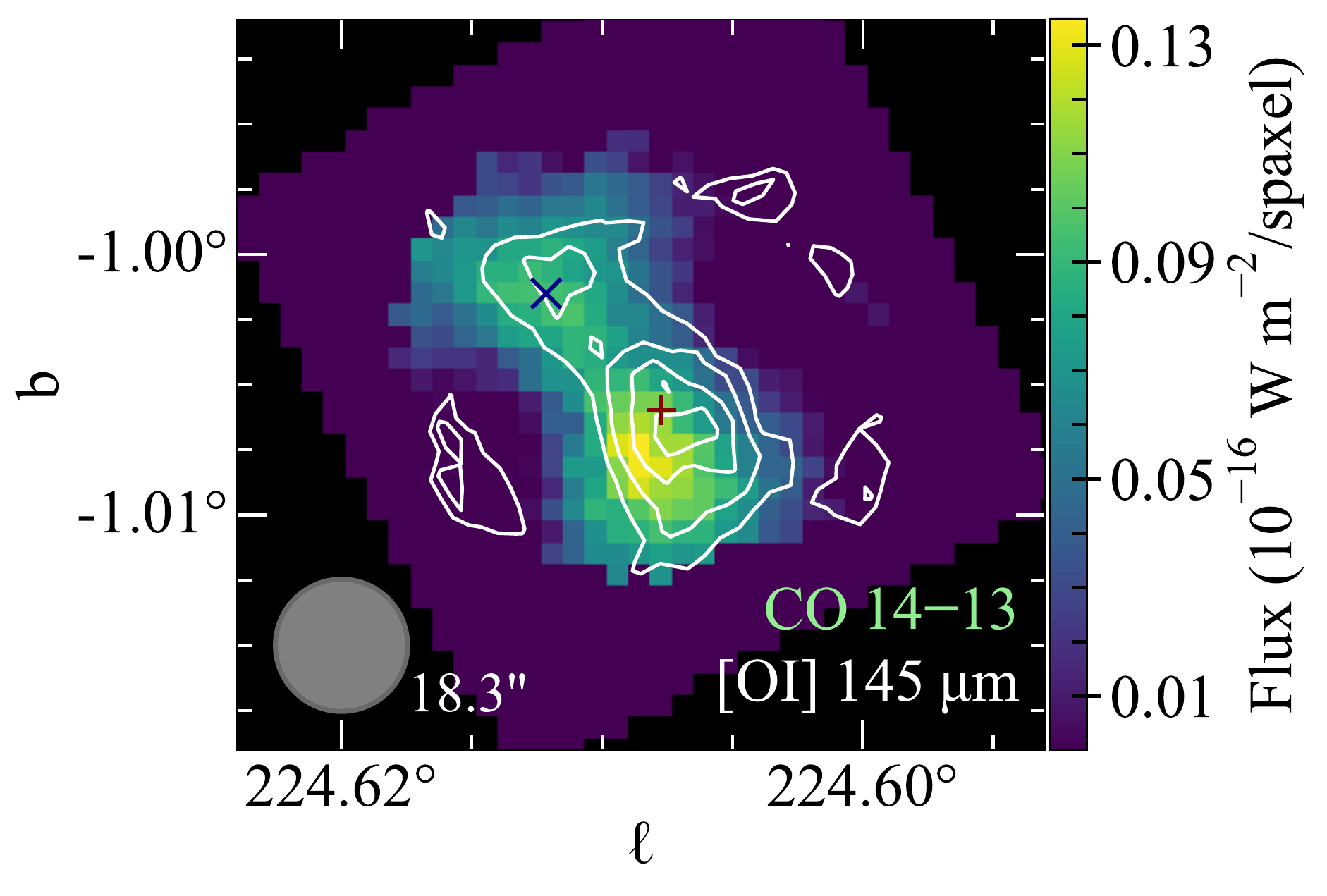}
\caption{Integrated intensity maps of selected pairs of FIR lines from FIFI-LS with the top line
shown in colors, and the bottom line in white contours. The contour levels are 25\%, 50\%, 75\%, and
95\% of the corresponding line emission peak. The "+" and "x" signs show the positions of the dense
cores A and B, respectively. The spaxel size in the emission map of the [\ion{O}{i}] line at
63~$\mu$m is $6\arcsec\times6\arcsec$ (blue channel) and in the map of other lines (CO~$14-13$,
[\ion{O}{i}] at 145~$\mu$m, and [\ion{C}{ii}] at 157.7~$\mu$m) is $12\arcsec\times12\arcsec$ (red
channel). Gray circles show the beam size for each color map.}\label{fig:sofia_overlaid_maps}
\end{figure*}

The FIFI-LS detector is composed of 5$\times$5 spatial pixels (hereafter \textit{spaxels}) with the
centers offset by 10$\arcsec$, similar to the PACS spectrometer on \textit{Herschel}.
The spaxel size is $6\arcsec\times6\arcsec$ in the blue channel (field of view, FOV, of
30$\arcsec\times30\arcsec$) and $12\arcsec\times12\arcsec$ in the red channel (FOV of
$1\arcmin\times1\arcmin$), providing an improvement over PACS by matching the actual
wavelength-dependent beam sizes. The FIR sky background was subtracted by symmetric chopping
around the telescope's optical axis with a matched telescope nod. The 300$\arcsec$ throw in the
east-west direction was chosen to avoid contaminated regions. 

Table \ref{table:sofialines} shows the full catalog of lines targeted with FIFI-LS. The maps of the
[\ion{O}{i}] line at 63.2~$\mu$m and the [\ion{C}{ii}] line at 157.7~$\mu$m were obtained
simultaneously as $2\times2$ mosaics, with a FOV of $60\arcsec\times60\arcsec$ in the blue channel
and $90\arcsec\times90\arcsec$ in the red channel, during the wall-clock time of 51~minutes. The
maps of the remaining lines were collected as a single FOV with dithering. The total observing time
was 239~minutes ($\sim$4~hours). 

The data were reduced with the SOFIA FIFI-LS pipeline \citep{Vacca2020}, which contains all the
necessary calibrations and flat-field corrections. For the telluric correction in the pipeline
reduction, we used water vapor values obtained with the method described in \citet{fischerPWV} and
\citet{iserlohePWV}. Then, the IDL-based software FLUXER
v.2.78\footnote{\url{http://ciserlohe.de/fluxer/fluxer.html} } was used to produce the continuum and
line emission maps. The continuum was fitted as a 0th order polynomial in spectral areas free of
line emission and the spectral line was fitted with a Gaussian. The selection of channels for the
baseline subtraction, the observed wavelength of the targeted line and its width were obtained over
the area with strong line detections, separately for each species. Subsequently, these values were
adopted for the entire datacube to obtain the integrated line fluxes for the entire map. Further
processing of the maps was performed with Python.

The overall calibration accuracy can be assumed to be within 20\%, including 
a 10\% calibration accuracy of the instrument, and the additional
uncertainty due to telluric effect, which is assumed to be well under 10\%.
For the [\ion{O}{i}] line at 63.2~$\mu$m, located close to the water absorption line, 
the water vapor overburden was determined between 3.5 and 3.7 $\mu$m. With an error of 10\% on this water
vapor range, this results in a transmission at the line location of 78-83\%.

\subsection{\textit{Herschel}/PACS}\label{ssec:obs-pacs}

We used the \textit{Herschel}/PACS 70~$\mu$m continuum map to verify the positions of the dense cores
in Gy~3--7 listed in the \textit{Herschel}/Hi-GAL compact source
catalogue\footnote{\url{https://vialactea.iaps.inaf.it/vialactea/public/HiGAL_360_clump_catalogue.
tar.gz}} \citep{elia21}. 
Figure~\ref{fig:tdust_nh2_ppmap} shows the 70~$\mu$m continuum map and the H$_2$ column density,
$N$(H$_2$), map obtained using the $\tt ppmap$ tool with the \textit{Herschel}/Hi-GAL survey
\citep{Marsh17}. Overall, there is a good agreement between the catalog positions and the peaks in
the continuum at 70~$\mu$m and $N$(H$_2$). In the subsequent analysis, we adopt the coordinates of
the 70~$\mu$m peaks as the dense core coordinates. We refer to the core at (RA, Dec) =
($7^{h}09^{m}20\fs 4,-10\degr 50\arcmin 28\farcs4$) as A and to that at ($7^{h}09^{m}21\fs
9,-10\degr 50\arcmin 35\arcsec$) as B; they correspond to the \textit{Herschel}/Hi-GAL catalog
sources HIGALBM224.6079--1.0065 and HIGALBM224.6128--1.0013, respectively \citep{elia21}. 

\subsection{RT4}\label{ssec:rt4}

We conducted a survey at 22~GHz using the Toru\'n 32-m radio telescope (RT4) toward the entire
CMa--l224 region. The full-beam width at half maximum of the antenna at 22~GHz is
$\sim$106$\arcsec$, with a pointing error of $\lesssim$12$\arcsec$ \citep{lew18}.

Two series of observations were performed from 2019 April to 2020 January and from 2020 January
to May 2020 in which a total of 205~positions were observed. We used the correlator with
2$\times$4096 channels and 8~MHz bandwidth operating in the frequency--switching mode, which
provided a local standard of rest velocity coverage from $-11.9$ to $41.9$~km~s$^{-1}$. 
The spectral resolution of the observations is 0.03 km~s$^{-1}$. The
observations were calibrated by the chopper wheel method and corrected for the gain elevation
effect. The system temperature varied from $\sim$120~K during winter to $\sim$200~K in summer.
Overall, the $3\sigma$ detection limit was between $\sim$3.8 and $\sim$7.5~Jy per 
0.03 km~s$^{-1}$ channel. The telescope
pointing was checked every $\sim$2~hrs by observing a nearby bright, point source. We successfully detected a variable water maser emission at (RA, Dec) = ($7^{h}09^{m}21\fs
05,-10\degr 50\arcmin 05\farcs4$), in direct vicinity of Gy 3-7, and
at (RA, Dec) = ($7^{h}09^{m}22\fs 66,-10\degr 30\arcmin 45\farcs6$), associated with IRAS
07069-1026 (see Appendix~\ref{app:maser}).

\section{Results}\label{sec:results}

Near-IR images of Gy~3--7 reveal an extended nebulosity associated with the two Hi-GAL dense cores
and several YSO candidates (see Figure~\ref{fig:multi}). Spatially-resolved FIR line emission
data obtained with FIFI-LS allows us to study key gas cooling lines at $\sim$10~000~au scales and
identify regions where processes responsible for the gas heating are at play.

\subsection{Line detections}\label{ssec:det}

Figure~\ref{fig:fifi_spectra} shows the FIR line emission toward the two Hi-GAL dense cores in
Gy~3--7 (see Section \ref{ssec:obs-pacs}; for the full list of targeted lines see Table
\ref{table:sofialines}). All lines are spectrally-unresolved with FIFI-LS and can be represented by
single Gaussian profiles (Section \ref{obs:fifi}).

The spectra show strong line emission in the [\ion{O}{i}] lines at 63~$\mu$m and 145~$\mu$m, as well
as in the high$-J$ CO lines: the CO~$14-13$ line at 186.0~$\mu$m, the CO~$16-15$ line at
162.8~$\mu$m, and the CO~$17-16$ line at 152.3~$\mu$m. Core~A shows also a detection of the
CO~$22-21$ line at 118.6~$\mu$m ($E_\mathrm{u}$ of $\sim$1400~K) and a tentative detection of
CO~$30-29$ line at 87.2~$\mu$m ($E_\mathrm{u}$ of $\sim$2600~K). The CO~$31-30$ line at 84.41~$\mu$m
is blended with the OH line at 84.42~$\mu$m, and due to the lack of baseline covering the OH
84.6~$\mu$m line from the doublet, its emission cannot be quantified. 
The OH doublet at $\sim$79.2~$\mu$m seems to be tentatively detected toward both cores, but is
severely affected by the rise of the baseline on its left side and its flux cannot be properly
measured (Appendix~\ref{app:spatial}). The OH doublet at 163.12 and 163.18~$\mu$m, located next to
the CO~16-15 line, is not detected in neither of the two cores.

The [\ion{O}{i}] line at 63.18~$\mu$m sits on the edge of a telluric water feature and the low
transmission at $\lambda\ga63.24$~$\mu$m increases the noise on the continuum. Since the
transmission at the spectral line location is well known and the S/N is very high, there is no
relevant effect on the uncertainty of the line flux. Possible self-absorption of this [\ion{O}{i}]
line cannot be identified at the spectral resolution of FIFI-LS. The higher-resolution spectra
collected toward other YSOs using the German REceiver for Astronomy at Terahertz Frequencies
\citep[GREAT; ][]{risacher2018} show that self-absorption could decrease the integrated emission of
the line by a factor of 2--3 \citep{Leurini15,Mookerjea21}. 

Similarly, self-absorption in the [\ion{C}{ii}] line at 157.7~$\mu$m might affect the FIFI-LS
spectra. Strong self-absorption unresolved by FIFI-LS may result in a non-detection
of this line toward core~A. In contrast, the line is clearly detected toward core-B. Decreases of
the flux of [\ion{C}{ii}] line due to self-absorption as high as a factor of 20 have been estimated
toward photodissociation regions \citep{guevara2020}.

In summary, FIFI-LS spectra provide detections of key FIR cooling lines toward two dense cores
in Gy~3--7: the high$-J$ CO, [\ion{O}{i}], and [\ion{C}{ii}]. Due to the atmospheric absorption, the
H$_2$O emission could not be targeted; the OH emission suffers from line blending and poor
baselines. In Section \ref{ssec:far-ir_emission}, we show the distribution of line emission in
various species toward the entire cluster. 

\subsection{Spatial extent of FIR line emission} 
\label{ssec:far-ir_emission}

The FIR range contains several important diagnostic lines, which provide information about the physical
conditions and processes that strongly contribute to the gas cooling \citep{GL78,kau96}. The
analysis of spatial extent of various FIR species pinpoints the presence of shocks and/or UV
radiation associated with star formation.

Figures \ref{fig:far-IRoverlaid_maps} and \ref{fig:near-IRoverlaid_maps} show the spatial extent of
the FIR lines detected toward Gy~3--7 (Section \ref{ssec:det}). The line emission is compared to
the FIR dust continuum emission at 70 or 160~$\mu$m from \textit{Herschel}/PACS
(Figure~\ref{fig:far-IRoverlaid_maps}) and the 4.5~$\mu$m continuum tracing warmer dust from
\textit{Spitzer}/IRAC (Figure~\ref{fig:near-IRoverlaid_maps}). Bright rotationally excited H$_2$
emission associated with shocks might also contribute to the flux in the 4.5~$\mu$m IRAC band
\citep{Cyganowski08,Cyganowski11}. Appendix \ref{app:spatial} shows additional maps of Gy~3--7 for
lines from various species. 

The high$-J$~CO emission distribution is elongated in the same direction as the IR continuum, but
its peaks are offset from the continuum peaks at wavelengths similar to those of the respective CO
lines. A similar pattern of emission is also seen for the [\ion{O}{i}] 63~$\mu$m and 145~$\mu$m
lines; yet, the peak of the [\ion{O}{i}] lines are almost co-spatial with the core positions, which
is not the case for the CO lines. Nevertheless, both the  CO and the [\ion{O}{i}] extend beyond the
core positions along the E--W direction. Such elongated high$-J$ CO morphologies have been commonly
interpreted as arising in shocked outflows from LM and IM protostars
\citep{goi12,he12,kri12,karska13,matuszak2015,green16,tobin16,kri17co}. Similarly, extended
[\ion{O}{i}] emission has been associated with embedded, atomic outflows \citep{karska13,nisini15},
as recently confirmed by detections of broad line wings in the 63 $\mu$m line with SOFIA/GREAT
\citep{Leurini15,kri17oxy,yang22}. Thus, the high$-J$ CO and [\ion{O}{i}] emission toward Gy~3--7
might also arise from outflows, where shocks and UV radiation both contribute to gas cooling (see
Section~\ref{sec:dis}).

The [\ion{C}{ii}] emission also follows the pattern of the continuum emission, but its two emission
peaks are offset by $\sim$11$\arcsec$ in different directions from the corresponding [\ion{O}{i}]
63~$\mu$m peaks. These differences are illustrated further in Figure~\ref{fig:sofia_overlaid_maps},
in which emission in various species is directly compared. Clearly, the [\ion{C}{ii}] emission
traces different regions of Gy~3--7 than the high$-J$ CO and [\ion{O}{i}] lines, which is at least
partly due to its lower critical density, $n_\mathrm{crit}\sim$(3.7--4.5)$\times10^3$ cm$^{-3}$ for
$T_\mathrm{kin}$ of 300-100 K \citep[assuming collisions with H$_2$, ][]{wie14}. As a result,
[\ion{C}{ii}] is excited in lower density regions and likely exposed to external UV radiation
creating a photodissociation region \citep{hol97}. To some extent, the pattern of [\ion{C}{ii}]
emission might be also affected by self-absorption, which is spectrally-unresolved in the FIFI-LS
data (Section \ref{ssec:det}). 

\section{Analysis}\label{sec:Analysis}
\begin{figure}[ht]
\includegraphics[width= 0.95\linewidth]{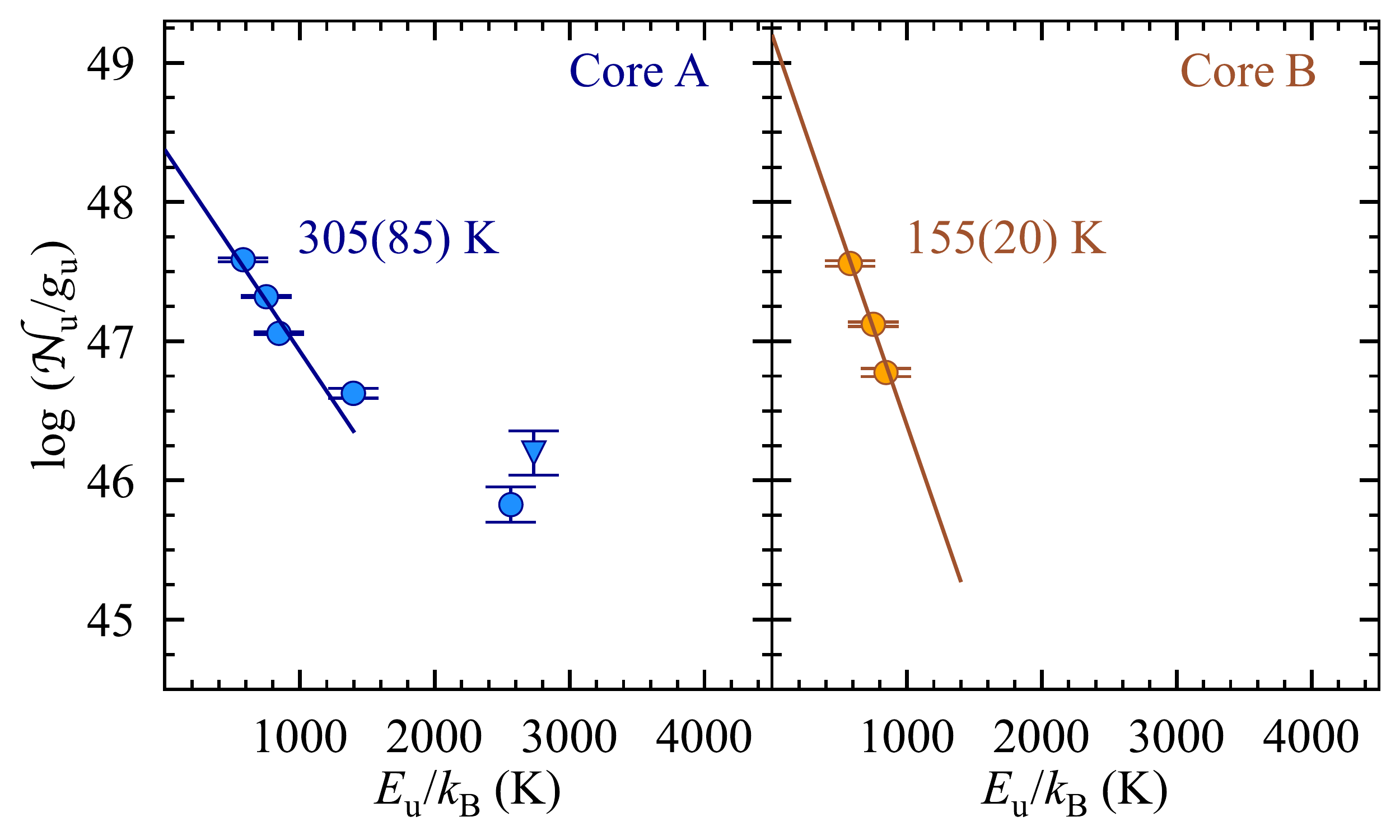} 
\caption{\label{diag} CO rotational diagrams toward core~A (left, in blue) and B (right, in orange).
Circles refer to values based on line detections and the triangle shows the measurement using the
upper limit of the CO 31-30 line, which is blended with OH. Solid lines show fits using transitions
belonging to the "warm" gas component; the CO~$30-29$ line at 87.19~$\mu$m is therefore not
included. The CO rotational temperature $T_\mathrm{rot}$ derived from the rotational diagram is
labeled in each panel and the value in parenthesis indicates its uncertainty.}
\end{figure}
The high$-J$ CO emission allows us to measure the CO rotational temperature of the warm molecular
gas and to estimate the total line cooling by the FIR CO lines. The mapping capabilities of
FIFI-LS provide information about the spatial distribution of the temperature and gas cooling across
the entire clump.

\subsection{CO rotational temperatures}\label{ssec:Trot}

We used Boltzmann (or rotational) diagrams to calculate the CO rotational temperature toward the two
dense cores in Gy~3--7 as a proxy of gas kinetic temperature. Assuming that all these lines are
optically thin and thermalized, their upper level column densities, $N_\mathrm{u}$, are estimated
using Eq.~\ref{eq:rot} following \cite{gl1999}: 
\begin{equation}
\label{eq:rot}
\ln \frac{N_\mathrm{u}}{g_\mathrm{u}}=
\ln
\frac{N_\mathrm{tot}}{\mathcal{Q}(T_{\mathrm{rot}})}-\frac{E_\mathrm{u}}{k_\mathrm{B}T_\mathrm{rot}}
,
\end{equation}
where $g_\mathrm{u}$ is the degeneracy of the upper level, $\mathcal{Q}(T_\mathrm{rot})$ is the
rotational partition function at a temperature, \Trot, $N_\mathrm{tot}$ is the total column density,
and $k_\mathrm{B}$ is the Boltzmann constant (see Table~\ref{table:sofialines}).

Due to the low spatial resolution of FIFI-LS, the emitting region of the highly-excited gas is
unresolved and thus we calculate instead the number of emitting molecules, $\mathcal{N}_\mathrm{u}$,
for each transition \citep[see e.g., ][]{her12,karska13}:
\begin{equation}
\label{eq:rot2}
\mathcal{N}_\mathrm{u}=\frac{4\pi d^{2}\lambda F_{\lambda}}{Ahc}
,\end{equation}
where $F_{\lambda}$ is the flux of the line at wavelength $\lambda$, $d$ is the distance to Gy~3--7,
$A$ is the Einstein coefficient, $c$ is the speed of light, and $h$ is the Planck's constant.
Consequently, the total number of emitting molecules $\mathcal{N}_\mathrm{tot}$ is derived instead
of the $N_\mathrm{tot}$ from Eq.~\ref{eq:rot}. 
\begin{table*}
\caption{CO rotational temperature, the number of emitting molecules, and total line
luminosities of CO and [\ion{O}{i}] lines toward dense cores in Gy~3--7 and IM~YSOs from
\cite{matuszak2015} \label{table:rot}} 
\centering 
\begin{tabular}{l | c c | r r r }
\hline \hline 
Source  &        $T_\mathrm{rot}$\tablefootmark{a}      &       
log$\mathcal{N}_\mathrm{tot}$\tablefootmark{a}  &        
\multicolumn{1}{c}{$L_\mathrm{CO}$(warm)\tablefootmark{b}}&     \multicolumn{1}{c}{$L_\mathrm{[\ion{O
}{i}]63\mu m}$}&
\multicolumn{1}{c}{$L_\mathrm{[\ion{O}{i}]145\mu m}$}\\
~       &        (K)    &        ~      &       \multicolumn{3}{c}{($10^{-3} L_\odot$)}      \\
\hline
\multicolumn{6}{c}{Dense cores in Gy~3--7}\\
\hline
Core~A &        $       305     \pm     85      $       &       $       50.41   \pm     0.34    $       &       $       73.31
        \pm     19.69   $       &       $       92.69   \pm     5.01    $       &       $       8.38    \pm     0.29    $ \\
Core~B &        $       155     \pm     20      $       &       $       50.96   \pm     0.24    $       &       $30.47\pm6.10$  &       
$       76.17   \pm     5.37    $       &       $       6.15    \pm     0.37    $ \\
\hline
\multicolumn{6}{c}{IM YSOs in the inner Milky Way}\\
\hline 
AFGL 490\tablefootmark{c}       &       $       255     \pm     25      $       &       $       51.11   \pm     0.13    $       &
        $       225.17  \pm     19.85   $       &       $       336.65  \pm     1.47    $       &       $       29.56 \pm     0.24$   \\
L1641   &       $       270     \pm     35      $       &       $       50.06   \pm     0.17    $       &       $       24.31 \pm     2.74    $       
&       $       14.69   \pm     0.14    $       &       $       1.20    \pm     0.05    $ \\
NGC 2071        &       $       295     \pm     20      $       &       $       51.23   \pm     0.10    $       &       $       425.38
        \pm     26.11   $       &       $       312.72  \pm     0.91    $       &       $       21.59   \pm     0.13    $ \\
Vela 17 &       $       215     \pm     25      $       &       $       51.08   \pm     0.17    $       &       $       125.56  
\pm     15.44   $       &       $       369.84  \pm     1.33    $       &       $       77.85   \pm     0.21    $        \\
Vela 19 &       $       255     \pm     20      $       &       $       50.66   \pm     0.10    $       &       $       86.00   \pm     9.27$
        &       $       130.69  \pm     0.70    $       &       $       13.60   \pm     0.09    $       \\
NGC 7129        &       $       295     \pm     35      $       &       $       50.84   \pm     0.14    $       &       $       187.26
        \pm     19.16   $       &       $       66.03   \pm     0.78    $       &       $       3.13    \pm     0.88    $ \\
\hline 
\end{tabular}
\begin{flushleft}
\tablefoot{
\tablefoottext{a}{$T_\mathrm{rot}$ and $\mathcal{N}_\mathrm{tot}$~calculated using only CO~$14-13$,
CO~$16-15$, CO~$18-17$, and CO~$22-21$ transitions for more direct comparisons with the cores in
Gy~3--7.}
\tablefoottext{b}{The CO line luminosity in the " warm" component is obtained using all
transitions reported in \cite{matuszak2015} with $J_\mathrm{up}$ from 14 to 24 (see Section
\ref{ssec:linecooling}).}
\tablefoottext{c}{We adopted the new distance of 0.97~kpc toward this source when calculating
$T_\mathrm{rot}$, $\mathcal{N}_\mathrm{tot}$, and luminositiy of CO and [\ion{O}{i}] lines (see
Table~\ref{tab:summary_distance_Luminosity_YSOs_MW_MCs} and references therein). }} 
\end{flushleft}
\end{table*}
\begin{figure*}[tb!]
\centering
\includegraphics[height= 0.34\linewidth]{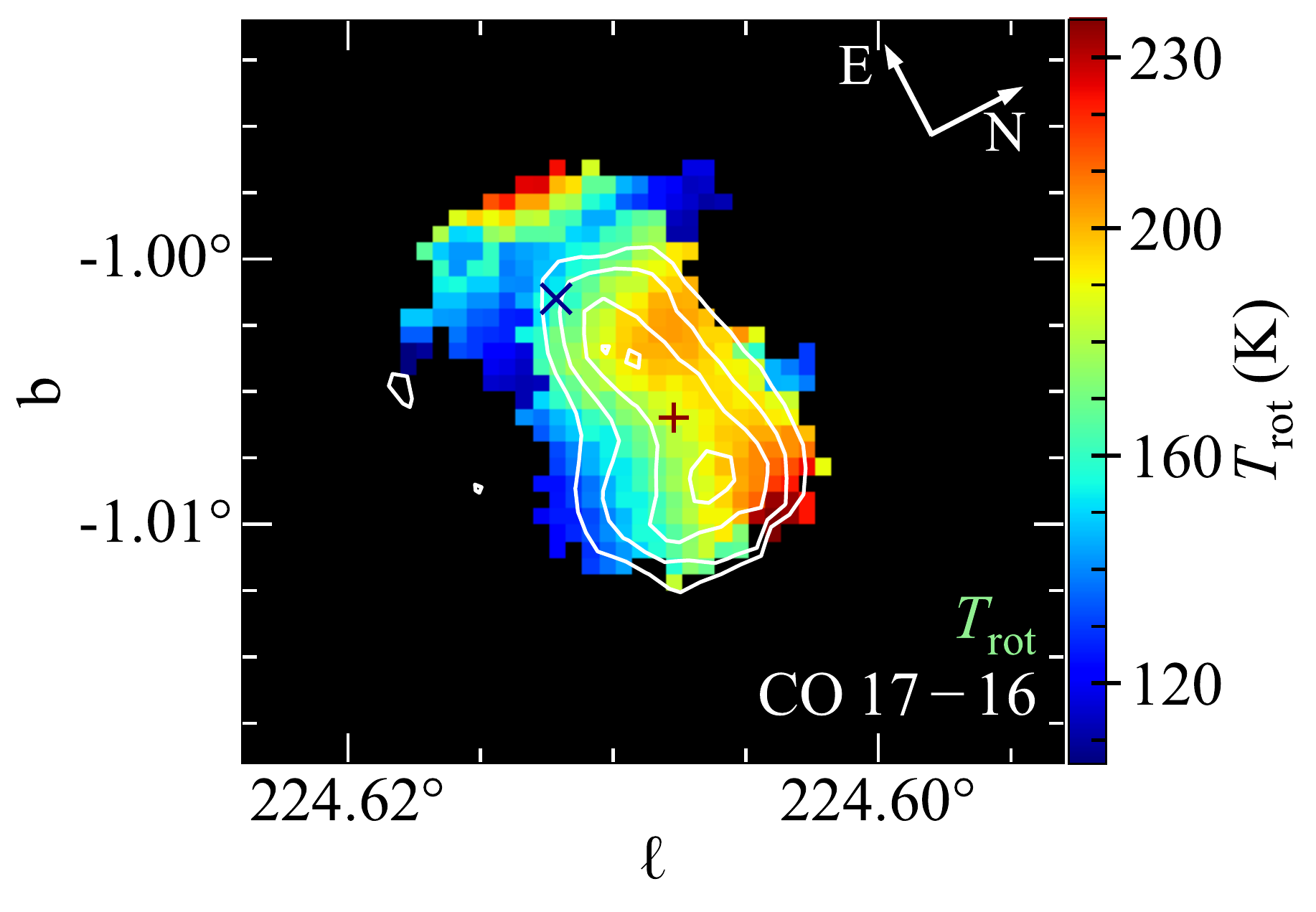}
\includegraphics[height= 0.34\linewidth]{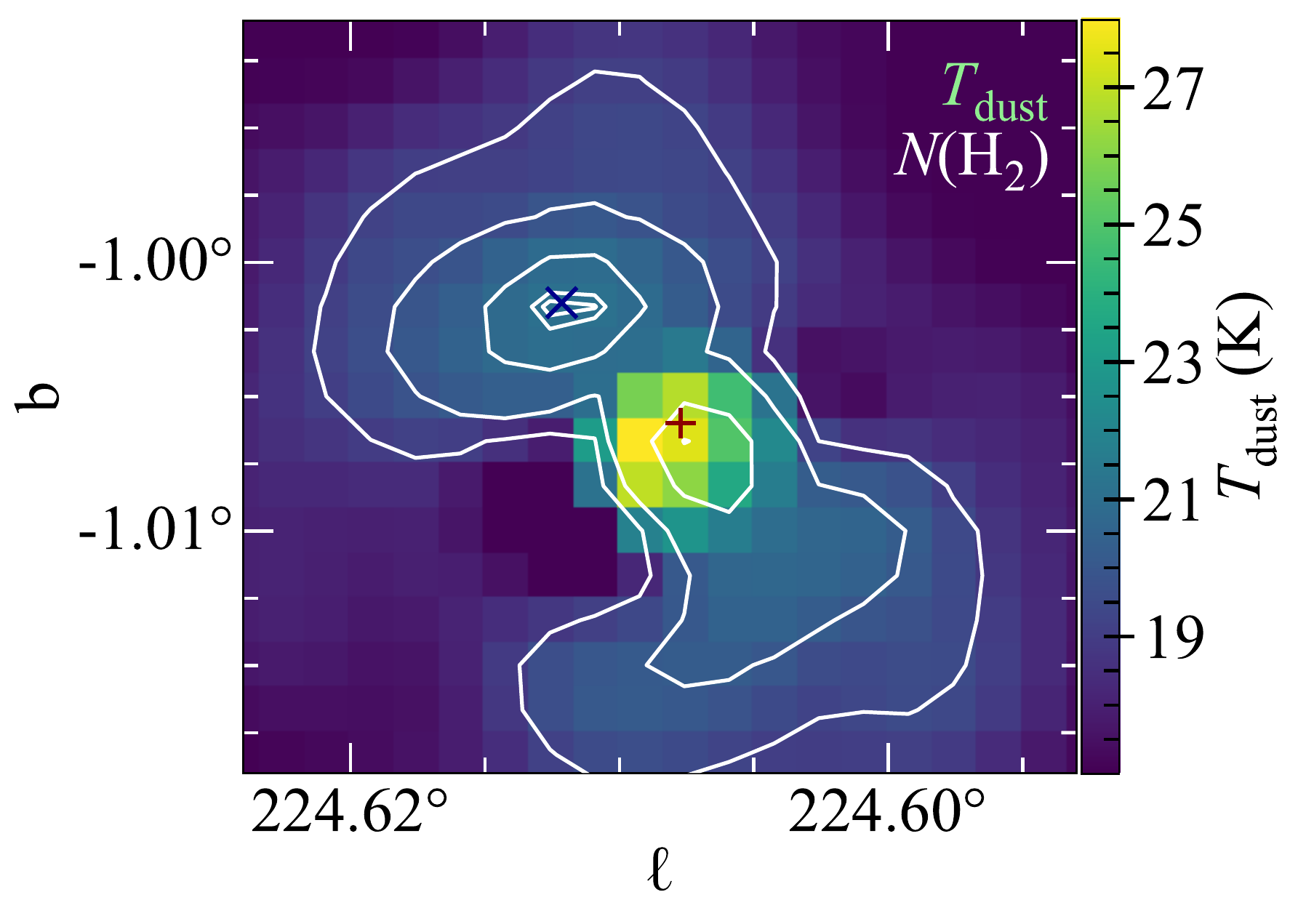}
\caption{Gas and dust temperatures toward Gy~3--7. Left: CO rotational temperature map obtained
using Boltzmann diagrams. White contour map shows the CO~$17-16$ emission at 25\%, 50\%, 75\%,
and 95\% of the emission peak. Right: H$_2$ column density contour map overlaid on the dust
temperature map derived from the $\tt ppmap$ tool with the \textit{Herschel}/Hi-GAL survey
\citep{Marsh17}. Contour levels of the $N$(H$_2$) are at (5, 10, 30, 50, 55)$\times
10^{21}$~cm$^{-2}$. }
\label{fig:Trot_maps}
\end{figure*} 
\begin{figure*}[tb]\centering
\includegraphics[height= 0.34\linewidth]{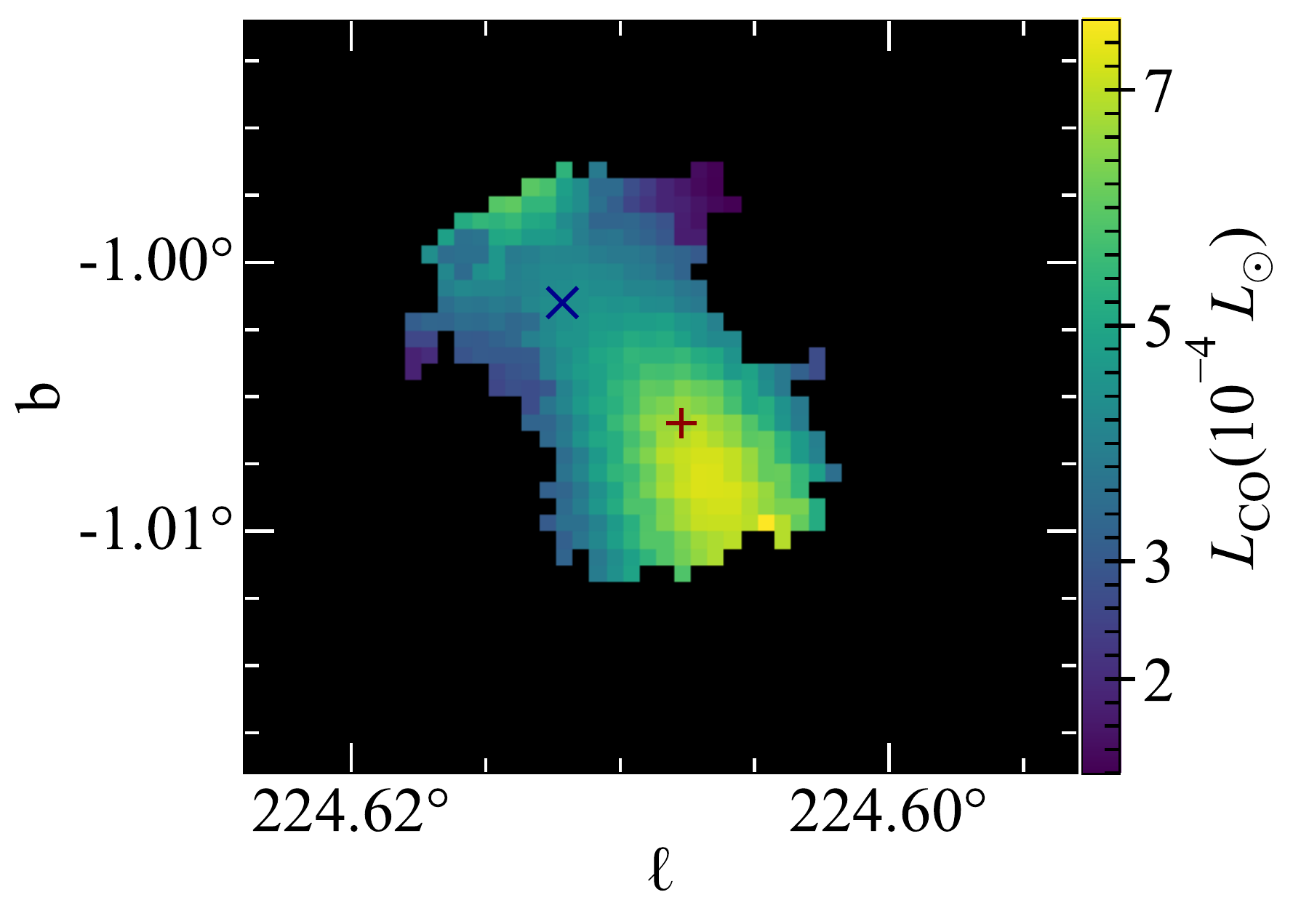}
\includegraphics[height= 0.34\linewidth]{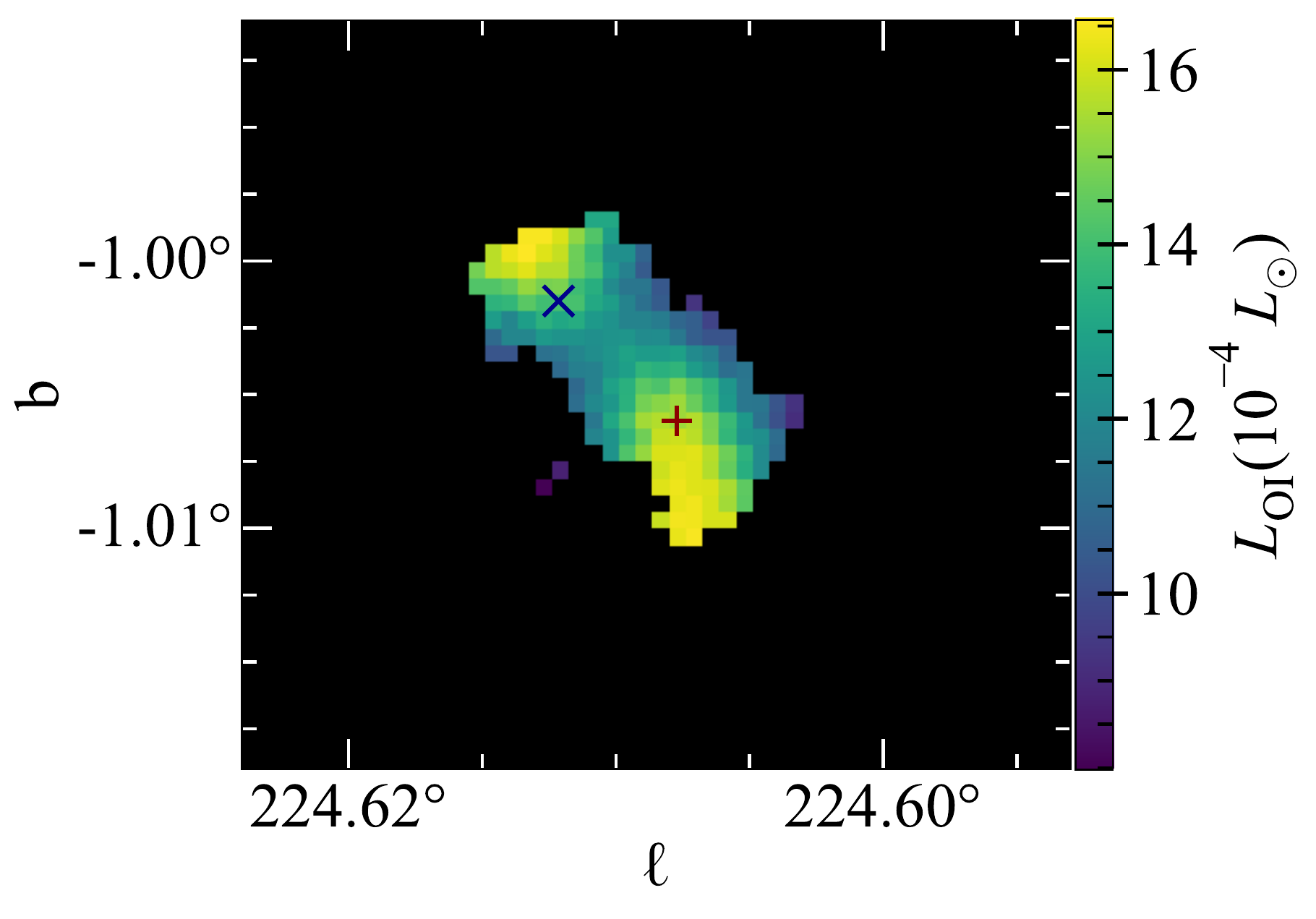}
\caption{Spatial extent of FIR line luminosities of CO (left) and [\ion{O}{i}] (right) toward
Gy~3--7. The calculation procedure is described in Section~\ref{ssec:linecooling}.}
\label{fig:total_luminosity_mole_atom_maps}
\end{figure*}
To measure $T_{\mathrm{rot}}$ over the same physical scales, we convolved the CO emission maps down
to the lowest spatial resolution corresponding to CO~$14-13$ observation (18.3$\arcsec$, see Table
\ref{table:sofialines}) and resampled the maps to the same pixel size. The flux of CO lines toward
cores A and B was calculated within a beam of 20$\arcsec$ (see Table~\ref{tab:flux_2cores}). 
We performed a linear fit on the rotational diagram using the {\tt curve$_{-}$fit} function in Python
($y = ax+b$). The \Trot~and $\mathcal{N}_\mathrm{tot}$ values are derived from the slope $a$ and
y-intercept $b$ of the fit (Eq.~\ref{eq:rot}). 

Figure~\ref{diag} shows the CO Boltzmann diagrams toward the two cores in Gy~3--7. We obtain
rotational temperatures of $305\pm85$~K and $155\pm20$~K toward dense cores A and B, respectively,
using CO lines with $J_\mathrm{up}$ of $14-22$ (see Table \ref{table:rot}). The same transitions
have been associated with the \lq\lq warm'' component detected on CO diagrams toward protostars in
the inner Galaxy and corresponding to the  
widely found $T_\mathrm{rot}$ of 300~K \citep{karska13,manoj2013,green2013}. The CO~$30-29$ and
CO~$31-30$ data at 87.2 and 84.4~$\mu$m, respectively, indicate the presence of the \lq\lq hot''
component toward core~A \citep{karska13}. However, the 84.4~$\mu$m line is blended with OH, which
clearly affects the CO line flux (see Figure~\ref{fig:fifi_spectra}). Consequently, we are not able
to constrain the \lq\lq hot'' component using only the CO~30-29 line. 

The spatial distribution of the \lq\lq warm'' component's CO rotational temperature toward the
entire Gy~3--7 clump is shown in the left panel of Figure~\ref{fig:Trot_maps}. Here, we calculate
$T_\mathrm{rot}$ using the three lowest transitions ($J_\mathrm{up}$ of $14-17$), which are detected
in a large part of the map, except for the map edges where even  CO~$17-16$ is not detected
(Figure~\ref{fig:far-IRoverlaid_maps}). The resulting $T_{\mathrm{rot}}$ values range from 105 to
230~K across the map, with a median value of 170(15)~$\pm$~30~K\footnote{The temperature in the
bracket shows the mean error of the \Trot~distribution and $\pm$~30 refers to the standard deviation
of the \Trot~values.}. 
The CO rotational temperatures are the highest in the vicinity of the CO~$17-16$ emission peaks,
which are offset by $\sim$17$\arcsec$ from core~A to the west 
(Figure~\ref{fig:far-IRoverlaid_maps}). The morphology of the $T_\mathrm{rot}$ distribution suggests
the origin of high$-J$ CO in a bipolar outflow driven by core~A. Significantly lower CO
temperatures, $\lesssim$150~K, are measured in the surroundings of core~B, without a clear outflow
signature from this object. The temperatures around 200~K at the eastern edge of the map are likely
caused by higher gas densities in this region, rather than higher gas kinetic temperatures. 

The spatial distribution of dust temperatures $T_\mathrm{dust}$ toward Gy~3--7, adopted from the
\textit{Herschel}/Hi-GAL survey, is also shown in the right panel of Figure~\ref{fig:Trot_maps}. The
morphology of regions with elevated temperatures is similar to the extent of $N$(H$_2$), and shows
two peaks at 27 and 21~K toward core~A and B, respectively. The pattern differs significantly from
the distribution of warm, $\sim$300~K gas, traced by CO lines, favoring the origin of the CO
emission in a bipolar outflow driven by core~A. Nevertheless, the lowest$-J$ CO transitions observed
with FIFI-LS might partly trace the extended continuum emission, for instance, on the eastern part of
Gy~3--7. We discuss this issue further in Section \ref{ssec:origin_far_ir_Gy37}.

\subsection{FIR line cooling}\label{ssec:linecooling}

The FIR line cooling budget in LM protostars is sensitive to a source's evolutionary stage and
$L_\mathrm{bol}$, and its contributions provide important information on the shock origin of the
FIR emission \citep{karska13, karska18}. Here, we quantify the luminosity of the FIR CO and
[\ion{O}{i}] lines toward the two cores in Gy~3--7 and a sample of IM protostars observed with
\textit{Herschel}/PACS \citep{matuszak2015}.

\subsubsection{Calculation procedure}\label{ssec:calc-Ltot}
We calculated the line cooling following procedures developed for \textit{Herschel}/PACS observations
in the same wavelength range \citep{karska18}. Briefly, we determined the line luminosity of CO lines
in the " warm"~component, $L_\mathrm{CO}$(warm), from the sum of the individual line
fluxes with $J_\mathrm{up}$ from 14 to 24, corresponding to $E\mathrm{_u}/k\mathrm{_B}$=580--1800~K.
Since not all CO transitions are observed, we use linear fits from the Boltzmann diagram to the
"warm"~component to recover the fluxes of the transitions not covered by FIFI-LS or PACS
observations. The flux uncertainties of those transitions were propagated from the 
parameters of the linear fit and their uncertainty. The same procedure could be applied to the
 "hot"~component toward Gy~3--7
cores due to the lack of data on a sufficient number of high$-J$ CO transitions. As a result, we did
not calculate the total FIR CO cooling, that is, we did not account for transitions with
$E_\mathrm{u}$>1800~K. The advantage of this approach is that we avoid significant source-to-source
variations in $L_\mathrm{CO}$(hot), which is reflected by the broad range of rotational temperatures
measured using the highest$-J$ CO lines \citep[see Figure~6 in ][]{karska18}. The total line
luminosity of [\ion{O}{i}], $L_\mathrm{[\ion{O}{i}]}$, is calculated by the addition of the fluxes
of the [\ion{O}{i}] 63 and 145~$\mu$m lines. Table \ref{table:rot} shows the FIR line
luminosities obtained for Gy~3--7, as well as those for six IM YSOs from \cite{matuszak2015}.
\begin{figure}[!t]
\begin{center}
\includegraphics[width=\linewidth]{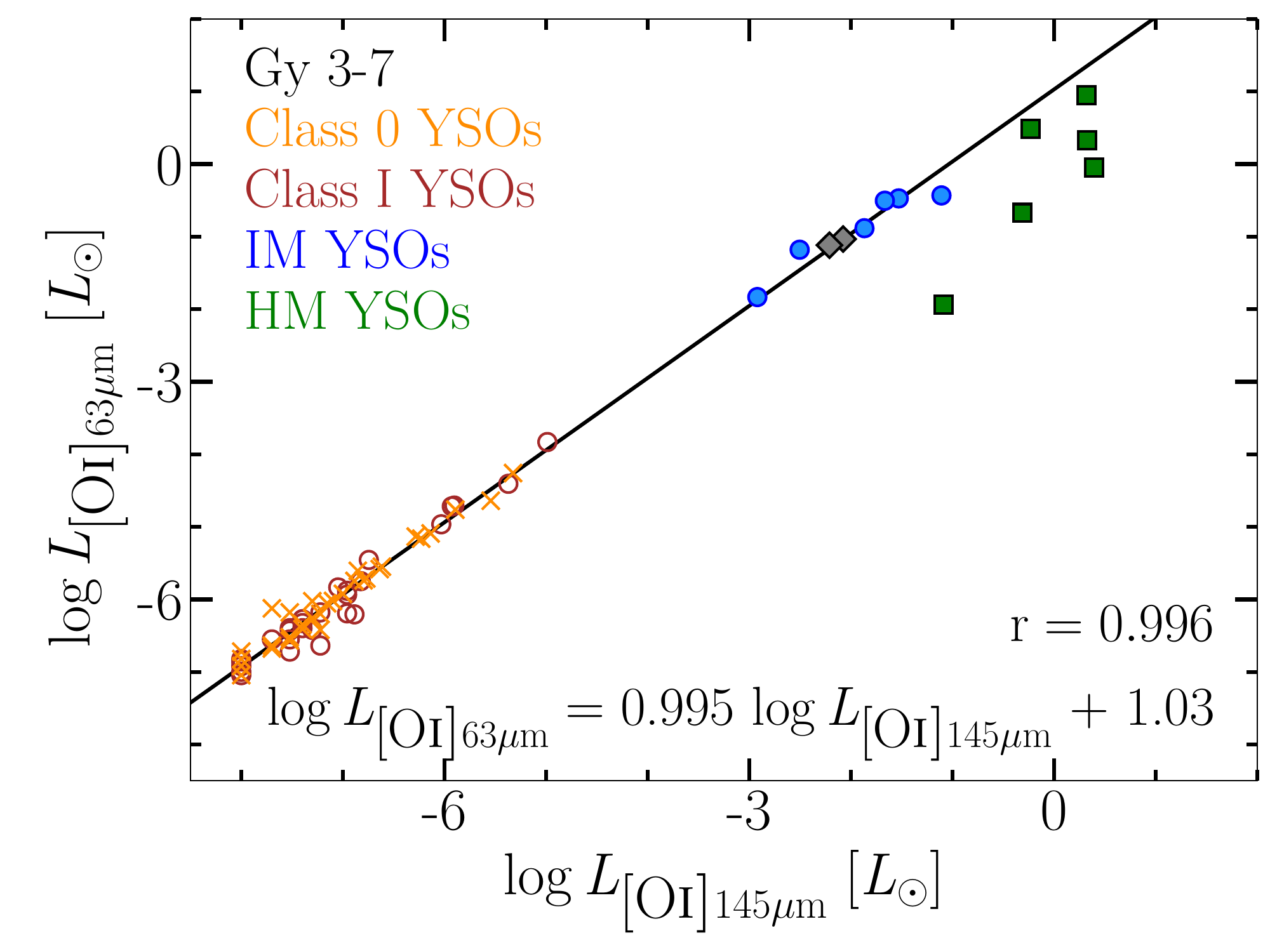}\\
\includegraphics[width=\linewidth]{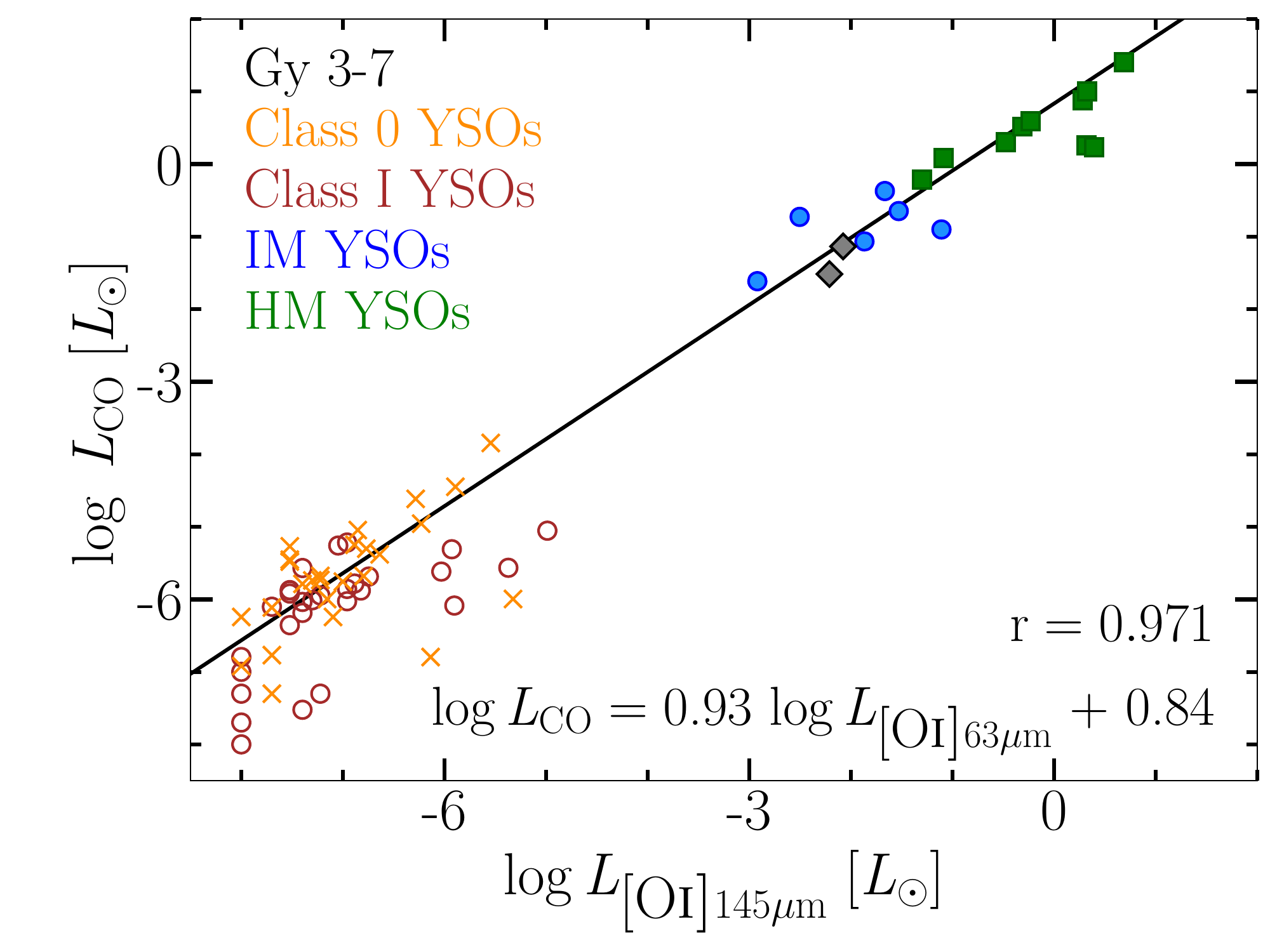}
\vspace{-2ex}
\caption{Correlations between luminosities of FIR CO and [\ion{O}{i}] lines. Top: Correlation
between the luminosities of the 63 and 145~$\mu$m [\ion{O}{i}] lines at  from LM to HM YSOs. Cores~A
and B in Gy~3--7 are marked as grey diamonds, Class 0 and Class I YSOs as 
orange "$\times$" signs and red circles \citep{karska18}, respectively, IM YSOs as blue circles \citep{matuszak2015}, and HM YSOs as green squares,
respectively \citep{karska14}. Black solid line is the linear fit to all sources except for the
HM~YSOs, showing a strong correlation between the two [\ion{O}{i}] line luminosities.
Bottom: Correlation between luminosities of the CO lines and the [\ion{O}{i}] line at 145~$\mu$m.
Black solid line shows the linear fit to all sources, including the HM~YSOs.
}\label{fig:correlation_vs_lumi_oxy}
\end{center}
\end{figure}

Figure~\ref{fig:total_luminosity_mole_atom_maps} shows the spatial distribution of the FIR line
luminosity of CO and [\ion{O}{i}] toward Gy~3--7. Most of the CO luminosity orginates in a region
west of core~A, suggesting an outflow origin (see also Section \ref{ssec:far-ir_emission}). A
similar region is characterized also by a high [\ion{O}{i}] luminosity, which typically follows the
pattern of high$-J$ CO emission around LM protostars \citep{karska13,nisini15,vD21}. Additionally, a
high [\ion{O}{i}] luminosity is measured to the east from core~B, toward the direction of the
[\ion{C}{ii}] peak (see Figures.~\ref{fig:near-IRoverlaid_maps} and \ref{fig:sofia_overlaid_maps}).
The lack of enhancement of CO line luminosity in this region could suggest an origin of the
[\ion{O}{i}] emission in the PDR. We explore this scenario further in Section \ref{ssec:pdr}.
\begin{figure}[ht!]
\includegraphics[angle=180,width=1\linewidth]{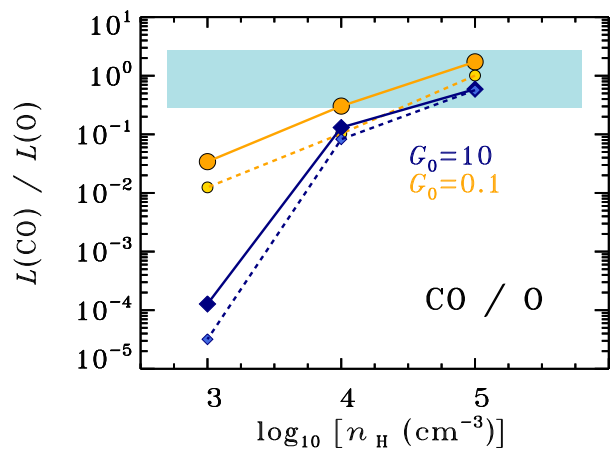}
 \vspace{-4ex}

\caption{Ratio of the CO and [\ion{O}{i}] luminosities as a function of pre-shock density for UV
irradiated $C-$shock models and observations of Gy~3--7 cores and IM YSOs from
\citet{matuszak2015} (light blue box). All models correspond to UV fields parameterized by $G_\mathrm{0}$ of 10 (in blue) and 0.1 (in
orange). Solid lines connect models with shock velocities $\varv _\mathrm{s}$
of 20~km~s$^\mathrm{-1}$, and dashed lines -- the models with $\varv _\mathrm{s}$ of 10~km~s$^\mathrm{-1}$.}\label{fig:shocks}
\end{figure}
\subsubsection{Flux correlations}

We compare the FIR line emission in Gy~3--7 with the data based on \textit{Herschel}/PACS
measurements toward LM Class 0 and Class I \citep[][]{karska18}, IM \citep[][]{matuszak2015}, and HM
YSOs \citep[][]{karska14}. CO luminosities are consistently calculated for the "warm" component 
on Boltzmann diagrams. The correlations between luminosities of the two [\ion{O}{i}] lines, as well as
the [\ion{O}{i}] and CO lines are shown in Figure~\ref{fig:correlation_vs_lumi_oxy}. 

A strong power-law correlation is found between the [\ion{O}{i}] line luminosities for all low- and
intermediate-mass sources including the two cores in Gy~3--7 (the Pearson coefficient of the
correlation is $r=0.996$, which corresponds to a significance of $8.2\sigma$).  
The observed ratios of the [\ion{O}{i}] 63~$\mu$m/145~$\mu$m lines span a range from 4 to 38 with a
median value of $\sim$12. These results are quantitatively similar to those shown in Figure~11 of
\cite{karska13} for a sub-sample of 18 LM~YSOs. 

The HM YSOs follow a similar trend, however, several sources show a flux deficit in the [\ion{O}{i}]
63~$\mu$m line, which is likely caused by line-of-sight contamination and optical depth effects
\citep[see, e.g.,][]{Liseau92,Leurini15}. A power-law fit to the sample including HM~YSOs shows a
shallower slope ($b=0.91$ versus $b=0.99$, when only LM and IM~YSOs are considered). The correlation 
strength is comparable to the one for LM and IM sources alone, with a Pearson coefficient of 0.991 corresponding to 8.5$\sigma$. 

The CO line luminosity in the  "warm" component shows a strong correlation with the 145~$\mu$m
[\ion{O}{i}] line luminosity. A power--law fit to the entire sample returns a slope of $b\sim$0.93
and the Pearson coefficient of $0.971$, corresponding to $\sim$8.2$\sigma$
(Figure~\ref{fig:correlation_vs_lumi_oxy}). Clearly, the line luminosity of the [\ion{O}{i}]
145~$\mu$m line shows a smaller scatter for the HM YSOs with respect to the 63~$\mu$m line. 
In case of LM YSOs, a significant scatter in the ratio of CO and [\ion{O}{i}] line 
luminosities is likely linked to their different evolutionary stages. The ratio of CO line luminosity 
over $L_\mathrm{bol}$ is $\sim$2.3 larger for Class 0 than Class I sources, whereas the [\ion{O}{i}] 
luminosities are similar for both groups \citep{karska18}. Thus, the molecular-to-atomic
line cooling is expected to be higher in Class 0 objects. Indeed, the linear fit using 
only Class 0 sources results in a shallower slope, but do not affect the general conclusions.

In summary, we find a strong correlation between the line luminosities of the [\ion{O}{i}] and CO
lines for YSOs in a broad mass range, consistent with previous results for LM YSOs \citep{karska13}.
Combined with the similar spatial extent of the [\ion{O}{i}] and CO lines
(Section~\ref{ssec:far-ir_emission}), the correlations suggest a similar physical origin of the two
species.

\subsection{Properties of a possible photodissociation region}\label{ssec:pdr}

Assuming that the [\ion{O}{i}] and [\ion{C}{ii}] lines predominately arise from a photodissociation
region (PDR), the UV field strengths, $G_\mathrm{0}$, and hydrogen nucleus number densities,
$n_\mathrm{H}$, across Gy~3--7 can be obtained from their ratios. These assumptions might be justified in
case of the eastern part of Gy~3--7 with relatively weak CO line luminosities (see
Section~\ref{ssec:linecooling}). On the contrary, the [\ion{O}{i}] line luminosity in the
surrounding of core~A closely follows the high$-J$ CO emission associated with outflow shocks.

We determined the physical properties of the PDR using the PDR Toolbox
2.1.1\footnote{\url{https://dustem.astro.umd.edu/}} \citep{Pound_Wolfire2011} based on the PDR
models provided by \citet{Kau06}. We used three line ratios involving the [\ion{O}{i}] lines at
63.2~$\mu$m and 145.5~$\mu$m, and [\ion{C}{ii}] line at 157.7~$\mu$m, and ran the code at each
spaxel of the FIFI-LS maps. We obtained gas densities of
10$^4$--10$^{5}$~cm$^{-3}$ and UV field strengths of the order of 10$^3$-10$^6$ times the average
interstellar UV radiation field \citep{hab68}.
These physical conditions are typical for dense,
star-forming clumps associated with HM YSOs \citep{oss10,Ben16,mirocha21}. However, given that
Gy~3--7 is associated with IM YSOs (see Section \ref{ssec:sed_fit}), UV radiation fields of 10$^3$
or higher are unlikely \citep[e.g.,][]{karska18}. 

The similar spatial extent of the [\ion{O}{i}] and high$-J$ CO emission
(Section~\ref{ssec:far-ir_emission}) and the strong line luminosity correlation between the two
species (Section~\ref{ssec:linecooling}) favor the origin of the bulk of [\ion{O}{i}] in the outflow
shocks rather than in the photodissociation region. We consider this scenario in
Section~\ref{ssec:shocks}.

\begin{table*}
\caption{Best-fit models of SED using the \citet{robitaille2017} classification \label{tab:models}} 
\centering 
\begin{tabular}{lclc}
\hline \hline 
Model Set & \# of sources & Components & Group \\ 
\hline 
s$---$s$-$i & 4 & star & -- \\
sp$--$s$-$i & 2 & star $+$ passive disk; $R_{\rm inner}$ = $R_{\rm sub}$ & $d$\\
sp$--$h$-$i & 4 & star $+$ passive disk; variable $R_{\rm inner}$ & $d$ \\
s$-$pbhmi & 1 & star $+$ power-law envelope $+$ cavity $+$ medium; variable $R_{\rm inner}$ & $e$\\
spubsmi & 1 & star $+$ passive disk $+$ Ulrich envelope $+$ cavity $+$ medium; $R_{\rm inner}$ =
$R_{\rm sub}$ & $d+e$\\
\hline 
\end{tabular} 
\begin{flushleft} \textbf{Notes}: Seven characters in the model set names indicate which component
is present; they are (in order): s: star; p: passive disk, p or u: power-law or Ulrich envelope; b:
bipolar cavities; h: inner hole; m: ambient medium; and i: interstellar dust. A dash ($-$) is used
when a component is absent. $R_{\rm inner}$ is the inner radius for the disk, envelope, and the
ambient medium - when one or more of these components are present. $R_{\rm sub}$ is the dust
sublimation radius.\\
\end{flushleft}
\end{table*}
\begin{figure*}[tbp]\vspace{-3cm}
\begin{center}
\includegraphics[scale=0.62]{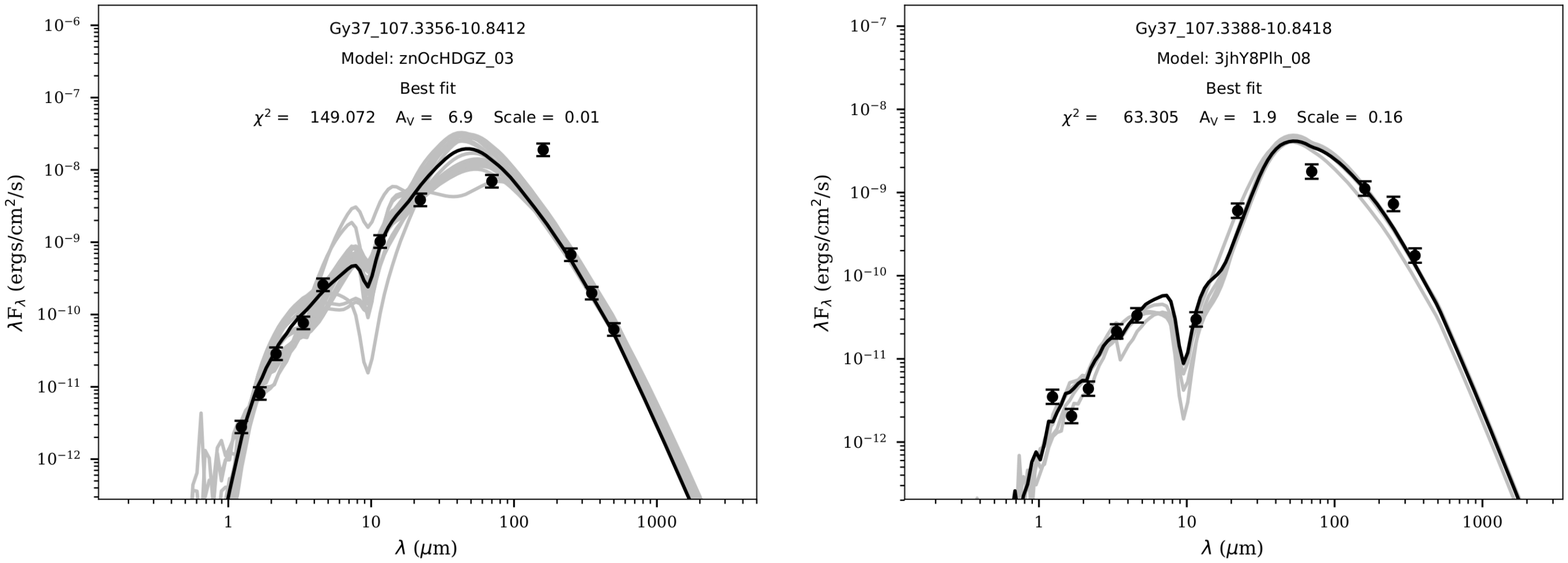} 
\end{center}\vspace{-3.5cm}
\caption{SEDs of YSOs in Gy~3--7 well-fitted with \cite{robitaille2017} models with envelopes. The
best fit model is indicated with the black solid line and gray lines show the YSO models with
$\chi^{2}$ between $\chi^{2}_\mathrm{best}$ and $\chi^{2}_\mathrm{best}$+ $F \times n$, where $n$ is
the number of data points and $F$ is a threshold parameter which we set to 3 \citep{sewilo19}.
Filled black circles are valid flux values with uncertainties. The values of a reduced $\chi^{2}$
and interstellar visual extinction for the best-fit model are indicated in the plots. Appendix
\ref{app:sed} shows the SEDs for the remaining YSOs in Gy~3--7.}
\label{fig:sed}
\end{figure*}

\subsection{Comparisons with UV-irradiated shocks}\label{ssec:shocks}

Bright FIR emission detected toward LM~YSOs has been interpreted in the context of continuous
($C-$type) shocks irradiated by UV photons \citep{karska14b,karska18,kri17co}. Figure
\ref{fig:shocks} shows a comparison of the FIR observations toward IM~YSOs, including two cores
in Gy~3--7, and the UV-irradiated shock models from \cite{mel15} and \cite{karska18}. 

Predictions of shock models were previously calculated for high$-J$ CO lines covered by
\textit{Herschel}/PACS. Here, we show the predictions for UV field strengths, $G_\mathrm{0}$, of 0.1 and
10, and shock velocities of 10 and 20~km~s$^{-1}$ \citep[see also Fig. 14 in][]{karska18}. For the
sake of comparison, we calculated the total FIR line luminosity of CO using the transitions in
the "warm" component (Section \ref{ssec:calc-Ltot}), as well as high$-J$ CO lines, both for
Gy~3--7 and for IM~YSOs \citep{matuszak2015}. The [\ion{O}{i}] line luminosity is calculated from
the sum of the two fine-structure [\ion{O}{i}] lines.

The models show a good match with observations for pre-shock H$_2$ number densities of
$10^{5}$~cm$^{-3}$ and the entire range of the considered UV radiation field strengths. 
For shock velocities, $\varv _\mathrm{s}$, of 20~km~s$^\mathrm{-1}$, a possible match 
is also found for pre-shock densities of 10$^4$~cm$^{-3}$ and G$_0$ of 0.1.
The compression factor $>$10 is expected in $C-$type shocks \citep{karska13}, so
the gas densities are of the order of 10$^5$-10$^{6}$~cm$^{-3}$, in agreement with those of LM~YSOs
\citep[e.g.,][]{kri12,mot17}.

\begin{table*}[h!]
\begin{center}
\tiny\tiny\tiny\tiny
\caption{Physical parameters for a subset of YSO candidates with at least five photometric data points
\label{t:physpar}}
\label{tab:params}
\setlength{\tabcolsep}{1.5 mm}
\begin{normalsize}
\begin{tabular}{cccrcrccc}
\hline
\hline
ID & Model & Class & $R_\ast$ & $T_\ast$ & \multicolumn{1}{c}{$L_\ast$} & $M_\ast^b$ & Age$^b$ &
Note$^c$\\
~ & set$^a$ & & ($R_\odot$) & (K) & \multicolumn{1}{c}{($L_\odot$)} & ($M_\odot$) & (Myr) & \\
\hline
1& sp--s-i      & II/III        & 5.2 & 3592 & 4.0 & ... & ...& \\
2& sp--s-i      & II/III        & 2.6 & 6791 & 13.2 & ... & ...& TP01\\
4& s---s-i      & star          & 1.2 & 2955 & 0.1 & ... & ...& \\
5& s---s-i      & star          & 1.0 & 3388 & 0.1 & ... & ...& \\
6& s---s-i      & star          & 1.0 & 5120 & 0.6 & ... & ...& TP03\\
9& sp--h-i      & II/III        & 3.3 & 3359 & 1.2 & ... & ...& \\
10& s-pbhmi     & 0                     & 25.8 & 6214 & 892.1 & 8.0 & 0.06& TP06\\
11& s---s-i     & star          & 1.2 & 2955 & 0.1 & ... & ...& \\
12& spubsmi     & I                     & 23.0 & 6194 & 702.1 & ... & ...& TP07\\
13& sp--h-i     & II/III        & 2.6 & 6112 & 8.4 & ... & ...& TP09\\ 
14& sp--h-i     & II/III        & 3.6 & 3736 & 2.2 & ... & ...& \\
15& sp--h-i     & II/III        & 10.3 & 7101 & 244.9 & ... & ...& TP08 \\
\hline
\end{tabular}
\begin{flushleft}
$^a$ See Table~\ref{tab:models} footnotes for the description of the model set names.\\
$^b$ We provide stellar masses ($M_{\rm \ast}$) and ages only for sources with reliable estimation
of these parameters (see text for details).\\
$^{c}$ IDs of the YSO candidates in Gy~3--7 identified by \cite{tapia97}, see
Figure~\ref{fig:multi}.
\end{flushleft}
\end{normalsize}
\end{center}
\end{table*}

Due to the lack of H$_2$O observations from FIFI-LS and non-detections of OH, we are limited to
comparisons between CO and [\ion{O}{i}] lines. Some of the [\ion{O}{i}] emission might arise in the
PDR (Section~\ref{ssec:pdr}), which would increase the observed line luminosity ratio; however, the
strong correlation of [\ion{O}{i}] and high$-J$ CO tracing outflow shocks does not support this
scenario. 

\subsection{Spectral energy distribution analysis}\label{ssec:sed_fit}
We investigate the physical properties of 15 individual YSO candidates from \cite{tapia97} and
\cite{sewilo19} in Gy~3--7 to understand their possible impact on the FIR and submillimeter line
emission. Spectral energy distribution (SED) models of YSOs from \citet{robitaille2017} are fitted
to the multi-wavelength photometry of YSOs using a dedicated fitting tool \citep{robitaille2007}.
Multi-wavelength photometry spanning from near- to mid-IR range of YSO candidates in the IRAS
field is presented in Appendix~\ref{app:sed}. 

We followed the procedures described in detail in \citet{karska22}. We used 18 sets of model SEDs
including various physical components of a YSO: star, disc, in-falling envelope, bipolar cavities,
and an ambient medium \citep{robitaille2017}. 
We used the PARSEC evolutionary tracks produced by the revised Padova code \citep{bressan2012,
chen2014, chen2015, tang2014} to quantify the results of SED model fitting. Models producing YSO
parameters outside of the PARSEC pre-main sequence (PMS) tracks were excluded. YSOs with models in
line with the PARSEC tracks are illustrated on the Hertzsprung-Russell diagram in
Appendix~\ref{app:sed}. For those YSOs, we calculate the stellar luminosity from the
Stefan-Boltzmann law, using the stellar radius and effective temperature from the SED fitting. The
masses and ages are determined from the closest PMS track; however, we only provide the stellar
masses and ages that are consistent with the SED fitting results (i.e., the evolutionary stage) and
YSO lifetimes from \citet{dunham2015}, respectively \citep[see Section 3.7 in ][]{karska22}.

Table~\ref{tab:models} shows the best-fit SED models for 12 YSOs in Gy~3--7. The SEDs of two sources
require an envelope contribution which is typical for deeply-embedded Class 0/I YSOs (see
Figure~\ref{fig:sed}), while six sources are successfully modeled with a passive disk and four are normal
stars (see Appendix~\ref{app:sed}). The resulting physical parameters determined from SED models are
shown in Table~\ref{t:physpar}. 

The two YSOs with envelopes, No. 10 and 12 in Table~\ref{tab:params}, are located in the center of
Gy~3--7 (see Figures~\ref{fig:multi} and \ref{fig:near-IRoverlaid_maps}), and their strong IR excess
has already been noted by \citet{tapia97}. The Class~0~YSO is co-spatial with the IRAS source and
the dense core~A \citep{elia21}, and might be the source of the outflow responsible for the FIR
emission. Both objects are in the IM~regime based on their luminosities obtained from SEDs. The four
objects that are modeled as stars with foreground extinction have photometry only from 1 to 5
$\mu$m. In this range, it is difficult to distinguish between stars with foreground extinction and
stars with disks, so we do not rule out the latter explanation.

\section{Discussion} \label{sec:dis}
Far-IR observations from FIFI-LS confirm the status of Gy~3--7 as a deeply-embedded cluster, as
originally proposed by \cite{tapia97}. Here, we discuss the likely origin of FIR emission in
Gy~3--7 and we search for any effects coming from metallicity by comparison with YSOs from the inner Galaxy.

\subsection{Origin of FIR emission in Gy~3--7}\label{ssec:origin_far_ir_Gy37}

A strong correlation of high$-J$ CO and [\ion{O}{i}] luminosities and their spatial extent in
Gy~3--7 provides a strong support toward the common origin of the two species. The modeling of envelopes
of HM YSOs showed that emission from shocks is necessary to reproduce line fluxes of high$-J$ CO
lines \citep{karska14}. The same is certainly  the case for LM and IM~YSOs as well, which are
characterized by lower envelope densities and temperatures. On the other hand, the [\ion{O}{i}]
emission could be associated with a photodissociation region \citep{kn96,hol97} or outflow shocks
\citep{karska13,nisini15}. Recent velocity-resolved profiles of the [\ion{O}{i}] line at 63 $\mu$m
with SOFIA/GREAT support the later scenario for LM YSOs, which does not suffer from strong
self-absorption \citep{kri17oxy,yang22}. Therefore, we assume that the entire high$-J$ CO and
[\ion{O}{i}] emission originates from outflow shocks. As shown in Section~\ref{ssec:shocks}, the line
luminosity ratio of CO and [\ion{O}{i}] is consistent with $C-$type shocks irradiated by UV fields
of 0.1--10~times the average interstellar radiation field and pre-shock densities of $10^4$--$10^5$
cm$^{-3}$ \citep[see also, ][]{mel15,karska18}. The detection of variable H$_2$O maser further
confirms the outflow activity in the region \citep{fu03}. We provide details in Appendix~\ref{app:maser}.

\begin{figure}[tb]
\includegraphics[width= 0.95\linewidth]{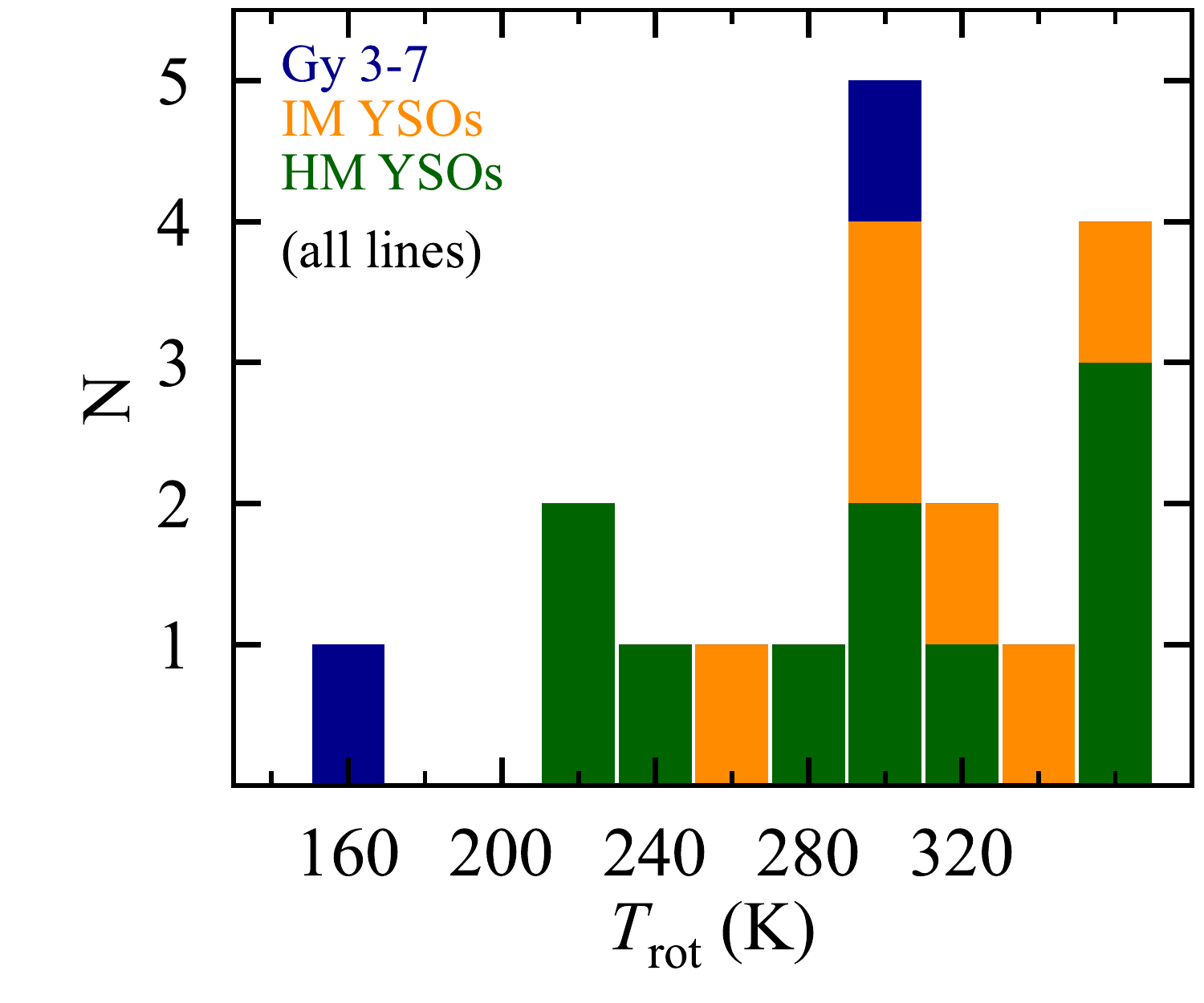}\\
\includegraphics[width= 0.95\linewidth]{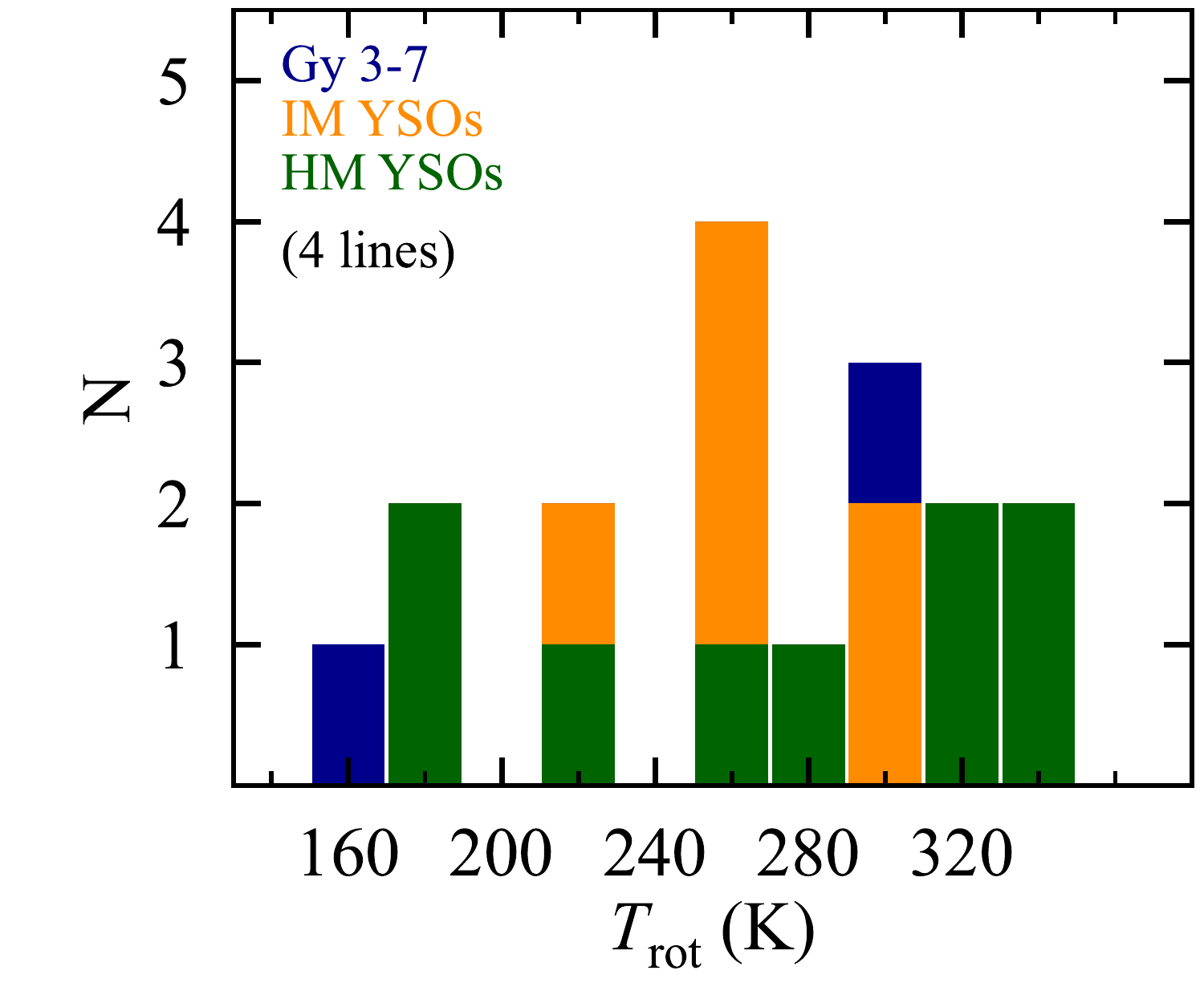}
\caption{CO rotational temperatures for dense cores in Gy~3--7 and intermediate- and high-mass YSOs
from the literature \citep{karska14,matuszak2015}. Top: Straightforward comparison with the
literature values.
Bottom: Comparison accounting for the number of observed CO lines considered in the rotational
diagrams.
\label{fig:hist_Trot}}
\end{figure}

The physical conditions in Gy~3--7 are also constrained by CO rotational temperatures, which provide
a good proxy of gas kinetic temperatures \citep{karska13,vD21}. The high$-J$ transitions are likely
optically thin and thermalized at gas densities >$10^5$~cm$^{-3}$ that are routinely measured for LM and
IM~YSOs \citep[e.g.,][]{kri12,mot17}. The top panel in Figure~\ref{fig:hist_Trot} shows that
$T_{\mathrm{rot,CO}}$ associated with the two dense cores in Gy~3--7 is either fully consistent
(core~A) or at the low-end of other IM YSOs (core~B), for which the median value from the literature
is $320(33)\pm35$~K \citep{matuszak2015}. 
Similar rotational temperatures have been measured for LM and HM YSOs
\citep{karska13,karska14,karska18,green2013,green16,manoj2013,yang18}, with average values of
$328(33)\pm63$~K \citep{karska18} and $300(23)\pm60$~K \citep{karska14}, respectively.

Some differences between the dense core~B in Gy~3--7 and the other YSOs studied in the literature
might result from a smaller number of transitions probed by FIFI-LS. The curvature seen in the
rotational diagrams causes the rotational temperature to be underestimated if high$-J$ CO
transitions are not observed. Therefore, we re-calculated all literature measurements of
$T_\mathrm{rot}$ for IM and HM~YSOs using the same or similar transitions as obtained for Gy~3--7
(Table~\ref{table:rot} and Appendix~\ref{app:Trot_MW}). As expected, CO rotational temperatures for
IM and HM~YSOs are now lower than reported in the literature and consistent even with core~B (see
bottom panel of Figure \ref{fig:hist_Trot}).
\subsection{Possible impact of metallicity on FIR line emission in the outer Galaxy}

Far-IR observations of HM~YSOs in the low-metallicity environments of the SMC and LMC show a lower
fraction of molecular-to-atomic emission with respect to Galactic YSOs \citep{Oliveira19}. Here, we
investigate the impact of bolometric luminosity, $L_\mathrm{bol}$, and source Galactocentric radius,
$R_\mathrm{GC}$, on the ratio of CO and [\ion{O}{i}] line luminosity.

\begin{figure}[tb]
\begin{center}
\includegraphics[width=0.95\linewidth]{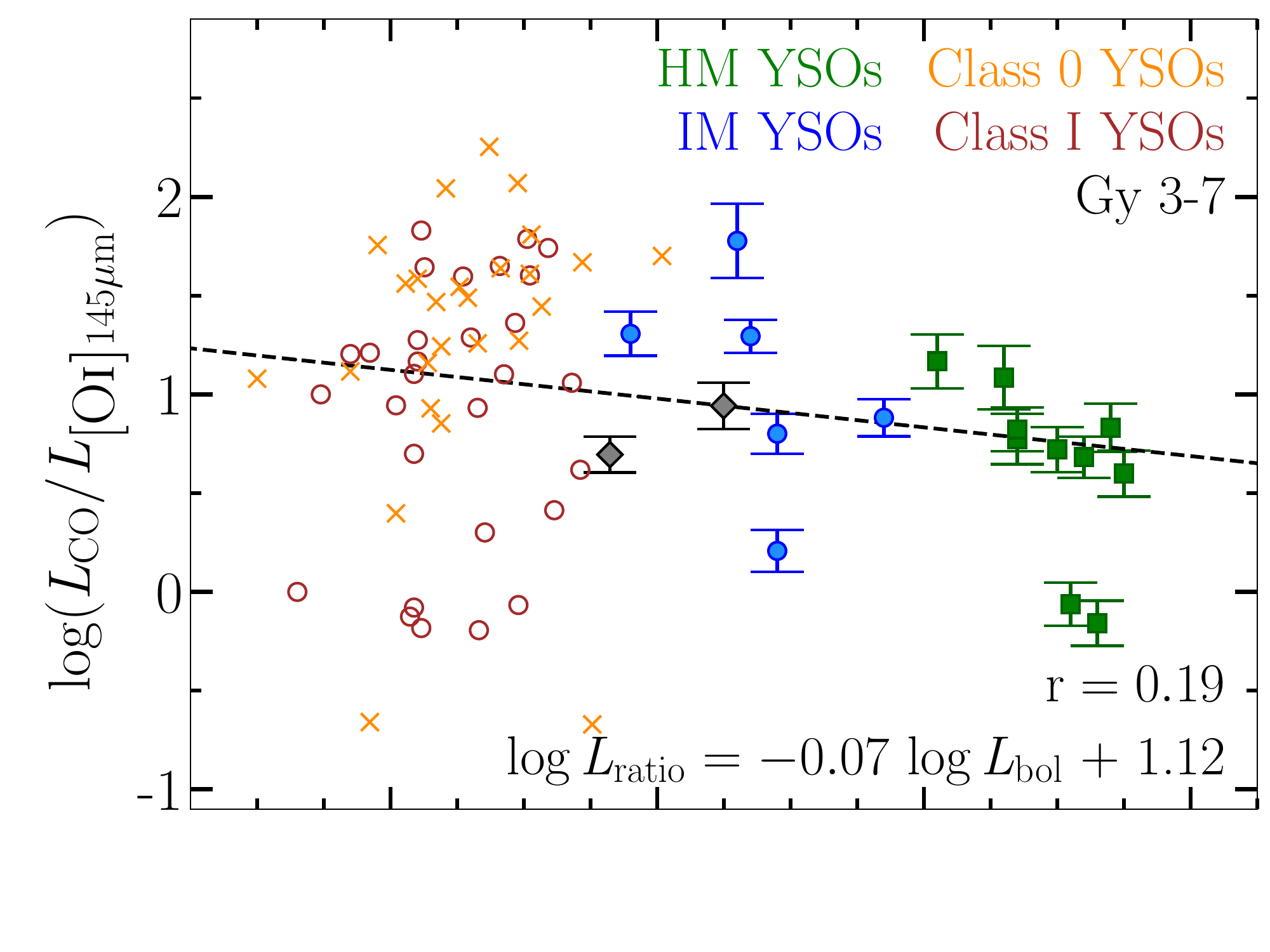}\\
\vspace{-5ex}
\includegraphics[width=0.95\linewidth]{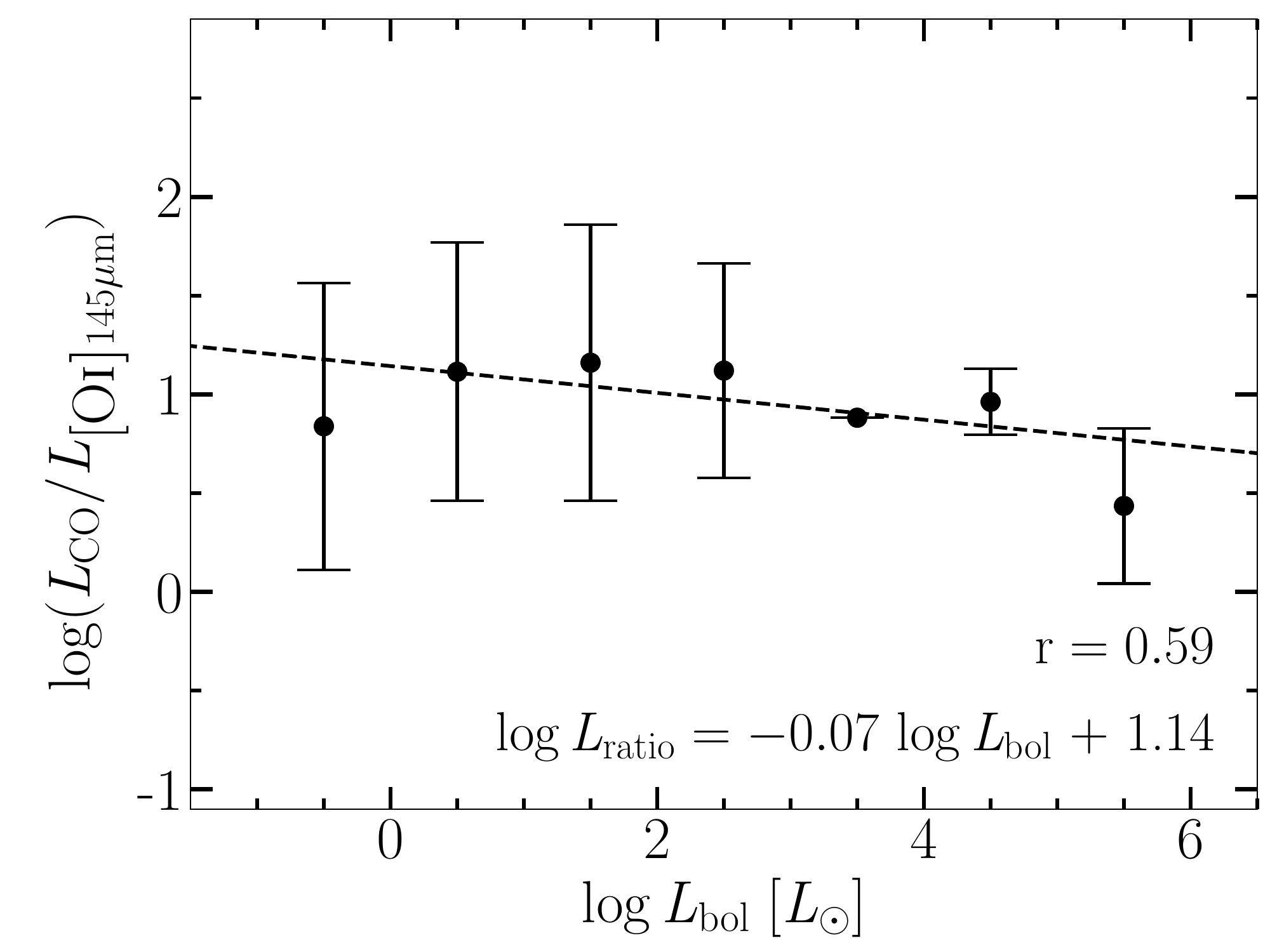} 
\end{center}
\vspace{-3ex}
\caption{Correlations between the ratio of CO and [\ion{O}{i}] line luminosities and bolometric
luminosity of YSOs in the two cores in Gy~3--7 (gray diamonds), 
Class 0 and Class I YSOs \citep[orange \lq$\times$' signs and red circles, respectively;][]{karska18},
 IM YSOs \citep[blue circles;][]{matuszak2015}, and HM~YSOs \citep[green
squares;][]{karska14}. The bolometric luminosities for Gy~3--7 cores are adopted from 
\cite{elia21}. The top panel shows a power-law fit to all individual data points, and the
bottom panel shows the fit to the data bins (both shown as black dashed line). 
\label{fig:correlation_plots}}
\end{figure}

Figure~\ref{fig:correlation_plots} shows the molecular-to-atomic ratio toward Gy~3--7 cores and
other Galactic IM and HM~YSOs as a function of $L_\mathrm{bol}$. The values of $L_\mathrm{bol}$ for
the two cores in Gy~3--7 are adopted from \cite{elia21}, and are equal to 75.9 and
324.2~$L_{\odot}$, respectively. The correlation is characterized by a Pearson coefficient of 0.19,
corresponding to 1.6$\sigma$. Since the number of LM~YSOs exceeds by far the number of IM and HM
sources, we also search for trends in the binned datasets. We bin sources in equal intervals of
log$L_\mathrm{bol}$=1, and adopt a 1$\sigma$ variance of the distribution as the uncertainty inside
the bin.
As a result, we confirm a weak correlation between the ratio of line luminosity and $L_\mathrm{bol}$
($r\sim$0.59, corresponding to 1.5$\sigma$). Thus, our large sample of YSOs starts to reveal a
relationship between the mass of YSOs and the fraction of molecular-to-atomic cooling. This has not
been possible with a sample of 18 LM~YSOs and 10 HM~YSOs analyzed in \cite{karska14}, where a ratio
of $\sim$4 was reported for YSOs in the entire mass regime. We note, however, that the
\textit{Herschel}/PACS measurements from the literature benefited from additional detections of
H$_2$O and OH, which were included in the molecular FIR cooling. 
\begin{figure}[tb]
\centering
\includegraphics[width=0.95\linewidth]{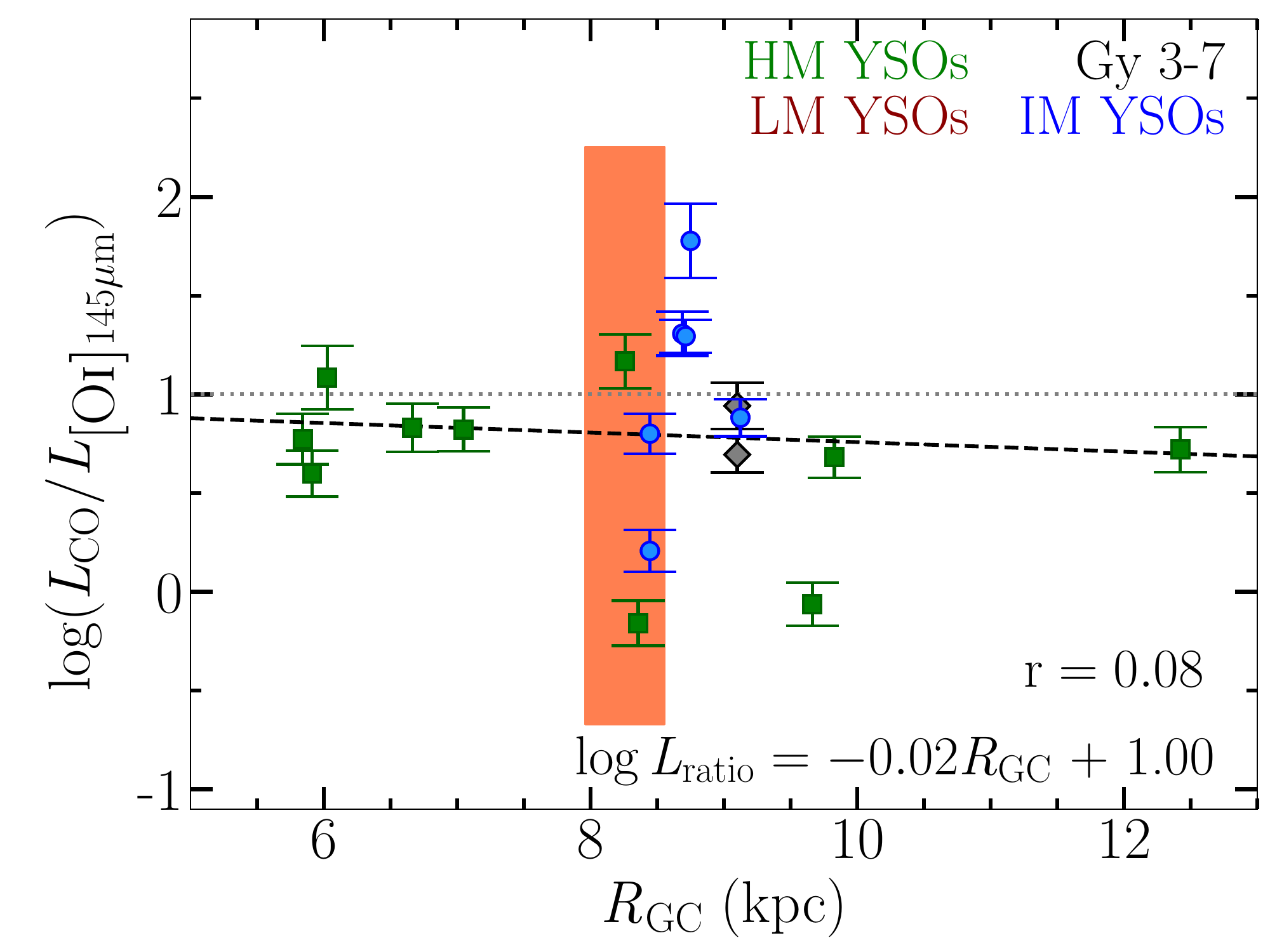}
\caption{Ratio of molecular to atomic line luminosities as a function of the Galactocentric
radius. The light coral box indicates the range of the line luminosity ratio and the Galactocentric
radius for LM~YSOs in the nearby clouds, which are excluded from the power-law fit to the remaining
YSOs (dashed black line). Gray dotted horizontal line indicates where the molecular luminosity
is equal to the atomic luminosity.} 
\label{fig:LCO_LOI145_vsRGC}
\end{figure}
\begin{figure}[tb]
\centering
\includegraphics[width=0.95\linewidth]{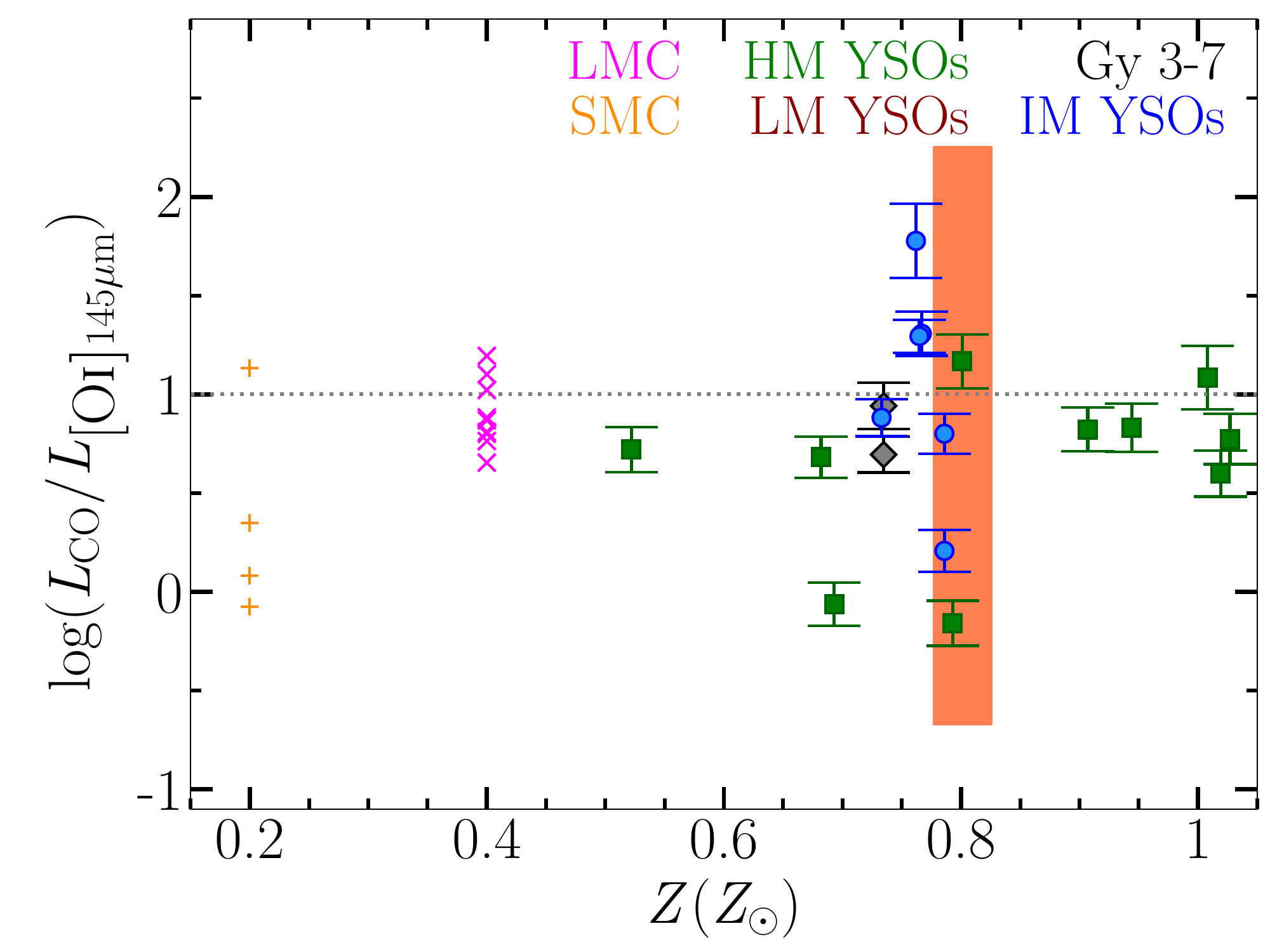}
\caption{Ratio of molecular to atomic line luminosities as a function of metallicity, $Z$.
HM~YSOs in the SMC and LMC are shown with cross and " $\times$" symbols in orange and purple
colors, respectively. The light coral box indicates the range of the line luminosity ratio and the
metallicity for LM~YSOs in the nearby clouds, which are excluded from the power-law fit to the
remaining YSOs (dashed black line). The gray dotted horizontal line indicates where the molecular
luminosity is equal to the atomic luminosity.} 
\label{fig:LCO_LOI145_vs_Z}
\end{figure}

Figure \ref{fig:LCO_LOI145_vsRGC} shows the molecular-to-atomic ratio as a function of
Galactocentric radius. In addition, Figure \ref{fig:LCO_LOI145_vs_Z} shows the same ratio as a
function of metallicity, $Z$, estimated assuming elemental abundance gradients derived from
\ion{H}{ii} regions \citep{balser11}. Appendix~\ref{app:Trot_MW} provides heliocentric
distances, Galactocentric radii, and metallicities of Galactic sources, as well the measurements of
CO and [\ion{O}{i}] for YSOs in both the Milky Way and the Magellanic Clouds.
 
The molecular-to-atomic ratio decreases very weakly with the Galactocentric radius
(Fig.~\ref{fig:LCO_LOI145_vsRGC}) but the Pearson coefficient is too small to confirm the
correlation ($r=0.08$, corresponding to 0.7$\sigma$). Additional observations of YSOs in the outer
Galaxy would be necessary to constrain the molecular-to-atomic ratio, in particular, in the gap
between 10 and 12~kpc. Due to the lack of such observations, we searched for trends as a function of
metallicity by including YSOs in the Magellanic Clouds (Fig. \ref{fig:LCO_LOI145_vs_Z}). We note,
however, that the [\ion{O}{i}] line at 145~$\mu$m was not observed toward sources in the Magellanic
Clouds, and the ratio of the two [\ion{O}{i}] lines calculated for LM~YSOs from \cite{karska18} was
adopted to estimate its fluxes \citep{Oliveira19}. As seen in Figure
\ref{fig:correlation_vs_lumi_oxy}, this ratio may differ for HM~YSOs due to self-absorption or
optical depth effects in the [\ion{O}{i}] 63~$\mu$m line. Additionally, the total CO luminosity of
YSOs in the SMC and LMC was estimated using the detections in SPIRE and the conversion factor from
\cite{yang18}, also based  on LM~YSOs. 

In conclusion, Gy~3--7 follows closely the correlations set by YSOs observed in the Milky Way and
the Magellanic Clouds. The ratio of molecular-to-atomic line emission is dominated by source
bolometric luminosities, and only a very weak decreasing trend with the Galactocentric radius was
detected.

\section{Conclusions}\label{conclusions}

We investigated the SOFIA/FIFI-LS maps of the CO transitions from $J=14-13$ to $J=31-30$, the
[\ion{O}{i}] lines at 63.2~$\mu$m and 145.5~$\mu$m, and the [\ion{C}{ii}] 158~$\mu$m line toward the
outer Galaxy cluster Gy~3--7. Spatial information on the FIR emission enabled us to quantify such 
physical parameters as temperatures, densities, and UV radiation fields, as well as to associate them
with identified YSOs. Our conclusions are as follows:

\begin{itemize}
 \item The CO $J=14-13$ to $J=16-15$ emission lines are detected in a significant part of Gy~3--7,
where \textit{Herschel}/PACS 160~$\mu$m continuum emission is also strong. Higher$-J$ CO lines up to
$31-30$ are clearly detected only toward the dense core~A.
 
 \item The spatial extent of the [\ion{O}{i}] emission at 63 and 145~$\mu$m is similar to that of
CO~$14-13$ and the 70 and 160~$\mu$m continuum emission. The [\ion{C}{ii}] emission is also
extended, but shows systematic shifts in the emission peaks away from the FIR continuum, tracing
lower density gas.
 
 \item The CO rotational diagrams show the warm components toward two cores with 
 \Trot~of 305 and 155~K, and a range of \Trot~from $\sim$105 to $230$~K throughout the cluster,
where only three high$-J$ CO lines are unambiguously detected. Similar rotational temperatures have
been detected toward IM and HM~YSOs in the inner Milky Way calculated using the same or similar CO
transitions.

\item A strong correlation of the CO and [\ion{O}{i}] line luminosities and their similar spatial
extent point at the common origin in the outflow shocks. The CO / [\ion{O}{i}] line luminosity ratio
of Gy~3--7 cores and other intermediate-mass YSOs is consistent with C-type shocks propagating at
pre-shock densities of $10^4 - 10^5$~cm$^{-3}$ and UV fields of 0.1-10 times the average
interstellar radiation field.
 
 \item Physical parameters for 15 YSO candidates in the Gy~3--7 cluster are obtained from a YSO SED
model fitting \citep{robitaille2017}. Two sources, corresponding to Hi-GAL dense cores from
\cite{elia21}, are well-fitted with YSO models including the envelope, confirming their early
evolutionary stage (Class 0/I). The location of the Class 0 source at the center of Gy~3--7 cluster
suggests that it might be the driving source of the outflow revealed by FIR emission.
 
 \item The ratio of warm CO and [\ion{O}{i}] at 145~$\mu$m line luminosities from protostellar
envelopes shows a weak decreasing trend with the bolometric luminosity and Galactocentric radius. We
do not identify any significant dependence of the line cooling in Gy~3--7 on metallicity. 
 
\end{itemize}

High-resolution submillimeter observations would be necessary 
to unambiguously associate FIR emission from FIFI-LS with candidate YSOs and 
their outflows. Alternatively, the efficient imaging of Gy 3-7 with the Mid-Infrared Instrument
 on board James Webb Space Telescope in F560W and/or F770 W filters tracing H$_2$ emission
  would unveil the details in the outflows \citep{yang22b}.

\begin{acknowledgements}
NL, AK, MF, MG, MK, and KK acknowledge support from the First TEAM grant of the Foundation for
Polish Science No. POIR.04.04.00-00-5D21/18-00 (PI: A. Karska). AK also acknowledges support from the
 Polish National Agency for Academic Exchange grant No. BPN/BEK/2021/1/00319/DEC/1.
This article has been supported by the Polish
National Science Center grant 2014/15/B/ST9/02111 and 2016/21/D/ST9/01098. 
The material is based upon work supported by NASA under award number 80GSFC21M0002 (MS).  The
research is supported by a research grant (19127) from VILLUM FONDEN (LEK). This work is based (in
part) on observations made with the NASA/DLR Stratospheric Observatory for Infrared Astronomy
(SOFIA). SOFIA is jointly operated by the Universities Space Research Association, Inc. (USRA),
under NASA contract NNA17BF53C, and the Deutsches SOFIA Institut (DSI) under DLR contract 50 OK 2002
to the University of Stuttgart. The 32~m radio telescope is operated by the Institute of Astronomy,
Nicolaus Copernicus University and supported by the Polish Ministry of Science and Higher Education
SpUB grant.

\end{acknowledgements}

\bibliographystyle{aa} 
\bibliography{gy37} 

\begin{thebibliography}{130}
\expandafter\ifx\csname natexlab\endcsname\relax\def\natexlab#1{#1}\fi

\bibitem[{{Arce} {et~al.}(2007){Arce}, {Shepherd}, {Gueth}, {Lee}, {Bachiller},
  {Rosen}, \& {Beuther}}]{arce07}
{Arce}, H.~G., {Shepherd}, D., {Gueth}, F., {et~al.} 2007, in Protostars and
  Planets V, ed. B.~{Reipurth}, D.~{Jewitt}, \& K.~{Keil}, 245

\bibitem[{{Balser} {et~al.}(2011){Balser}, {Rood}, {Bania}, \&
  {Anderson}}]{balser11}
{Balser}, D.~S., {Rood}, R.~T., {Bania}, T.~M., \& {Anderson}, L.~D. 2011,
  \apj, 738, 27

\bibitem[{{Benedettini} {et~al.}(2020){Benedettini}, {Molinari}, {Baldeschi},
  {Beltr{\'a}n}, {Brand}, {Cesaroni}, {Elia}, {Fontani}, {Merello}, {Olmi},
  {Pezzuto}, {Rygl}, {Schisano}, {Testi}, \& {Traficante}}]{bene20}
{Benedettini}, M., {Molinari}, S., {Baldeschi}, A., {et~al.} 2020, \aap, 633,
  A147

\bibitem[{{Benz} {et~al.}(2016){Benz}, {Bruderer}, {van Dishoeck}, {Melchior},
  {Wampfler}, {van der Tak}, {Goicoechea}, {Indriolo}, {Kristensen}, {Lis},
  {Mottram}, {Bergin}, {Caselli}, {Herpin}, {Hogerheijde}, {Johnstone},
  {Liseau}, {Nisini}, {Tafalla}, {Visser}, \& {Wyrowski}}]{Ben16}
{Benz}, A.~O., {Bruderer}, S., {van Dishoeck}, E.~F., {et~al.} 2016, \aap, 590,
  A105

\bibitem[{{Bica} {et~al.}(2003){Bica}, {Dutra}, \& {Barbuy}}]{bica03}
{Bica}, E., {Dutra}, C.~M., \& {Barbuy}, B. 2003, \aap, 397, 177

\bibitem[{{Bressan} {et~al.}(2012){Bressan}, {Marigo}, {Girardi}, {Salasnich},
  {Dal Cero}, {Rubele}, \& {Nanni}}]{bressan2012}
{Bressan}, A., {Marigo}, P., {Girardi}, L., {et~al.} 2012, \mnras, 427, 127

\bibitem[{{Bronfman} {et~al.}(1996){Bronfman}, {Nyman}, \& {May}}]{Bronfman96}
{Bronfman}, L., {Nyman}, L.~A., \& {May}, J. 1996, \aaps, 115, 81

\bibitem[{{Bruderer} {et~al.}(2009){Bruderer}, {Benz}, {Doty}, {van Dishoeck},
  \& {Bourke}}]{bru09}
{Bruderer}, S., {Benz}, A.~O., {Doty}, S.~D., {van Dishoeck}, E.~F., \&
  {Bourke}, T.~L. 2009, \apj, 700, 872

\bibitem[{{Butner} {et~al.}(1990){Butner}, {Evans}, {Harvey}, {Mundy}, {Natta},
  \& {Randich}}]{Butner90}
{Butner}, H.~M., {Evans}, Neal~J., I., {Harvey}, P.~M., {et~al.} 1990, \apj,
  364, 164

\bibitem[{{Chen} {et~al.}(2015){Chen}, {Bressan}, {Girardi}, {Marigo}, {Kong},
  \& {Lanza}}]{chen2015}
{Chen}, Y., {Bressan}, A., {Girardi}, L., {et~al.} 2015, \mnras, 452, 1068

\bibitem[{{Chen} {et~al.}(2014){Chen}, {Girardi}, {Bressan}, {Marigo},
  {Barbieri}, \& {Kong}}]{chen2014}
{Chen}, Y., {Girardi}, L., {Bressan}, A., {et~al.} 2014, \mnras, 444, 2525

\bibitem[{{Cyganowski} {et~al.}(2011){Cyganowski}, {Brogan}, {Hunter},
  {Churchwell}, \& {Zhang}}]{Cyganowski11}
{Cyganowski}, C.~J., {Brogan}, C.~L., {Hunter}, T.~R., {Churchwell}, E., \&
  {Zhang}, Q. 2011, \apj, 729, 124

\bibitem[{{Cyganowski} {et~al.}(2008){Cyganowski}, {Whitney}, {Holden},
  {Braden}, {Brogan}, {Churchwell}, {Indebetouw}, {Watson}, {Babler},
  {Benjamin}, {Gomez}, {Meade}, {Povich}, {Robitaille}, \&
  {Watson}}]{Cyganowski08}
{Cyganowski}, C.~J., {Whitney}, B.~A., {Holden}, E., {et~al.} 2008, \aj, 136,
  2391

\bibitem[{{Dunham} {et~al.}(2015){Dunham}, {Allen}, {Evans},
  {Broekhoven-Fiene}, {Cieza}, {Di Francesco}, {Gutermuth}, {Harvey},
  {Hatchell}, {Heiderman}, {Huard}, {Johnstone}, {Kirk}, {Matthews}, {Miller},
  {Peterson}, \& {Young}}]{dunham2015}
{Dunham}, M.~M., {Allen}, L.~E., {Evans}, Neal~J., I., {et~al.} 2015, \apjs,
  220, 11

\bibitem[{{Elia} {et~al.}(2021){Elia}, {Merello}, {Molinari}, {Schisano},
  {Zavagno}, {Russeil}, {M{\`e}ge}, {Martin}, {Olmi}, {Pestalozzi}, {Plume},
  {Ragan}, {Benedettini}, {Eden}, {Moore}, {Noriega-Crespo}, {Paladini},
  {Palmeirim}, {Pezzuto}, {Pilbratt}, {Rygl}, {Schilke}, {Strafella}, {Tan},
  {Traficante}, {Baldeschi}, {Bally}, {di Giorgio}, {Fiorellino}, {Liu},
  {Piazzo}, \& {Polychroni}}]{elia21}
{Elia}, D., {Merello}, M., {Molinari}, S., {et~al.} 2021, \mnras, 504, 2742

\bibitem[{{Elia} {et~al.}(2013){Elia}, {Molinari}, {Fukui}, {Schisano}, {Olmi},
  {Veneziani}, {Hayakawa}, {Pestalozzi}, {Schneider}, {Benedettini}, {di
  Giorgio}, {Ikhenaode}, {Mizuno}, {Onishi}, {Pezzuto}, {Piazzo}, {Polychroni},
  {Rygl}, {Yamamoto}, \& {Maruccia}}]{Elia13}
{Elia}, D., {Molinari}, S., {Fukui}, Y., {et~al.} 2013, \apj, 772, 45

\bibitem[{{Elitzur} {et~al.}(1989){Elitzur}, {Hollenbach}, \&
  {McKee}}]{elitzur1989}
{Elitzur}, M., {Hollenbach}, D.~J., \& {McKee}, C.~F. 1989, \apj, 346, 983

\bibitem[{{Esteban} \& {Garc{\'\i}a-Rojas}(2018)}]{esteban18}
{Esteban}, C. \& {Garc{\'\i}a-Rojas}, J. 2018, \mnras, 478, 2315

\bibitem[{{Fa{\'u}ndez} {et~al.}(2004){Fa{\'u}ndez}, {Bronfman}, {Garay},
  {Chini}, {Nyman}, \& {May}}]{Faundez04}
{Fa{\'u}ndez}, S., {Bronfman}, L., {Garay}, G., {et~al.} 2004, \aap, 426, 97

\bibitem[{{Fern{\'a}ndez-Mart{\'\i}n}
  {et~al.}(2017){Fern{\'a}ndez-Mart{\'\i}n}, {P{\'e}rez-Montero},
  {V{\'\i}lchez}, \& {Mampaso}}]{fernandez2017}
{Fern{\'a}ndez-Mart{\'\i}n}, A., {P{\'e}rez-Montero}, E., {V{\'\i}lchez},
  J.~M., \& {Mampaso}, A. 2017, \aap, 597, A84

\bibitem[{{Fischer} {et~al.}(2018){Fischer}, {Beckmann}, {Bryant}, {Colditz},
  {Fumi}, {Geis}, {Hamidouche}, {Henning}, {H{\"o}nle}, {Iserlohe}, {Klein},
  {Krabbe}, {Looney}, {Poglitsch}, {Raab}, {Rebell}, {Rosenthal}, {Savage},
  {Schweitzer}, {Trinh}, \& {Vacca}}]{fischer18}
{Fischer}, C., {Beckmann}, S., {Bryant}, A., {et~al.} 2018, Journal of
  Astronomical Instrumentation, 7, 1840003

\bibitem[{{Fischer} {et~al.}(2021){Fischer}, {Iserlohe}, {Vacca}, {Fadda},
  {Colditz}, {Fischer}, \& {Krabbe}}]{fischerPWV}
{Fischer}, C., {Iserlohe}, C., {Vacca}, W., {et~al.} 2021, \pasp, 133, 055001

\bibitem[{{Flower} \& {Pineau des For{\^e}ts}(2012)}]{fl12}
{Flower}, D.~R. \& {Pineau des For{\^e}ts}, G. 2012, \mnras, 421, 2786

\bibitem[{{Frank} {et~al.}(2014){Frank}, {Ray}, {Cabrit}, {Hartigan}, {Arce},
  {Bacciotti}, {Bally}, {Benisty}, {Eisl{\"o}ffel}, {G{\"u}del}, {Lebedev},
  {Nisini}, \& {Raga}}]{frank14}
{Frank}, A., {Ray}, T.~P., {Cabrit}, S., {et~al.} 2014, in Protostars and
  Planets VI, ed. H.~{Beuther}, R.~S. {Klessen}, C.~P. {Dullemond}, \&
  T.~{Henning}, 451

\bibitem[{{Furuya} {et~al.}(2003{\natexlab{a}}){Furuya}, {Kitamura}, {Wootten},
  {Claussen}, \& {Kawabe}}]{fu03}
{Furuya}, R.~S., {Kitamura}, Y., {Wootten}, A., {Claussen}, M.~J., \& {Kawabe},
  R. 2003{\natexlab{a}}, \apjs, 144, 71

\bibitem[{{Furuya} {et~al.}(2003{\natexlab{b}}){Furuya}, {Kitamura}, {Wootten},
  {Claussen}, \& {Kawabe}}]{furuya2003}
{Furuya}, R.~S., {Kitamura}, Y., {Wootten}, A., {Claussen}, M.~J., \& {Kawabe},
  R. 2003{\natexlab{b}}, \apjs, 144, 71

\bibitem[{{Giannini} {et~al.}(2005){Giannini}, {Massi}, {Podio}, {Lorenzetti},
  {Nisini}, {Caratti o Garatti}, {Liseau}, {Lo Curto}, \&
  {Vitali}}]{Giannini05}
{Giannini}, T., {Massi}, F., {Podio}, L., {et~al.} 2005, \aap, 433, 941

\bibitem[{{Goicoechea} {et~al.}(2012){Goicoechea}, {Cernicharo}, {Karska},
  {Herczeg}, {Polehampton}, {Wampfler}, {Kristensen}, {van Dishoeck},
  {Etxaluze}, {Bern{\'e}}, \& {Visser}}]{goi12}
{Goicoechea}, J.~R., {Cernicharo}, J., {Karska}, A., {et~al.} 2012, \aap, 548,
  A77

\bibitem[{{Goldsmith} \& {Langer}(1978)}]{GL78}
{Goldsmith}, P.~F. \& {Langer}, W.~D. 1978, \apj, 222, 881

\bibitem[{{Goldsmith} \& {Langer}(1999)}]{gl1999}
{Goldsmith}, P.~F. \& {Langer}, W.~D. 1999, \apj, 517, 209

\bibitem[{{Graczyk} {et~al.}(2014){Graczyk}, {Pietrzy{\'n}ski}, {Thompson},
  {Gieren}, {Pilecki}, {Konorski}, {Udalski}, {Soszy{\'n}ski}, {Villanova},
  {G{\'o}rski}, {Suchomska}, {Karczmarek}, {Kudritzki}, {Bresolin}, \&
  {Gallenne}}]{Graczyk14}
{Graczyk}, D., {Pietrzy{\'n}ski}, G., {Thompson}, I.~B., {et~al.} 2014, \apj,
  780, 59

\bibitem[{{Green} {et~al.}(2013){Green}, {Evans}, {J{\o}rgensen}, {Herczeg},
  {Kristensen}, {Lee}, {Dionatos}, {Yildiz}, {Salyk}, {Meeus}, {Bouwman},
  {Visser}, {Bergin}, {van Dishoeck}, {Rascati}, {Karska}, {van Kempen},
  {Dunham}, {Lindberg}, {Fedele}, \& {DIGIT Team}}]{green2013}
{Green}, J.~D., {Evans}, Neal~J., I., {J{\o}rgensen}, J.~K., {et~al.} 2013,
  \apj, 770, 123

\bibitem[{{Green} {et~al.}(2016){Green}, {Yang}, {Evans}, {Karska}, {Herczeg},
  {van Dishoeck}, {Lee}, {Larson}, \& {Bouwman}}]{green16}
{Green}, J.~D., {Yang}, Y.-L., {Evans}, Neal~J., I., {et~al.} 2016, \aj, 151,
  75

\bibitem[{{Griffin} {et~al.}(2010){Griffin}, {Abergel}, {Abreu}, {Ade},
  {Andr{\'e}}, {Augueres}, {Babbedge}, {Bae}, {Baillie}, {Baluteau}, {Barlow},
  {Bendo}, {Benielli}, {Bock}, {Bonhomme}, {Brisbin}, {Brockley-Blatt},
  {Caldwell}, {Cara}, {Castro-Rodriguez}, {Cerulli}, {Chanial}, {Chen},
  {Clark}, {Clements}, {Clerc}, {Coker}, {Communal}, {Conversi}, {Cox},
  {Crumb}, {Cunningham}, {Daly}, {Davis}, {de Antoni}, {Delderfield}, {Devin},
  {di Giorgio}, {Didschuns}, {Dohlen}, {Donati}, {Dowell}, {Dowell}, {Duband},
  {Dumaye}, {Emery}, {Ferlet}, {Ferrand}, {Fontignie}, {Fox}, {Franceschini},
  {Frerking}, {Fulton}, {Garcia}, {Gastaud}, {Gear}, {Glenn}, {Goizel},
  {Griffin}, {Grundy}, {Guest}, {Guillemet}, {Hargrave}, {Harwit}, {Hastings},
  {Hatziminaoglou}, {Herman}, {Hinde}, {Hristov}, {Huang}, {Imhof}, {Isaak},
  {Israelsson}, {Ivison}, {Jennings}, {Kiernan}, {King}, {Lange}, {Latter},
  {Laurent}, {Laurent}, {Leeks}, {Lellouch}, {Levenson}, {Li}, {Li},
  {Lilienthal}, {Lim}, {Liu}, {Lu}, {Madden}, {Mainetti}, {Marliani}, {McKay},
  {Mercier}, {Molinari}, {Morris}, {Moseley}, {Mulder}, {Mur}, {Naylor},
  {Nguyen}, {O'Halloran}, {Oliver}, {Olofsson}, {Olofsson}, {Orfei}, {Page},
  {Pain}, {Panuzzo}, {Papageorgiou}, {Parks}, {Parr-Burman}, {Pearce},
  {Pearson}, {P{\'e}rez-Fournon}, {Pinsard}, {Pisano}, {Podosek}, {Pohlen},
  {Polehampton}, {Pouliquen}, {Rigopoulou}, {Rizzo}, {Roseboom}, {Roussel},
  {Rowan-Robinson}, {Rownd}, {Saraceno}, {Sauvage}, {Savage}, {Savini},
  {Sawyer}, {Scharmberg}, {Schmitt}, {Schneider}, {Schulz}, {Schwartz},
  {Shafer}, {Shupe}, {Sibthorpe}, {Sidher}, {Smith}, {Smith}, {Smith},
  {Spencer}, {Stobie}, {Sudiwala}, {Sukhatme}, {Surace}, {Stevens}, {Swinyard},
  {Trichas}, {Tourette}, {Triou}, {Tseng}, {Tucker}, {Turner}, {Vaccari},
  {Valtchanov}, {Vigroux}, {Virique}, {Voellmer}, {Walker}, {Ward}, {Waskett},
  {Weilert}, {Wesson}, {White}, {Whitehouse}, {Wilson}, {Winter}, {Woodcraft},
  {Wright}, {Xu}, {Zavagno}, {Zemcov}, {Zhang}, \& {Zonca}}]{griffin10}
{Griffin}, M.~J., {Abergel}, A., {Abreu}, A., {et~al.} 2010, \aap, 518, L3

\bibitem[{{Guevara} {et~al.}(2020){Guevara}, {Stutzki}, {Ossenkopf-Okada},
  {Simon}, {P{\'e}rez-Beaupuits}, {Beuther}, {Bihr}, {Higgins}, {Graf}, \&
  {G{\"u}sten}}]{guevara2020}
{Guevara}, C., {Stutzki}, J., {Ossenkopf-Okada}, V., {et~al.} 2020, \aap, 636,
  A16

\bibitem[{{Gyulbudaghian}(2012)}]{gy12}
{Gyulbudaghian}, A.~L. 2012, Astrophysics, 55, 92

\bibitem[{{Habing}(1968)}]{hab68}
{Habing}, H.~J. 1968, \bain, 19, 421

\bibitem[{{Hachisuka} {et~al.}(2006){Hachisuka}, {Brunthaler}, {Menten},
  {Reid}, {Imai}, {Hagiwara}, {Miyoshi}, {Horiuchi}, \& {Sasao}}]{Hachisuka06}
{Hachisuka}, K., {Brunthaler}, A., {Menten}, K.~M., {et~al.} 2006, \apj, 645,
  337

\bibitem[{{Hatchell} \& {van der Tak}(2003)}]{Hatchell03}
{Hatchell}, J. \& {van der Tak}, F.~F.~S. 2003, \aap, 409, 589

\bibitem[{{He} {et~al.}(2012){He}, {Takahashi}, \& {Chen}}]{he12}
{He}, J.~H., {Takahashi}, S., \& {Chen}, X. 2012, \apjs, 202, 1

\bibitem[{{Herczeg} {et~al.}(2012){Herczeg}, {Karska}, {Bruderer},
  {Kristensen}, {van Dishoeck}, {J{\o}rgensen}, {Visser}, {Wampfler}, {Bergin},
  {Y{\i}ld{\i}z}, {Pontoppidan}, \& {Gracia-Carpio}}]{her12}
{Herczeg}, G.~J., {Karska}, A., {Bruderer}, S., {et~al.} 2012, \aap, 540, A84

\bibitem[{{Heyer} \& {Dame}(2015)}]{hd15}
{Heyer}, M. \& {Dame}, T.~M. 2015, \araa, 53, 583

\bibitem[{{Hollenbach} \& {McKee}(1989)}]{Hol89}
{Hollenbach}, D. \& {McKee}, C.~F. 1989, \apj, 342, 306

\bibitem[{{Hollenbach} \& {Tielens}(1997)}]{hol97}
{Hollenbach}, D.~J. \& {Tielens}, A.~G.~G.~M. 1997, \araa, 35, 179

\bibitem[{{Immer} {et~al.}(2013){Immer}, {Reid}, {Menten}, {Brunthaler}, \&
  {Dame}}]{Immer13}
{Immer}, K., {Reid}, M.~J., {Menten}, K.~M., {Brunthaler}, A., \& {Dame}, T.~M.
  2013, \aap, 553, A117

\bibitem[{{Iserlohe} {et~al.}(2021){Iserlohe}, {Fischer}, {Vacca}, {Fischer},
  {Colditz}, \& {Krabbe}}]{iserlohePWV}
{Iserlohe}, C., {Fischer}, C., {Vacca}, W.~D., {et~al.} 2021, \pasp, 133,
  055002

\bibitem[{{Jakob} {et~al.}(2007){Jakob}, {Kramer}, {Simon}, {Schneider},
  {Ossenkopf}, {Bontemps}, {Graf}, \& {Stutzki}}]{Jakob07}
{Jakob}, H., {Kramer}, C., {Simon}, R., {et~al.} 2007, \aap, 461, 999

\bibitem[{{Jim{\'e}nez-Donaire} {et~al.}(2017){Jim{\'e}nez-Donaire}, {Meeus},
  {Karska}, {Montesinos}, {Bouwman}, {Eiroa}, \& {Henning}}]{jim17}
{Jim{\'e}nez-Donaire}, M.~J., {Meeus}, G., {Karska}, A., {et~al.} 2017, \aap,
  605, A62

\bibitem[{{Johnstone} {et~al.}(2010){Johnstone}, {Fich}, {McCoey}, {van
  Kempen}, {Fuente}, {Kristensen}, {Cernicharo}, {Caselli}, {Visser}, {Plume},
  {Herczeg}, {van Dishoeck}, {Wampfler}, {Bachiller}, {Baudry}, {Benedettini},
  {Bergin}, {Benz}, {Bjerkeli}, {Blake}, {Bontemps}, {Braine}, {Bruderer},
  {Codella}, {Daniel}, {di Giorgio}, {Dominik}, {Doty}, {Encrenaz}, {Giannini},
  {Goicoechea}, {de Graauw}, {Helmich}, {Herpin}, {Hogerheijde}, {Jacq},
  {J{\o}rgensen}, {Larsson}, {Lis}, {Liseau}, {Marseille}, {Melnick},
  {Neufeld}, {Nisini}, {Olberg}, {Parise}, {Pearson}, {Risacher},
  {Santiago-Garc{\'\i}a}, {Saraceno}, {Shipman}, {Tafalla}, {van der Tak},
  {Wyrowski}, {Y{\i}ld{\i}z}, {Caux}, {Honingh}, {Jellema}, {Schieder},
  {Teyssier}, \& {Whyborn}}]{Johnstone10}
{Johnstone}, D., {Fich}, M., {McCoey}, C., {et~al.} 2010, \aap, 521, L41

\bibitem[{{Karska} {et~al.}(2013){Karska}, {Herczeg}, {van Dishoeck},
  {Wampfler}, {Kristensen}, {Goicoechea}, {Visser}, {Nisini}, {San
  Jos{\'e}-Garc{\'\i}a}, {Bruderer}, {{\'S}niady}, {Doty}, {Fedele},
  {Y{\i}ld{\i}z}, {Benz}, {Bergin}, {Caselli}, {Herpin}, {Hogerheijde},
  {Johnstone}, {J{\o}rgensen}, {Liseau}, {Tafalla}, {van der Tak}, \&
  {Wyrowski}}]{karska13}
{Karska}, A., {Herczeg}, G.~J., {van Dishoeck}, E.~F., {et~al.} 2013, \aap,
  552, A141

\bibitem[{{Karska} {et~al.}(2014{\natexlab{a}}){Karska}, {Herpin}, {Bruderer},
  {Goicoechea}, {Herczeg}, {van Dishoeck}, {San Jos{\'e}-Garc{\'\i}a},
  {Contursi}, {Feuchtgruber}, {Fedele}, {Baudry}, {Braine}, {Chavarr{\'\i}a},
  {Cernicharo}, {van der Tak}, \& {Wyrowski}}]{karska14}
{Karska}, A., {Herpin}, F., {Bruderer}, S., {et~al.} 2014{\natexlab{a}}, \aap,
  562, A45

\bibitem[{{Karska} {et~al.}(2018){Karska}, {Kaufman}, {Kristensen}, {van
  Dishoeck}, {Herczeg}, {Mottram}, {Tychoniec}, {Lindberg}, {Evans}, {Green},
  {Yang}, {Gusdorf}, {Itrich}, \& {Si{\'o}dmiak}}]{karska18}
{Karska}, A., {Kaufman}, M.~J., {Kristensen}, L.~E., {et~al.} 2018, \apjs, 235,
  30

\bibitem[{{Karska} {et~al.}(2022){Karska}, {Koprowski}, {Solarz}, {Szczerba},
  {Sewi{\l}o}, {Si{\'o}dmiak}, {Elia}, {Gawro{\'n}ski}, {Grzesiak}, {Yung},
  {Fischer}, \& {Kristensen}}]{karska22}
{Karska}, A., {Koprowski}, M., {Solarz}, A., {et~al.} 2022, arXiv e-prints,
  arXiv:2204.11687

\bibitem[{{Karska} {et~al.}(2014{\natexlab{b}}){Karska}, {Kristensen}, {van
  Dishoeck}, {Drozdovskaya}, {Mottram}, {Herczeg}, {Bruderer}, {Cabrit},
  {Evans}, {Fedele}, {Gusdorf}, {J{\o}rgensen}, {Kaufman}, {Melnick},
  {Neufeld}, {Nisini}, {Santangelo}, {Tafalla}, \& {Wampfler}}]{karska14b}
{Karska}, A., {Kristensen}, L.~E., {van Dishoeck}, E.~F., {et~al.}
  2014{\natexlab{b}}, \aap, 572, A9

\bibitem[{{Kaufman} \& {Neufeld}(1996{\natexlab{a}})}]{kn96}
{Kaufman}, M.~J. \& {Neufeld}, D.~A. 1996{\natexlab{a}}, \apj, 456, 611

\bibitem[{{Kaufman} \& {Neufeld}(1996{\natexlab{b}})}]{kau96}
{Kaufman}, M.~J. \& {Neufeld}, D.~A. 1996{\natexlab{b}}, \apj, 456, 611

\bibitem[{{Kaufman} {et~al.}(2006){Kaufman}, {Wolfire}, \&
  {Hollenbach}}]{Kau06}
{Kaufman}, M.~J., {Wolfire}, M.~G., \& {Hollenbach}, D.~J. 2006, \apj, 644, 283

\bibitem[{{Klein} {et~al.}(2014){Klein}, {Beckmann}, {Bryant}, {Colditz},
  {Fischer}, {Fumi}, {Geis}, {H{\"o}nle}, {Krabbe}, {Looney}, {Poglitsch},
  {Raab}, {Rebell}, \& {Savage}}]{fifi}
{Klein}, R., {Beckmann}, S., {Bryant}, A., {et~al.} 2014, in Society of
  Photo-Optical Instrumentation Engineers (SPIE) Conference Series, Vol. 9147,
  Ground-based and Airborne Instrumentation for Astronomy V, ed. S.~K.
  {Ramsay}, I.~S. {McLean}, \& H.~{Takami}, 91472X

\bibitem[{{Kristensen} {et~al.}(2017{\natexlab{a}}){Kristensen}, {Gusdorf},
  {Mottram}, {Karska}, {Visser}, {Wiesemeyer}, {G{\"u}sten}, \&
  {Simon}}]{kri17oxy}
{Kristensen}, L.~E., {Gusdorf}, A., {Mottram}, J.~C., {et~al.}
  2017{\natexlab{a}}, \aap, 601, L4

\bibitem[{{Kristensen} {et~al.}(2012){Kristensen}, {van Dishoeck}, {Bergin},
  {Visser}, {Y{\i}ld{\i}z}, {San Jose-Garcia}, {J{\o}rgensen}, {Herczeg},
  {Johnstone}, {Wampfler}, {Benz}, {Bruderer}, {Cabrit}, {Caselli}, {Doty},
  {Harsono}, {Herpin}, {Hogerheijde}, {Karska}, {van Kempen}, {Liseau},
  {Nisini}, {Tafalla}, {van der Tak}, \& {Wyrowski}}]{kri12}
{Kristensen}, L.~E., {van Dishoeck}, E.~F., {Bergin}, E.~A., {et~al.} 2012,
  \aap, 542, A8

\bibitem[{{Kristensen} {et~al.}(2017{\natexlab{b}}){Kristensen}, {van
  Dishoeck}, {Mottram}, {Karska}, {Y{\i}ld{\i}z}, {Bergin}, {Bjerkeli},
  {Cabrit}, {Doty}, {Evans}, {Gusdorf}, {Harsono}, {Herczeg}, {Johnstone},
  {J{\o}rgensen}, {van Kempen}, {Lee}, {Maret}, {Tafalla}, {Visser}, \&
  {Wampfler}}]{kri17co}
{Kristensen}, L.~E., {van Dishoeck}, E.~F., {Mottram}, J.~C., {et~al.}
  2017{\natexlab{b}}, \aap, 605, A93

\bibitem[{{Kuchar} \& {Bania}(1994)}]{Kuchar_Bania94}
{Kuchar}, T.~A. \& {Bania}, T.~M. 1994, \apj, 436, 117

\bibitem[{{Ladd} {et~al.}(1993){Ladd}, {Deane}, {Sanders}, \&
  {Wynn-Williams}}]{Ladd93}
{Ladd}, E.~F., {Deane}, J.~R., {Sanders}, D.~B., \& {Wynn-Williams}, C.~G.
  1993, \apj, 419, 186

\bibitem[{{Ladeyschikov} {et~al.}(2022){Ladeyschikov}, {Gong}, {Sobolev},
  {Menten}, {Urquhart}, {Breen}, {Shakhvorostova}, {Bayandina}, \&
  {Tsivilev}}]{ladeyschikov2022}
{Ladeyschikov}, D.~A., {Gong}, Y., {Sobolev}, A.~M., {et~al.} 2022, \apjs, 261,
  14

\bibitem[{{Leurini} {et~al.}(2015){Leurini}, {Wyrowski}, {Wiesemeyer},
  {Gusdorf}, {G{\"u}sten}, {Menten}, {Gerin}, {Levrier}, {H{\"u}bers},
  {Jacobs}, {Ricken}, \& {Richter}}]{Leurini15}
{Leurini}, S., {Wyrowski}, F., {Wiesemeyer}, H., {et~al.} 2015, \aap, 584, A70

\bibitem[{{Lew}(2018)}]{lew18}
{Lew}, B. 2018, Experimental Astronomy, 45, 81

\bibitem[{{Liseau} {et~al.}(1992){Liseau}, {Lorenzetti}, {Nisini}, {Spinoglio},
  \& {Moneti}}]{Liseau92}
{Liseau}, R., {Lorenzetti}, D., {Nisini}, B., {Spinoglio}, L., \& {Moneti}, A.
  1992, \aap, 265, 577

\bibitem[{{Litvak}(1969)}]{litvak1969}
{Litvak}, M.~M. 1969, Science, 165, 855

\bibitem[{{Lombardi} {et~al.}(2011){Lombardi}, {Alves}, \& {Lada}}]{Lombardi11}
{Lombardi}, M., {Alves}, J., \& {Lada}, C.~J. 2011, \aap, 535, A16

\bibitem[{{Manoj} {et~al.}(2013){Manoj}, {Watson}, {Neufeld}, {Megeath},
  {Vavrek}, {Yu}, {Visser}, {Bergin}, {Fischer}, {Tobin}, {Stutz}, {Ali},
  {Wilson}, {Di Francesco}, {Osorio}, {Maret}, \& {Poteet}}]{manoj2013}
{Manoj}, P., {Watson}, D.~M., {Neufeld}, D.~A., {et~al.} 2013, \apj, 763, 83

\bibitem[{{Marsh} {et~al.}(2017){Marsh}, {Whitworth}, {Lomax}, {Ragan},
  {Becciani}, {Cambr{\'e}sy}, {Di Giorgio}, {Eden}, {Elia}, {Kacsuk},
  {Molinari}, {Palmeirim}, {Pezzuto}, {Schneider}, {Sciacca}, \&
  {Vitello}}]{Marsh17}
{Marsh}, K.~A., {Whitworth}, A.~P., {Lomax}, O., {et~al.} 2017, \mnras, 471,
  2730

\bibitem[{{Matuszak} {et~al.}(2015){Matuszak}, {Karska}, {Kristensen},
  {Herczeg}, {Tychoniec}, {van Kempen}, \& {Fuente}}]{matuszak2015}
{Matuszak}, M., {Karska}, A., {Kristensen}, L.~E., {et~al.} 2015, \aap, 578,
  A20

\bibitem[{{Melnick} \& {Kaufman}(2015)}]{mel15}
{Melnick}, G.~J. \& {Kaufman}, M.~J. 2015, \apj, 806, 227

\bibitem[{{Menten} {et~al.}(1992){Menten}, {Reid}, {Pratap}, {Moran}, \&
  {Wilson}}]{menten92}
{Menten}, K.~M., {Reid}, M.~J., {Pratap}, P., {Moran}, J.~M., \& {Wilson},
  T.~L. 1992, \apjl, 401, L39

\bibitem[{{Minier} {et~al.}(2009){Minier}, {Andr{\'e}}, {Bergman}, {Motte},
  {Wyrowski}, {Le Pennec}, {Rodriguez}, {Boulade}, {Doumayrou}, {Dubreuil},
  {Gallais}, {Hamon}, {Lagage}, {Lortholary}, {Martignac}, {Rev{\'e}ret},
  {Roussel}, {Talvard}, {Willmann}, \& {Olofsson}}]{Minier09}
{Minier}, V., {Andr{\'e}}, P., {Bergman}, P., {et~al.} 2009, \aap, 501, L1

\bibitem[{{Mirocha} {et~al.}(2021){Mirocha}, {Karska}, {Gronowski},
  {Kristensen}, {Tychoniec}, {Harsono}, {Figueira}, {G{\l}adkowski}, \&
  {{\.Z}{\'o}{\l}towski}}]{mirocha21}
{Mirocha}, A., {Karska}, A., {Gronowski}, M., {et~al.} 2021, \aap, 656, A146

\bibitem[{{Molinari} {et~al.}(2010){Molinari}, {Swinyard}, {Bally}, {Barlow},
  {Bernard}, {Martin}, {Moore}, {Noriega-Crespo}, {Plume}, {Testi}, {Zavagno},
  {Abergel}, {Ali}, {Anderson}, {Andr{\'e}}, {Baluteau}, {Battersby},
  {Beltr{\'a}n}, {Benedettini}, {Billot}, {Blommaert}, {Bontemps}, {Boulanger},
  {Brand}, {Brunt}, {Burton}, {Calzoletti}, {Carey}, {Caselli}, {Cesaroni},
  {Cernicharo}, {Chakrabarti}, {Chrysostomou}, {Cohen}, {Compiegne}, {de
  Bernardis}, {de Gasperis}, {di Giorgio}, {Elia}, {Faustini}, {Flagey},
  {Fukui}, {Fuller}, {Ganga}, {Garcia-Lario}, {Glenn}, {Goldsmith}, {Griffin},
  {Hoare}, {Huang}, {Ikhenaode}, {Joblin}, {Joncas}, {Juvela}, {Kirk},
  {Lagache}, {Li}, {Lim}, {Lord}, {Marengo}, {Marshall}, {Masi}, {Massi},
  {Matsuura}, {Minier}, {Miville-Desch{\^e}nes}, {Montier}, {Morgan}, {Motte},
  {Mottram}, {M{\"u}ller}, {Natoli}, {Neves}, {Olmi}, {Paladini}, {Paradis},
  {Parsons}, {Peretto}, {Pestalozzi}, {Pezzuto}, {Piacentini}, {Piazzo},
  {Polychroni}, {Pomar{\`e}s}, {Popescu}, {Reach}, {Ristorcelli}, {Robitaille},
  {Robitaille}, {Rod{\'o}n}, {Roy}, {Royer}, {Russeil}, {Saraceno}, {Sauvage},
  {Schilke}, {Schisano}, {Schneider}, {Schuller}, {Schulz}, {Sibthorpe},
  {Smith}, {Smith}, {Spinoglio}, {Stamatellos}, {Strafella}, {Stringfellow},
  {Sturm}, {Taylor}, {Thompson}, {Traficante}, {Tuffs}, {Umana}, {Valenziano},
  {Vavrek}, {Veneziani}, {Viti}, {Waelkens}, {Ward-Thompson}, {White},
  {Wilcock}, {Wyrowski}, {Yorke}, \& {Zhang}}]{molinari2010}
{Molinari}, S., {Swinyard}, B., {Bally}, J., {et~al.} 2010, \aap, 518, L100

\bibitem[{{Mookerjea} {et~al.}(2021){Mookerjea}, {Sandell}, {Veena},
  {G{\"u}sten}, {Riquelme}, {Wiesemeyer}, {Wyrowski}, \&
  {Mertens}}]{Mookerjea21}
{Mookerjea}, B., {Sandell}, G., {Veena}, V.~S., {et~al.} 2021, \aap, 648, A40

\bibitem[{{Moscadelli} {et~al.}(2009){Moscadelli}, {Reid}, {Menten},
  {Brunthaler}, {Zheng}, \& {Xu}}]{Moscadelli09}
{Moscadelli}, L., {Reid}, M.~J., {Menten}, K.~M., {et~al.} 2009, \apj, 693, 406

\bibitem[{{Motogi} {et~al.}(2011){Motogi}, {Sorai}, {Habe}, {Honma},
  {Kobayashi}, \& {Sato}}]{Motogi11}
{Motogi}, K., {Sorai}, K., {Habe}, A., {et~al.} 2011, \pasj, 63, 31

\bibitem[{{Mottram} {et~al.}(2017){Mottram}, {van Dishoeck}, {Kristensen},
  {Karska}, {San Jos{\'e}-Garc{\'\i}a}, {Khanna}, {Herczeg}, {Andr{\'e}},
  {Bontemps}, {Cabrit}, {Carney}, {Drozdovskaya}, {Dunham}, {Evans}, {Fedele},
  {Green}, {Harsono}, {Johnstone}, {J{\o}rgensen}, {K{\"o}nyves}, {Nisini},
  {Persson}, {Tafalla}, {Visser}, \& {Y{\i}ld{\i}z}}]{mot17}
{Mottram}, J.~C., {van Dishoeck}, E.~F., {Kristensen}, L.~E., {et~al.} 2017,
  \aap, 600, A99

\bibitem[{{Navarete} {et~al.}(2015){Navarete}, {Damineli}, {Barbosa}, \&
  {Blum}}]{Navarete15}
{Navarete}, F., {Damineli}, A., {Barbosa}, C.~L., \& {Blum}, R.~D. 2015,
  \mnras, 450, 4364

\bibitem[{{Neckel}(1978)}]{Neckel78}
{Neckel}, T. 1978, \aap, 69, 51

\bibitem[{{Nisini} {et~al.}(2015){Nisini}, {Santangelo}, {Giannini},
  {Antoniucci}, {Cabrit}, {Codella}, {Davis}, {Eisl{\"o}ffel}, {Kristensen},
  {Herczeg}, {Neufeld}, \& {van Dishoeck}}]{nisini15}
{Nisini}, B., {Santangelo}, G., {Giannini}, T., {et~al.} 2015, \apj, 801, 121

\bibitem[{{Oliveira} {et~al.}(2019){Oliveira}, {van Loon}, {Sewi{\l}o}, {Lee},
  {Lebouteiller}, {Chen}, {Cormier}, {Filipovi{\'c}}, {Carlson}, {Indebetouw},
  {Madden}, {Meixner}, {Sargent}, \& {Fukui}}]{Oliveira19}
{Oliveira}, J.~M., {van Loon}, J.~T., {Sewi{\l}o}, M., {et~al.} 2019, \mnras,
  490, 3909

\bibitem[{{Ossenkopf} {et~al.}(2010){Ossenkopf}, {R{\"o}llig}, {Simon},
  {Schneider}, {Okada}, {Stutzki}, {Gerin}, {Akyilmaz}, {Beintema}, {Benz},
  {Berne}, {Boulanger}, {Bumble}, {Coeur-Joly}, {Dedes}, {Diez-Gonzalez},
  {France}, {Fuente}, {Gallego}, {Goicoechea}, {G{\"u}sten}, {Harris},
  {Higgins}, {Jackson}, {Jarchow}, {Joblin}, {Klein}, {Kramer}, {Lord},
  {Martin}, {Martin-Pintado}, {Mookerjea}, {Neufeld}, {Phillips}, {Rizzo}, {van
  der Tak}, {Teyssier}, \& {Yorke}}]{oss10}
{Ossenkopf}, V., {R{\"o}llig}, M., {Simon}, R., {et~al.} 2010, \aap, 518, L79

\bibitem[{{Pickett} {et~al.}(1998){Pickett}, {Poynter}, {Cohen}, {Delitsky},
  {Pearson}, \& {M{\"u}ller}}]{Pic98}
{Pickett}, H.~M., {Poynter}, R.~L., {Cohen}, E.~A., {et~al.} 1998, \jqsrt, 60,
  883

\bibitem[{{Pietrzy{\'n}ski} {et~al.}(2013){Pietrzy{\'n}ski}, {Graczyk},
  {Gieren}, {Thompson}, {Pilecki}, {Udalski}, {Soszy{\'n}ski}, {Koz{\l}owski},
  {Konorski}, {Suchomska}, {Bono}, {Moroni}, {Villanova}, {Nardetto},
  {Bresolin}, {Kudritzki}, {Storm}, {Gallenne}, {Smolec}, {Minniti}, {Kubiak},
  {Szyma{\'n}ski}, {Poleski}, {Wyrzykowski}, {Ulaczyk}, {Pietrukowicz},
  {G{\'o}rski}, \& {Karczmarek}}]{Pietrzynski13}
{Pietrzy{\'n}ski}, G., {Graczyk}, D., {Gieren}, W., {et~al.} 2013, \nat, 495,
  76

\bibitem[{{Pilbratt} {et~al.}(2010){Pilbratt}, {Riedinger}, {Passvogel},
  {Crone}, {Doyle}, {Gageur}, {Heras}, {Jewell}, {Metcalfe}, {Ott}, \&
  {Schmidt}}]{Pi10}
{Pilbratt}, G.~L., {Riedinger}, J.~R., {Passvogel}, T., {et~al.} 2010, \aap,
  518, L1

\bibitem[{{Poglitsch} {et~al.}(2010){Poglitsch}, {Waelkens}, {Geis},
  {Feuchtgruber}, {Vandenbussche}, {Rodriguez}, {Krause}, {Renotte}, {van
  Hoof}, {Saraceno}, {Cepa}, {Kerschbaum}, {Agn{\`e}se}, {Ali}, {Altieri},
  {Andreani}, {Augueres}, {Balog}, {Barl}, {Bauer}, {Belbachir}, {Benedettini},
  {Billot}, {Boulade}, {Bischof}, {Blommaert}, {Callut}, {Cara}, {Cerulli},
  {Cesarsky}, {Contursi}, {Creten}, {De Meester}, {Doublier}, {Doumayrou},
  {Duband}, {Exter}, {Genzel}, {Gillis}, {Gr{\"o}zinger}, {Henning},
  {Herreros}, {Huygen}, {Inguscio}, {Jakob}, {Jamar}, {Jean}, {de Jong},
  {Katterloher}, {Kiss}, {Klaas}, {Lemke}, {Lutz}, {Madden}, {Marquet},
  {Martignac}, {Mazy}, {Merken}, {Montfort}, {Morbidelli}, {M{\"u}ller},
  {Nielbock}, {Okumura}, {Orfei}, {Ottensamer}, {Pezzuto}, {Popesso},
  {Putzeys}, {Regibo}, {Reveret}, {Royer}, {Sauvage}, {Schreiber}, {Stegmaier},
  {Schmitt}, {Schubert}, {Sturm}, {Thiel}, {Tofani}, {Vavrek}, {Wetzstein},
  {Wieprecht}, \& {Wiezorrek}}]{pacs}
{Poglitsch}, A., {Waelkens}, C., {Geis}, N., {et~al.} 2010, \aap, 518, L2

\bibitem[{{Pound} \& {Wolfire}(2011)}]{Pound_Wolfire2011}
{Pound}, M.~W. \& {Wolfire}, M.~G. 2011, {PDRT: Photo Dissociation Region
  Toolbox}

\bibitem[{{Purser} {et~al.}(2021){Purser}, {Lumsden}, {Hoare}, \&
  {Kurtz}}]{Purser21}
{Purser}, S.~J.~D., {Lumsden}, S.~L., {Hoare}, M.~G., \& {Kurtz}, S. 2021,
  \mnras, 504, 338

\bibitem[{{Risacher} {et~al.}(2018){Risacher}, {G{\"u}sten}, {Stutzki},
  {H{\"u}bers}, {Aladro}, {Bell}, {Buchbender}, {B{\"u}chel}, {Csengeri},
  {Duran}, {Graf}, {Higgins}, {Honingh}, {Jacobs}, {Justen}, {Klein},
  {Mertens}, {Okada}, {Parikka}, {P{\"u}tz}, {Reyes}, {Richter}, {Ricken},
  {Riquelme}, {Rothbart}, {Schneider}, {Simon}, {Wienold}, {Wiesemeyer},
  {Ziebart}, {Fusco}, {Rosner}, \& {Wohler}}]{risacher2018}
{Risacher}, C., {G{\"u}sten}, R., {Stutzki}, J., {et~al.} 2018, Journal of
  Astronomical Instrumentation, 7, 1840014

\bibitem[{{Robitaille}(2017)}]{robitaille2017}
{Robitaille}, T.~P. 2017, \aap, 600, A11

\bibitem[{{Robitaille} {et~al.}(2007){Robitaille}, {Whitney}, {Indebetouw}, \&
  {Wood}}]{robitaille2007}
{Robitaille}, T.~P., {Whitney}, B.~A., {Indebetouw}, R., \& {Wood}, K. 2007,
  \apjs, 169, 328

\bibitem[{{Roman-Duval} {et~al.}(2010){Roman-Duval}, {Jackson}, {Heyer},
  {Rathborne}, \& {Simon}}]{roman2010}
{Roman-Duval}, J., {Jackson}, J.~M., {Heyer}, M., {Rathborne}, J., \& {Simon},
  R. 2010, \apj, 723, 492

\bibitem[{{Russell} \& {Dopita}(1992)}]{rd92}
{Russell}, S.~C. \& {Dopita}, M.~A. 1992, \apj, 384, 508

\bibitem[{{Rygl} {et~al.}(2012){Rygl}, {Brunthaler}, {Sanna}, {Menten}, {Reid},
  {van Langevelde}, {Honma}, {Torstensson}, \& {Fujisawa}}]{Rygl12}
{Rygl}, K.~L.~J., {Brunthaler}, A., {Sanna}, A., {et~al.} 2012, \aap, 539, A79

\bibitem[{{Sandell}(2000)}]{Sandell2000}
{Sandell}, G. 2000, \aap, 358, 242

\bibitem[{{Sandell} \& {Sievers}(2004)}]{Sandell04}
{Sandell}, G. \& {Sievers}, A. 2004, \apj, 600, 269

\bibitem[{{Sch{\"o}ier} {et~al.}(2005){Sch{\"o}ier}, {van der Tak}, {van
  Dishoeck}, \& {Black}}]{schoier05}
{Sch{\"o}ier}, F.~L., {van der Tak}, F.~F.~S., {van Dishoeck}, E.~F., \&
  {Black}, J.~H. 2005, \aap, 432, 369

\bibitem[{{Sewi{\l}o} {et~al.}(2019){Sewi{\l}o}, {Whitney}, {Yung},
  {Robitaille}, {Elia}, {Indebetouw}, {Schisano}, {Szczerba}, {Karska},
  {Wiseman}, {Babler}, {Boyer}, {Fischer}, {Meade}, {Olmi}, {Padgett}, \&
  {Si{\'o}dmiak}}]{sewilo19}
{Sewi{\l}o}, M., {Whitney}, B.~A., {Yung}, B. H.~K., {et~al.} 2019, \apjs, 240,
  26

\bibitem[{{Shevchenko} \& {Yakubov}(1989)}]{Shevchenko_Yakubov89}
{Shevchenko}, V.~S. \& {Yakubov}, S.~D. 1989, \sovast, 33, 370

\bibitem[{{Soares} \& {Bica}(2002)}]{Soares02}
{Soares}, J.~B. \& {Bica}, E. 2002, \aap, 388, 172

\bibitem[{{Soares} \& {Bica}(2003)}]{Soares03}
{Soares}, J.~B. \& {Bica}, E. 2003, \aap, 404, 217

\bibitem[{{Sodroski} {et~al.}(1997){Sodroski}, {Odegard}, {Arendt}, {Dwek},
  {Weiland}, {Hauser}, \& {Kelsall}}]{sodroski1997}
{Sodroski}, T.~J., {Odegard}, N., {Arendt}, R.~G., {et~al.} 1997, \apj, 480,
  173

\bibitem[{{Stanke} {et~al.}(2000){Stanke}, {McCaughrean}, \&
  {Zinnecker}}]{Stanke2000}
{Stanke}, T., {McCaughrean}, M.~J., \& {Zinnecker}, H. 2000, \aap, 355, 639

\bibitem[{{Szymczak} {et~al.}(2000){Szymczak}, {Hrynek}, \& {Kus}}]{szymczak}
{Szymczak}, M., {Hrynek}, G., \& {Kus}, A.~J. 2000, \aaps, 143, 269

\bibitem[{{Szymczak} {et~al.}(2016){Szymczak}, {Olech}, {Wolak}, {Bartkiewicz},
  \& {Gawro{\'n}ski}}]{szym16}
{Szymczak}, M., {Olech}, M., {Wolak}, P., {Bartkiewicz}, A., \&
  {Gawro{\'n}ski}, M. 2016, \mnras, 459, L56

\bibitem[{{Tang} {et~al.}(2014){Tang}, {Bressan}, {Rosenfield}, {Slemer},
  {Marigo}, {Girardi}, \& {Bianchi}}]{tang2014}
{Tang}, J., {Bressan}, A., {Rosenfield}, P., {et~al.} 2014, \mnras, 445, 4287

\bibitem[{{Tapia} {et~al.}(1997){Tapia}, {Persi}, {Bohigas}, \&
  {Ferrari-Toniolo}}]{tapia97}
{Tapia}, M., {Persi}, P., {Bohigas}, J., \& {Ferrari-Toniolo}, M. 1997, \aj,
  113, 1769

\bibitem[{{Tobin} {et~al.}(2016){Tobin}, {Stutz}, {Manoj}, {Megeath}, {Karska},
  {Nagy}, {Wyrowski}, {Fischer}, {Watson}, \& {Stanke}}]{tobin16}
{Tobin}, J.~J., {Stutz}, A.~M., {Manoj}, P., {et~al.} 2016, \apj, 831, 36

\bibitem[{{Urquhart} {et~al.}(2012){Urquhart}, {Hoare}, {Lumsden}, {Oudmaijer},
  {Moore}, {Mottram}, {Cooper}, {Mottram}, \& {Rogers}}]{Urquhart12}
{Urquhart}, J.~S., {Hoare}, M.~G., {Lumsden}, S.~L., {et~al.} 2012, \mnras,
  420, 1656

\bibitem[{{Urquhart} {et~al.}(2011){Urquhart}, {Morgan}, {Figura}, {Moore},
  {Lumsden}, {Hoare}, {Oudmaijer}, {Mottram}, {Davies}, \&
  {Dunham}}]{Urquhart11}
{Urquhart}, J.~S., {Morgan}, L.~K., {Figura}, C.~C., {et~al.} 2011, \mnras,
  418, 1689

\bibitem[{{Vacca} {et~al.}(2020){Vacca}, {Clarke}, {Perera}, {Fadda}, \&
  {Holt}}]{Vacca2020}
{Vacca}, W., {Clarke}, M., {Perera}, D., {Fadda}, D., \& {Holt}, J. 2020, in
  Astronomical Society of the Pacific Conference Series, Vol. 527, Astronomical
  Data Analysis Software and Systems XXIX, ed. R.~{Pizzo}, E.~R. {Deul}, J.~D.
  {Mol}, J.~{de Plaa}, \& H.~{Verkouter}, 547

\bibitem[{{Valdettaro} {et~al.}(2001){Valdettaro}, {Palla}, {Brand},
  {Cesaroni}, {Comoretto}, {Di Franco}, {Felli}, {Natale}, {Palagi}, {Panella},
  \& {Tofani}}]{Valdettaro01}
{Valdettaro}, R., {Palla}, F., {Brand}, J., {et~al.} 2001, \aap, 368, 845

\bibitem[{{van der Tak}(2012)}]{vandertak12}
{van der Tak}, F.~F.~S. 2012, Philosophical Transactions of the Royal Society
  of London Series A, 370, 5186

\bibitem[{{van der Tak} {et~al.}(1999){van der Tak}, {van Dishoeck}, {Evans},
  {Bakker}, \& {Blake}}]{vandertak99}
{van der Tak}, F. F.~S., {van Dishoeck}, E.~F., {Evans}, Neal~J., I., {Bakker},
  E.~J., \& {Blake}, G.~A. 1999, \apj, 522, 991

\bibitem[{{van Dishoeck} {et~al.}(2011){van Dishoeck}, {Kristensen}, {Benz},
  {Bergin}, {Caselli}, {Cernicharo}, {Herpin}, {Hogerheijde}, {Johnstone},
  {Liseau}, {Nisini}, {Shipman}, {Tafalla}, {van der Tak}, {Wyrowski},
  {Aikawa}, {Bachiller}, {Baudry}, {Benedettini}, {Bjerkeli}, {Blake},
  {Bontemps}, {Braine}, {Brinch}, {Bruderer}, {Chavarr{\'\i}a}, {Codella},
  {Daniel}, {de Graauw}, {Deul}, {di Giorgio}, {Dominik}, {Doty}, {Dubernet},
  {Encrenaz}, {Feuchtgruber}, {Fich}, {Frieswijk}, {Fuente}, {Giannini},
  {Goicoechea}, {Helmich}, {Herczeg}, {Jacq}, {J{\o}rgensen}, {Karska},
  {Kaufman}, {Keto}, {Larsson}, {Lefloch}, {Lis}, {Marseille}, {McCoey},
  {Melnick}, {Neufeld}, {Olberg}, {Pagani}, {Pani{\'c}}, {Parise}, {Pearson},
  {Plume}, {Risacher}, {Salter}, {Santiago-Garc{\'\i}a}, {Saraceno},
  {St{\"a}uber}, {van Kempen}, {Visser}, {Viti}, {Walmsley}, {Wampfler}, \&
  {Y{\i}ld{\i}z}}]{Dishoeck11}
{van Dishoeck}, E.~F., {Kristensen}, L.~E., {Benz}, A.~O., {et~al.} 2011,
  \pasp, 123, 138

\bibitem[{{van Dishoeck} {et~al.}(2021){van Dishoeck}, {Kristensen}, {Mottram},
  {Benz}, {Bergin}, {Caselli}, {Herpin}, {Hogerheijde}, {Johnstone}, {Liseau},
  {Nisini}, {Tafalla}, {van der Tak}, {Wyrowski}, {Baudry}, {Benedettini},
  {Bjerkeli}, {Blake}, {Braine}, {Bruderer}, {Cabrit}, {Cernicharo}, {Choi},
  {Coutens}, {de Graauw}, {Dominik}, {Fedele}, {Fich}, {Fuente}, {Furuya},
  {Goicoechea}, {Harsono}, {Helmich}, {Herczeg}, {Jacq}, {Karska}, {Kaufman},
  {Keto}, {Lamberts}, {Larsson}, {Leurini}, {Lis}, {Melnick}, {Neufeld},
  {Pagani}, {Persson}, {Shipman}, {Taquet}, {van Kempen}, {Walsh}, {Wampfler},
  {Y{\i}ld{\i}z}, \& {WISH Team}}]{vD21}
{van Dishoeck}, E.~F., {Kristensen}, L.~E., {Mottram}, J.~C., {et~al.} 2021,
  \aap, 648, A24

\bibitem[{{Visser} {et~al.}(2012){Visser}, {Kristensen}, {Bruderer}, {van
  Dishoeck}, {Herczeg}, {Brinch}, {Doty}, {Harsono}, \& {Wolfire}}]{vi12}
{Visser}, R., {Kristensen}, L.~E., {Bruderer}, S., {et~al.} 2012, \aap, 537,
  A55

\bibitem[{{White} {et~al.}(2010){White}, {Abergel}, {Spencer}, {Schneider},
  {Naylor}, {Anderson}, {Joblin}, {Ade}, {Andr{\'e}}, {Arab}, {Baluteau},
  {Bernard}, {Blagrave}, {Bontemps}, {Boulanger}, {Cohen}, {Compiegne}, {Cox},
  {Dartois}, {Davis}, {Emery}, {Fulton}, {Gom}, {Griffin}, {Gry}, {Habart},
  {Huang}, {Jones}, {Kirk}, {Lagache}, {Leeks}, {Lim}, {Madden}, {Makiwa},
  {Martin}, {Miville-Desch{\^e}nes}, {Molinari}, {Moseley}, {Motte}, {Okumura},
  {Pinheiro Gon{\c{c}}alves}, {Polehampton}, {Rodet}, {Rod{\'o}n}, {Russeil},
  {Saraceno}, {Sidher}, {Swinyard}, {Ward-Thompson}, \& {Zavagno}}]{whi10}
{White}, G.~J., {Abergel}, A., {Spencer}, L., {et~al.} 2010, \aap, 518, L114

\bibitem[{{Wiesenfeld} \& {Goldsmith}(2014)}]{wie14}
{Wiesenfeld}, L. \& {Goldsmith}, P.~F. 2014, \apj, 780, 183

\bibitem[{{Wilson} {et~al.}(2005){Wilson}, {Dame}, {Masheder}, \&
  {Thaddeus}}]{Wilson05}
{Wilson}, B.~A., {Dame}, T.~M., {Masheder}, M.~R.~W., \& {Thaddeus}, P. 2005,
  \aap, 430, 523

\bibitem[{{Wood} \& {Churchwell}(1989)}]{WC89}
{Wood}, D. O.~S. \& {Churchwell}, E. 1989, \apj, 340, 265

\bibitem[{{Wouterloot} \& {Brand}(1989)}]{Wouterloot89}
{Wouterloot}, J.~G.~A. \& {Brand}, J. 1989, \aaps, 80, 149

\bibitem[{{Xu} {et~al.}(2009){Xu}, {Reid}, {Menten}, {Brunthaler}, {Zheng}, \&
  {Moscadelli}}]{Xu09}
{Xu}, Y., {Reid}, M.~J., {Menten}, K.~M., {et~al.} 2009, \apj, 693, 413

\bibitem[{{Yang} {et~al.}(2022{\natexlab{a}}){Yang}, {Evans}, {Karska},
  {Kristensen}, {Aladro}, {Ramsey}, {Green}, \& {Lee}}]{yang22}
{Yang}, Y.-L., {Evans}, N.~J., {Karska}, A., {et~al.} 2022{\natexlab{a}}, \apj,
  925, 93

\bibitem[{{Yang} {et~al.}(2018){Yang}, {Green}, {Evans}, {Lee}, {J{\o}rgensen},
  {Kristensen}, {Mottram}, {Herczeg}, {Karska}, {Dionatos}, {Bergin},
  {Bouwman}, {van Dishoeck}, {van Kempen}, {Larson}, \&
  {Y{\i}ld{\i}z}}]{yang18}
{Yang}, Y.-L., {Green}, J.~D., {Evans}, Neal~J., I., {et~al.} 2018, \apj, 860,
  174

\bibitem[{{Yang} {et~al.}(2022{\natexlab{b}}){Yang}, {Green}, {Pontoppidan},
  {Bergner}, {Cleeves}, {Evans}, {Garrod}, {Jin}, {Kim}, {Kim}, {Lee}, {Sakai},
  {Shingledecker}, {Shope}, {Tobin}, \& {van Dishoeck}}]{yang22b}
{Yang}, Y.-L., {Green}, J.~D., {Pontoppidan}, K.~M., {et~al.}
  2022{\natexlab{b}}, \apjl, 941, L13

\end{thebibliography}

\begin{appendix} 

\section{Water masers in CMa$-l$224 }\label{app:maser}

\subsection{Survey results}

\begin{figure}[b!]
\centering
\includegraphics[width = 0.49 \textwidth]{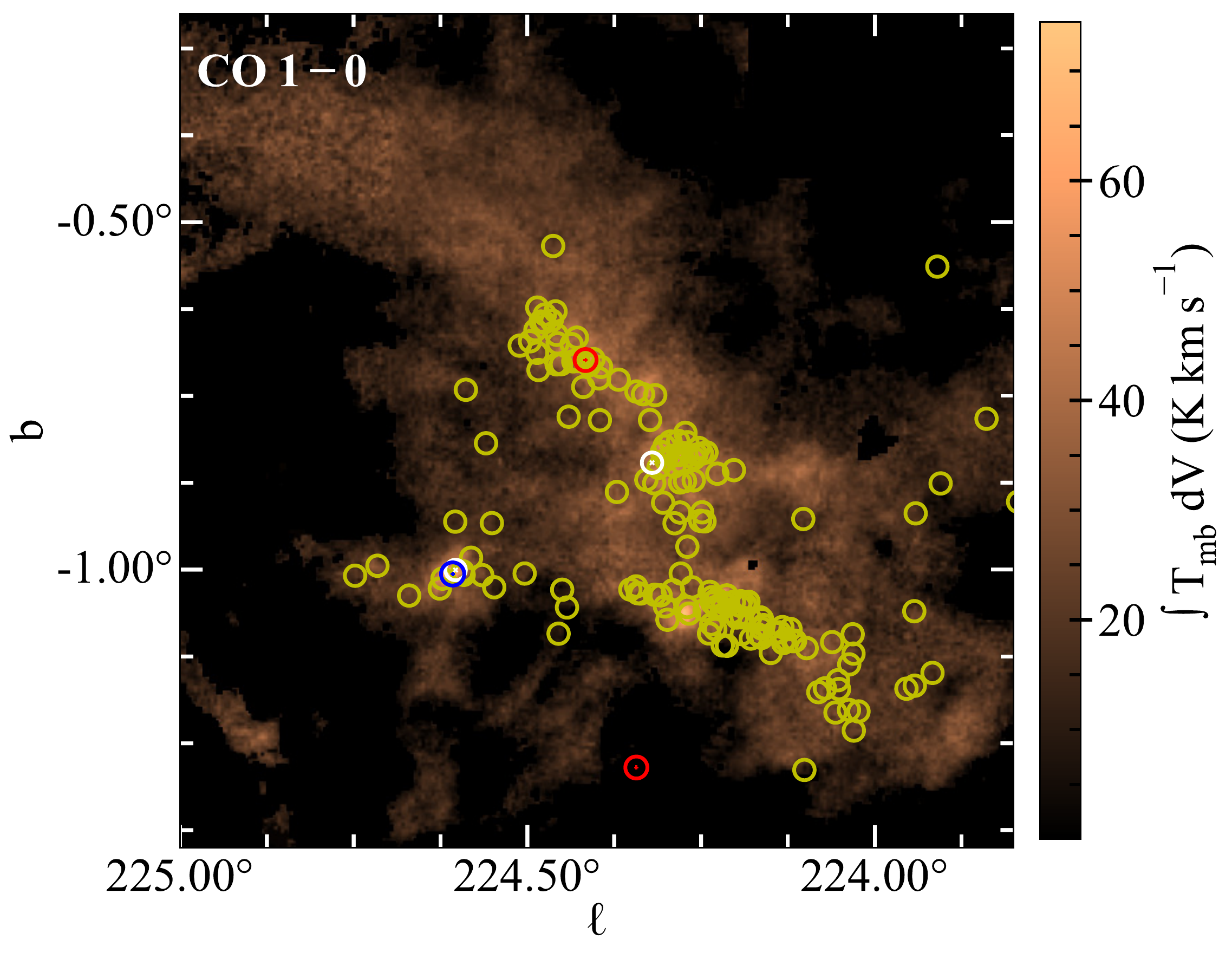}
\caption{CO~$1-0$ integrated intensity map of CMa$-l$224. Yellow circles show
 all RT4 pointings, while white circles indicate those with the 22 GHz water
  maser detections: toward Gy 3--7 and source IRAS 07069$-$1026 in the main filament. Blue
   and red circles indicate the positions of water masers detected toward CMa$-l$224 with
    the GBT \citep{Urquhart11} and the Medicina telescope \citep{Valdettaro01}, respectively
     (see text for details).}
\label{fig:CO10_maser_pointting}
\end{figure}

Water masers are important signposts of both low- and high-mass star formation;
 they are collisionally excited, tracing warm molecular gas behind shock waves in the
  environment of YSOs and H\,{\sc ii} regions (e.g., \citealt{litvak1969};  \citealt{elitzur1989};
   \citealt{furuya2003}; \citealt{ladeyschikov2022} and references therein).
    Figure~\ref{fig:CO10_maser_pointting} shows the pointings of our 22 GHz water maser survey
     of the CMa$-l$224 star-forming region with the 32-m radio telescope in Toru{\'n} (RT4;
      a half-power beam width, HPBW$\sim$106$''$, corresponding to $\sim$0.5 pc at 1 kpc;
       see Section \ref{ssec:rt4}), overlaid on the CO 1--0 integrated intensity image from the Forgotten
        Quadrant Survey (FQS, \citealt{bene20}; HPBW$\sim$55$''$ at 115 GHz). The RT4 water maser
         survey targeted the fields harboring YSO candidates identified by \citet{sewilo19}.
          Out of 205 observed fields, (185, 18, 2) were observed in (2, 3, 6) epochs. Water
           masers were detected toward two fields, one centered on Gy 3--7 and the other on
            source IRAS 07069$-$1026 in the main filament in CMa$-l$224. The very low
             detection rate is likely the result of the low sensitivity of our observations
              and the variability of the maser emission (see Section \ref{ssec:rt4}).

The detection of the water maser toward IRAS 07069$-$1026 constitutes the first maser
 detection toward this source, while the detection toward Gy 3--7 has previously been reported
  in the literature. \citet{Urquhart11} detected the 22 GHz water maser emission toward Gy 3--7
   (their source G224.6075$-$01.0063) with the 100-m Green Bank Telescope (GBT; HPBW$\sim$30$''$).
    Water masers were detected toward two other locations in CMa$-l$224 by \citet{Valdettaro01}
     with the Medicina 32-m radio telescope (a similar angular resolution as for RT4), toward
      IRAS 07077$-$1026 and IRAS 07054$-$1039 (see Figure~\ref{fig:CO10_maser_pointting}).

\begin{figure}
\centering
\includegraphics[width = 0.48\textwidth]{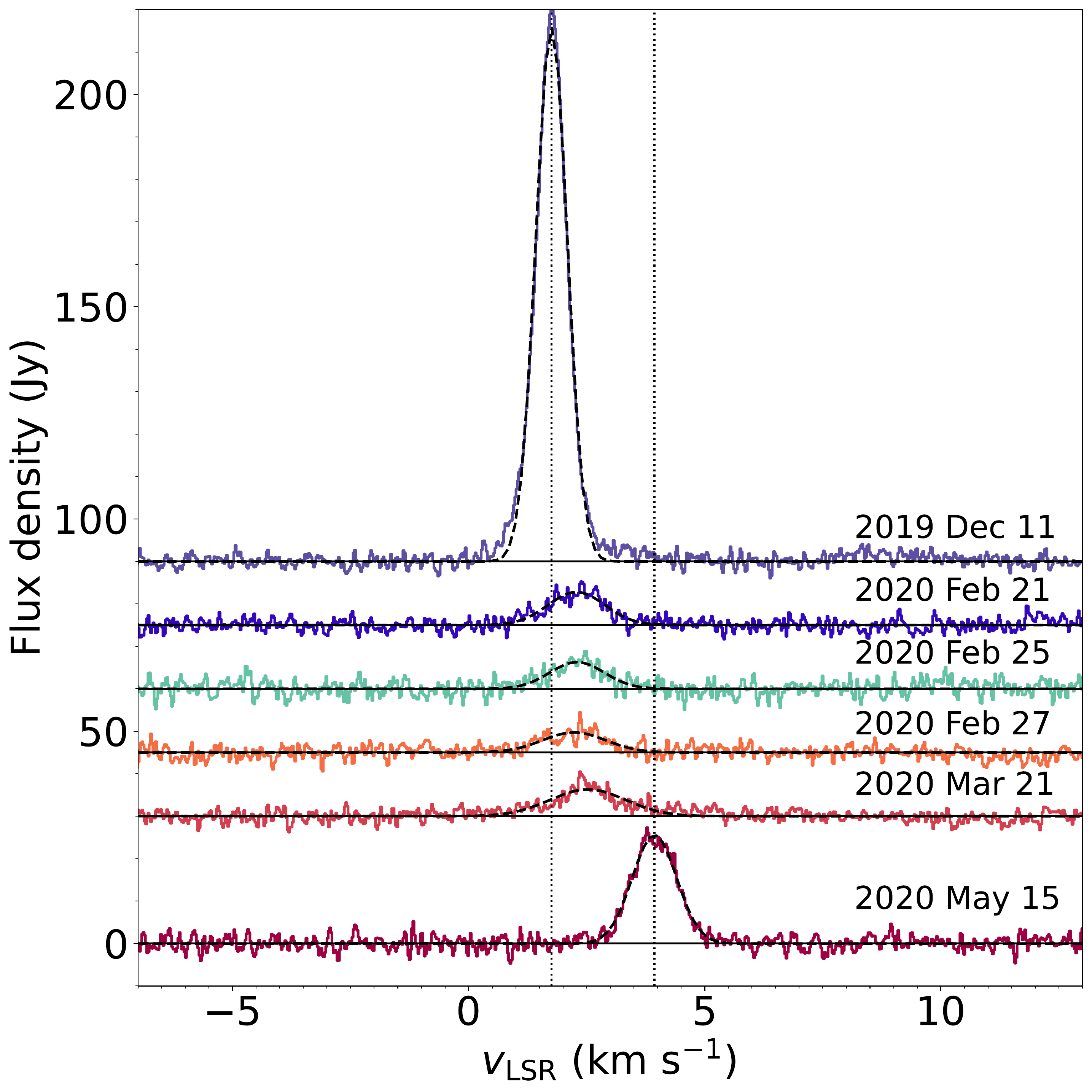}
\caption{22 GHz H$_2$O maser spectra obtained toward Gy 3--7 with the 32-m radio telescope
 in Toru{\'n} from December 2019 to May 2020. The date of observations is indicated above each spectrum. 
 Except for the spectrum corresponding to the observation on May 15, 2020, spectra are shifted
 vertically by 30~Jy (for the observation on March 21, 2020) and 15~Jy (for the remaining observations)
  to improve the clarity of the figure. Dashed vertical lines show central velocities of the
   lines detected in the latest (May 15, 2020) and earliest (December 11, 2019) of the six
    epochs of observations (see also Table \ref{tab:maser}).}
\label{fig:maser}
\end{figure}
\begin{figure}
\centering
\includegraphics[width = 0.48\textwidth]{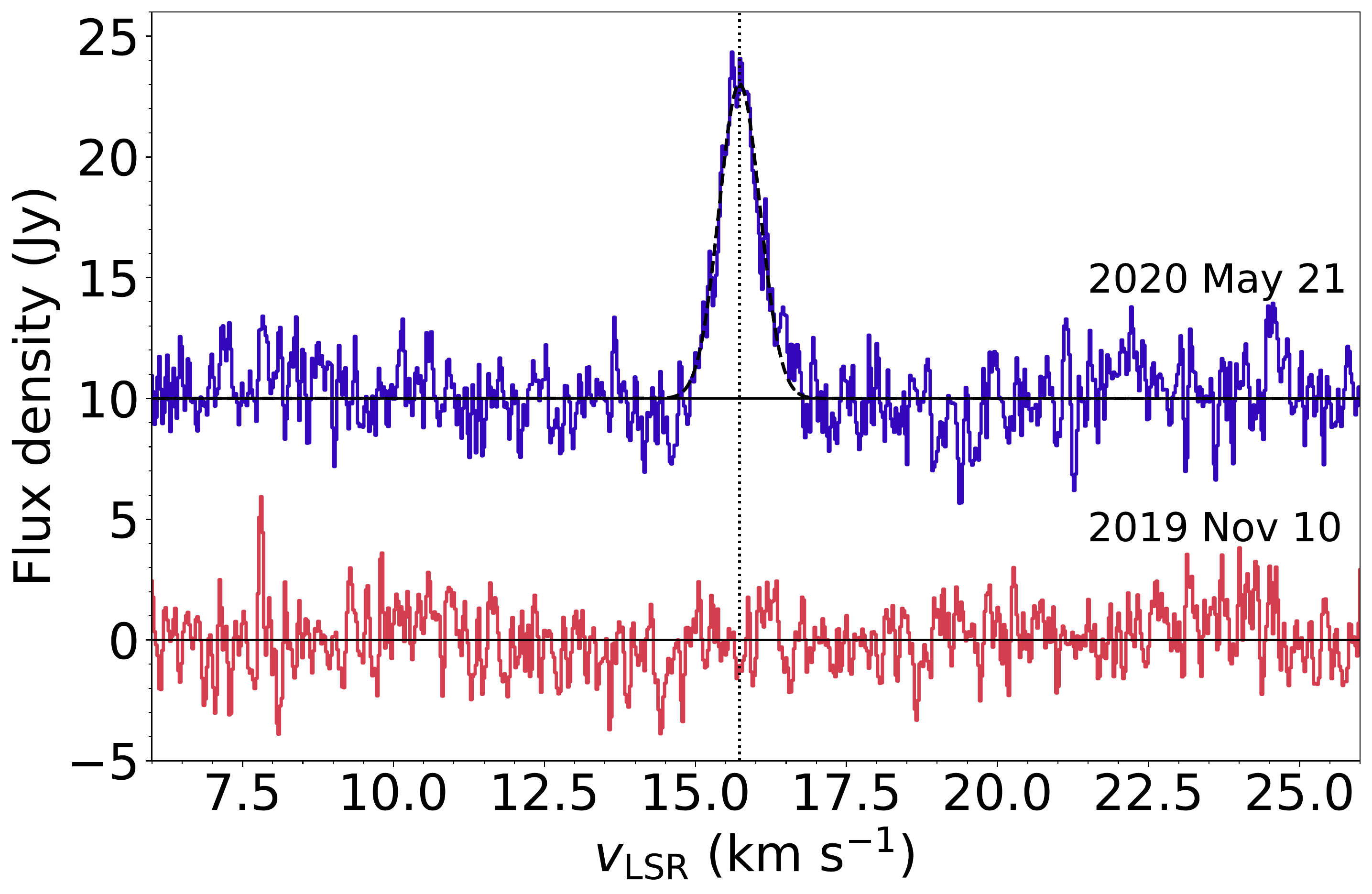}
\caption{22 GHz H$_2$O maser emission detected toward IRAS 07069$-$1026 (see also Table \ref{tab:maser}).
 The H$_2$O maser was detected only in one of the two observing epochs.
The spectrum observed on May 21, 2020 is shifted vertically by 10 Jy to improve the clarity of the figure.}
\label{fig:maser2}
\end{figure}
       
\begin{table*}[ht!]
\centering
\caption{Parameters of the 22 GHz H$_2$O maser lines detected toward CMa$-l224$ with RT4}
\label{tab:maser}
\begin{tabular}{l c r c r}
\hline\hline
\multicolumn{1}{c}{Obs. date} & $V_\mathrm{p}$ & \multicolumn{1}{c}{$S_\mathrm{p}$} & FWHM  &\multicolumn{1}{c}{$\int S dV$}   \\
~ & (km s$^{-1}$) & \multicolumn{1}{c}{(Jy)} & (km s$^{-1}$) & \multicolumn{1}{c}{(Jy km s$^{-1}$)}   \\
\hline
\multicolumn{5}{c}{Gy 3--7} \\
\hline
2019 Dec 11 & $1.76\pm0.01$ & $117.33\pm0.43$ & 0.79$\pm$0.003& $41.65\pm0.32$\\
2020 Feb 21 & $2.28\pm0.03$ & $7.27\pm0.25$ & $1.42\pm0.06$& $4.67\pm0.33$   \\
2020 Feb 25 & $2.30\pm0.04$ & $5.87\pm0.35$ & $1.36\pm0.09$&$3.61\pm0.45$  \\
2020 Feb 27 & $2.23\pm0.08$ & $4.41\pm0.43$ &$1.59\pm0.18$&$3.16\pm0.66$ \\
2020 Mar 21 & $2.53\pm0.06$ & $5.87\pm0.33$ &$1.90\pm0.12$&$5.04\pm0.60$  \\
2020 May 15 & $3.94\pm0.01$ & $23.80\pm0.33$ &$1.11\pm0.02$&$11.93\pm0.35$\\
\hline
\multicolumn{5}{c}{IRAS 07069$-$1026} \\
\hline
2020 May 21 & $15.73\pm0.02$ & $12.15\pm0.40$ &$0.80\pm0.03$&$4.38\pm0.31 $ \\
\hline
\end{tabular} \\
\end{table*}

The water maser spectra for Gy 3–7 and IRAS 07069-1026 for all epochs of the RT4 observations
 are shown in Figure \ref{fig:maser} and Figure \ref{fig:maser2}, respectively.
Gy 3--7 was observed with RT4 six times over five months with the time interval
 between subsequent epochs ranging from 2 days to over two months  (see Table~\ref{tab:maser}
  for observing dates). The water maser emission was detected in all epochs. IRAS 07069$-$1026
   was observed in two epochs separated by about six months (see Table~\ref{tab:maser}). 
   The water maser emission toward IRAS 07069$-$1026 was only detected in the second epoch,
    illustrating a high variability of the water maser phenomenon.  The RT4 spectra of 
    Gy~3--7 also show clear evidence for a variable nature of the maser emission in this
     region. Both the peak flux density and central velocity of the maser spots varied
      significantly between April 2019 and May 2020, covering the flux density and velocity
       ranges of 4.4--117.2 Jy and 1.8--3.9 km s$^{-1}$, respectively.

The Gauss functions were fitted to the 22 GHz water maser lines observed
 towards Gy 3--7 and IRAS 07069$-$1026. The Gauss fitting results are shown Table \ref{tab:maser}:
  the central velocities ($V_\mathrm{p}$) of the water maser lines, their peak flux densities
   ($S_\mathrm{p}$), full-widths at half maximum (FWHM), and the integrated flux densities
    ($\int S dV$). The measured flux densities are likely underestimated by 10--20\% due
     to the atmospheric conditions at the observatory site; no correction for atmospheric
      attenuation was applied. 

\subsection{Water maser emission in Gy 3--7}

\cite{Urquhart11} report the detection of a single water maser spot toward Gy~3--7
 (G224.6075$-$01.0063) with the peak flux density of 2.57 Jy and velocity of 28.7~km s$^{-1}$
  ($\sim$25~km s$^{-1}$ higher than the emission detected with RT4). However, the inspection
   of the spectrum provided by \cite{Urquhart11} reveals a second, much fainter and broader
    water maser line centered at the velocity of $\sim$0~km s$^{-1}$, only $\sim$2--4 ~km s$^{-1}$
     lower than the peak velocities of the lines detected in our observations. No water maser
      emission at higher velocities was detected in any of the six RT4 epochs. 

The systemic velocity of  Gy~3--7  is 16.7~km s$^{-1}$ based on the NH$_3$ observations
 of \cite{Urquhart11}, in a very good agreement with the CO 1--0 velocity of $\sim$16 km s$^{-1}$
  \citep{bene20}, indicating that NH$_3$ and CO trace the same molecular gas. The water maser
   emission toward Gy 3--7 is blueshifted (this study) and redshifted \citep{Urquhart11} with
    respect to the systemic velocity.  The blueshifted emission has the similar velocity offset
     ($\sim$13--16~km s$^{-1}$) from the systemic velocity as the redshifted emission (12~km s$^{-1}$),
      extending the velocity range over which the maser spots are found toward Gy 3--7 from $\sim$2~km s$^{-1}$
       reported in \cite{Urquhart11} to $\sim$30~km s$^{-1}$. 

About 60\% of the sources from the \cite{Urquhart11} sample have total velocity ranges
 of the water maser emission of $\lesssim$20~km s$^{-1}$.  The mean/median velocity range of
  the entire sample is 24.7/15.2~km s$^{-1}$. Six sources have velocity ranges of $>$100~km s$^{-1}$.
    The Gy 3--7's water maser velocity range of $\sim$30~km s$^{-1}$ is larger than the velocity
     ranges of the majority of the sources from the \cite{Urquhart11} sample, but it is well
      within the observed values.

The water maser data currently available for Gy 3--7 show that the blueshifted water
 maser spots are brighter than the redshitfed ones (and thus detected more often), in agreement
  with the results obtained by \cite{Urquhart11} for a large sample of $\sim$300 YSOs and H\,{\sc ii}
   regions. The higher relative velocities of blueshifted masers also agree with the \cite{Urquhart11}'
    results.

The difference of $\sim$12--16~km s$^{-1}$ between the systemic velocity of Gy 3--7
 and the velocities of maser spots is small enough to assume that there is the physical
  association between the molecular gas traced by NH$_3$ and CO and the maser emission
   (possibly originating in outflow shocks); it is less likely that the maser emission
    is arising from a different region located along the same line of sight. The mean
     difference between the maser and systemic velocities for the \cite{Urquhart11} sample
      of YSOs and H\,{\sc ii} regions is $-$3.8~km s$^{-1}$, but with the large standard
       deviation ($\sim$20~km s$^{-1}$). It is possible that the water maser emission with
        smaller offsets from the systemic velocity of Gy 3--7 exists, but remained undetected
         due to the maser variability and/or limited sensitivity of the observations.

The spatial resolution of the existing water maser observations of Gy 3--7 is too
 low to accurately pinpoint the location of the maser spots in the region. Higher resolution
  observations are needed to investigate the distribution of the maser spots toward Gy 3--7
   and the origin of the maser emission.  Systematic monitoring of the water maser activity
    at high angular resolution would be necessary to constrain any periodicity in the maser
     line intensity in one or more velocity components \citep{szym16}.

\section{Spatial extent of FIR line emission}\label{app:spatial}

\begin{figure*}[htp]
\centering
\includegraphics[height=0.28\linewidth]{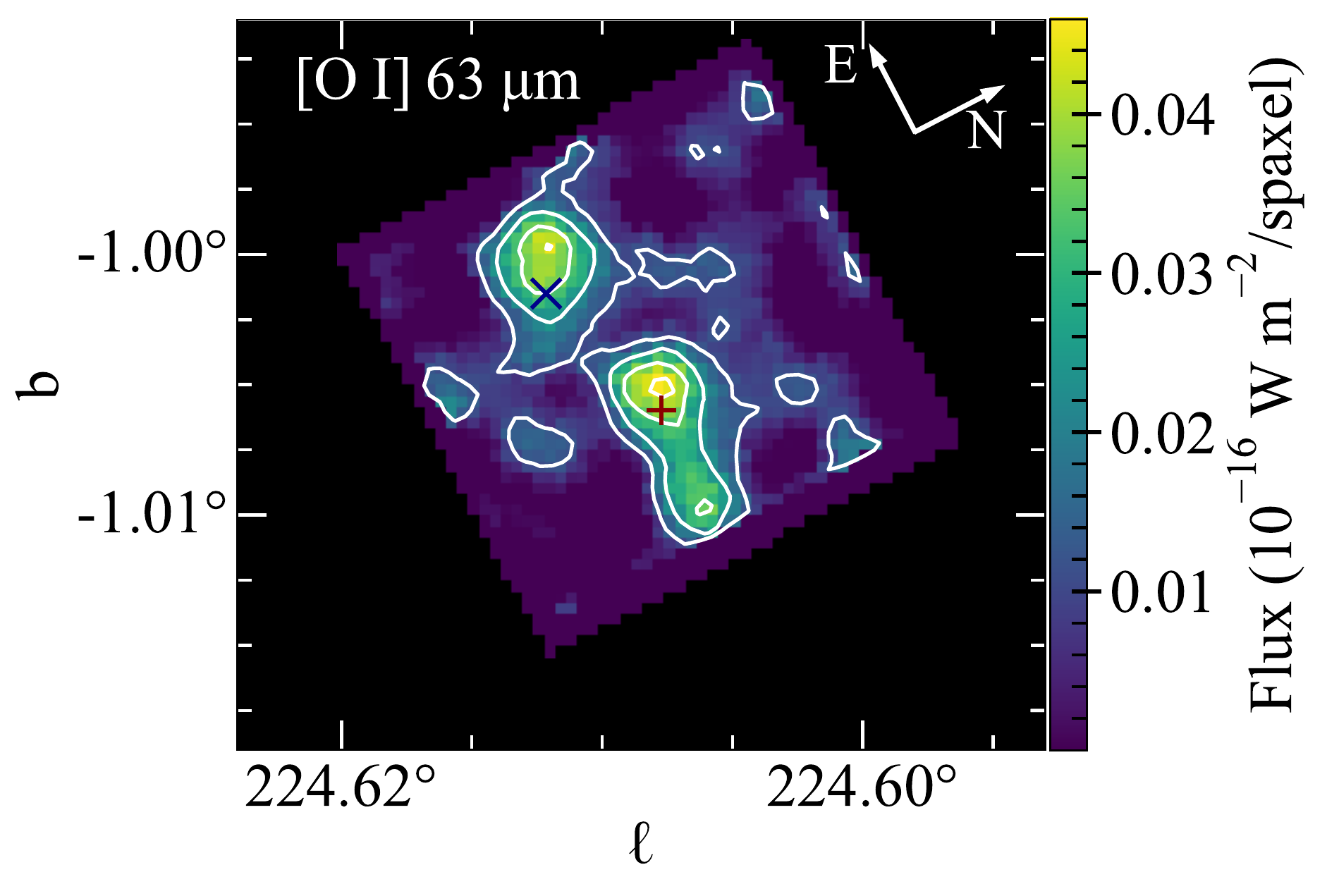}
\includegraphics[height=0.28\linewidth]{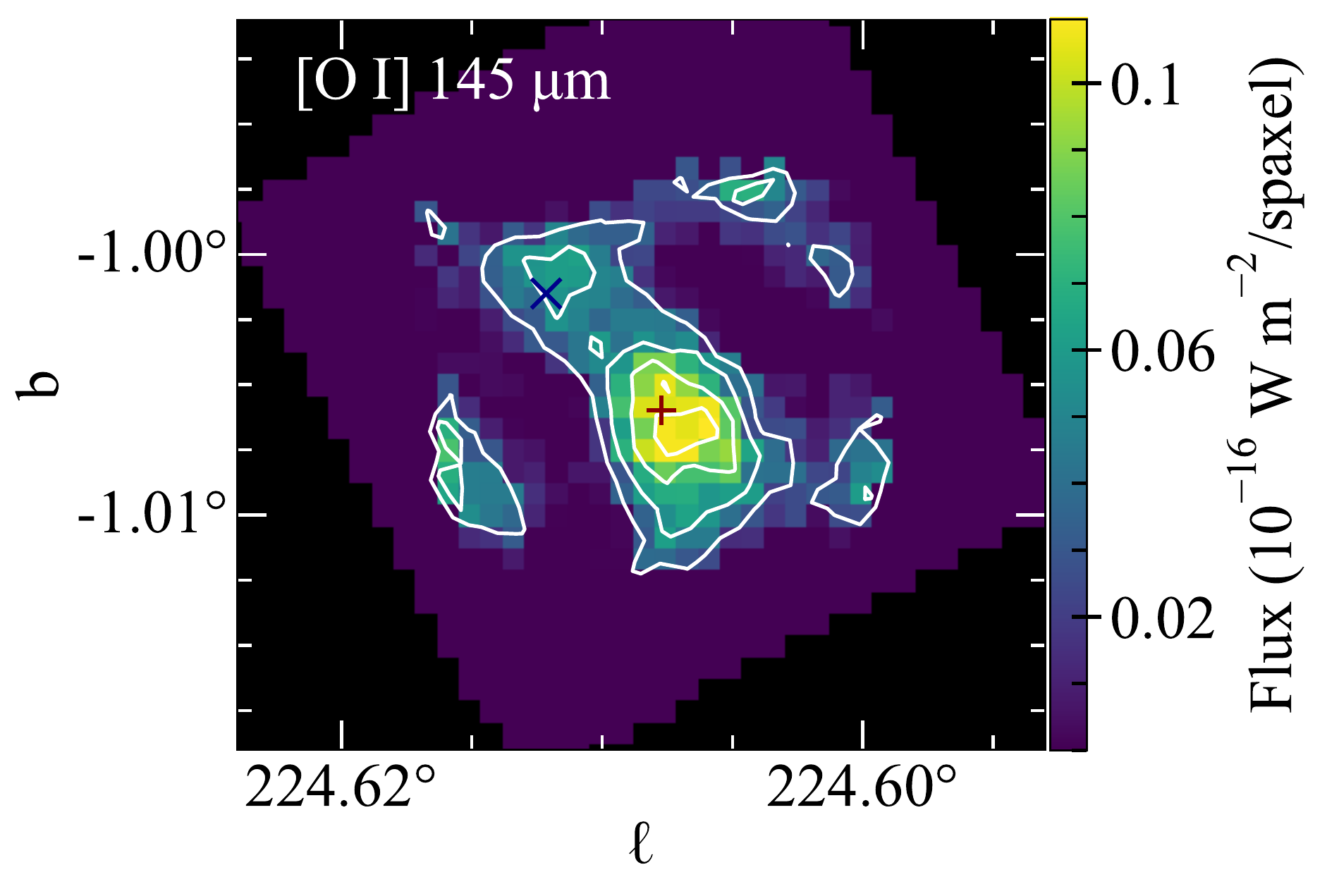}
\includegraphics[height=0.28\linewidth]{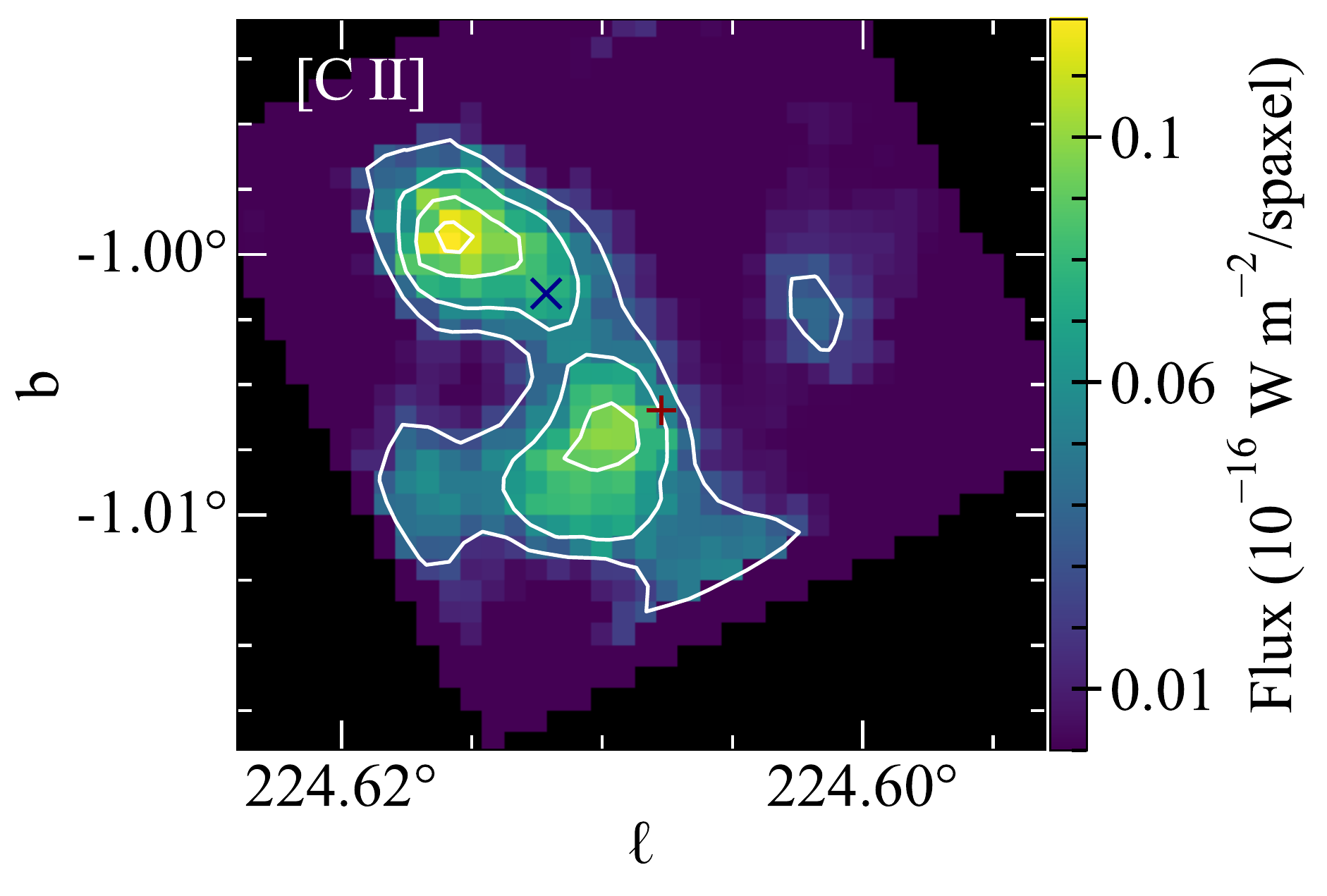}
\includegraphics[height=0.28\linewidth]{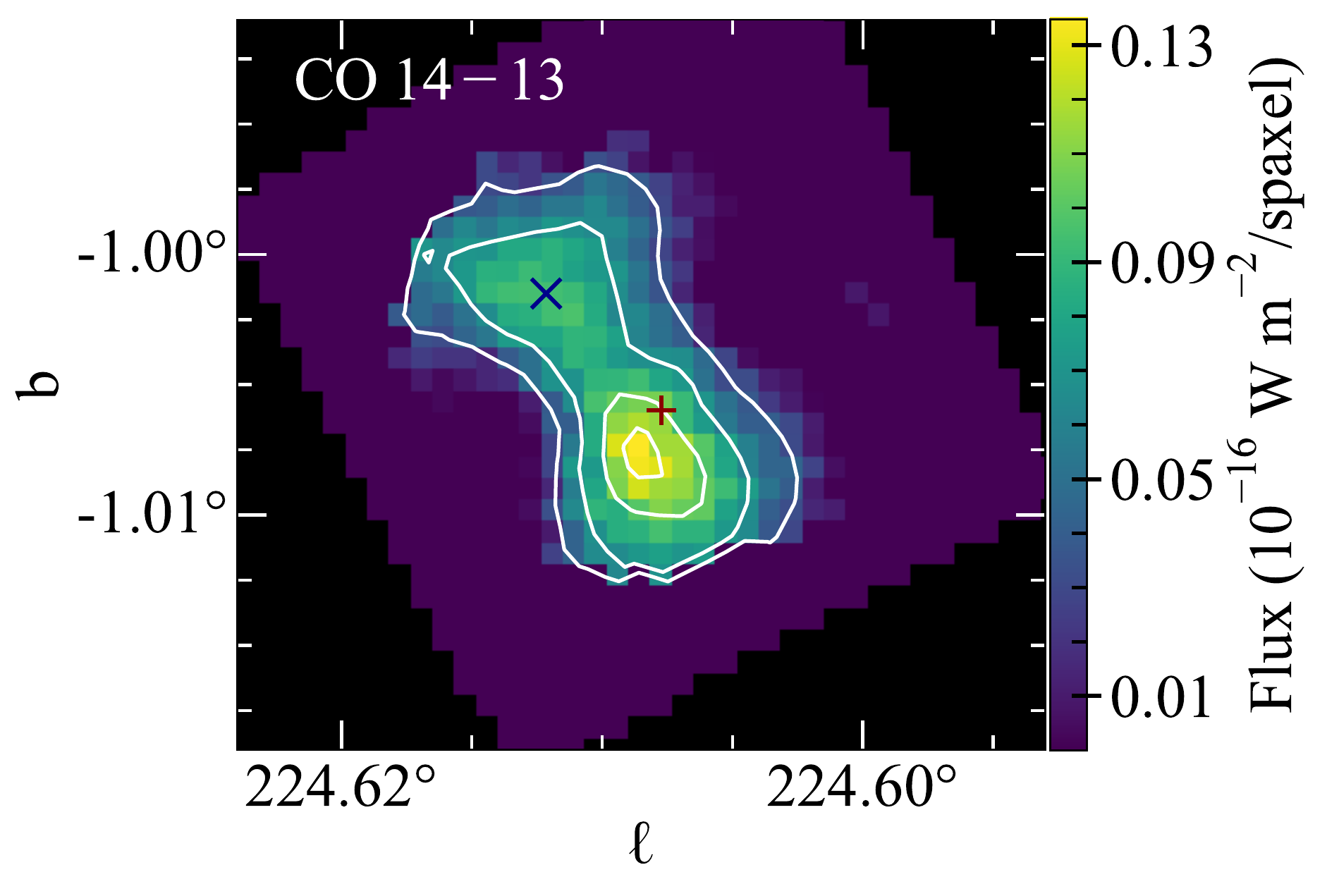}
\includegraphics[height=0.28\linewidth]{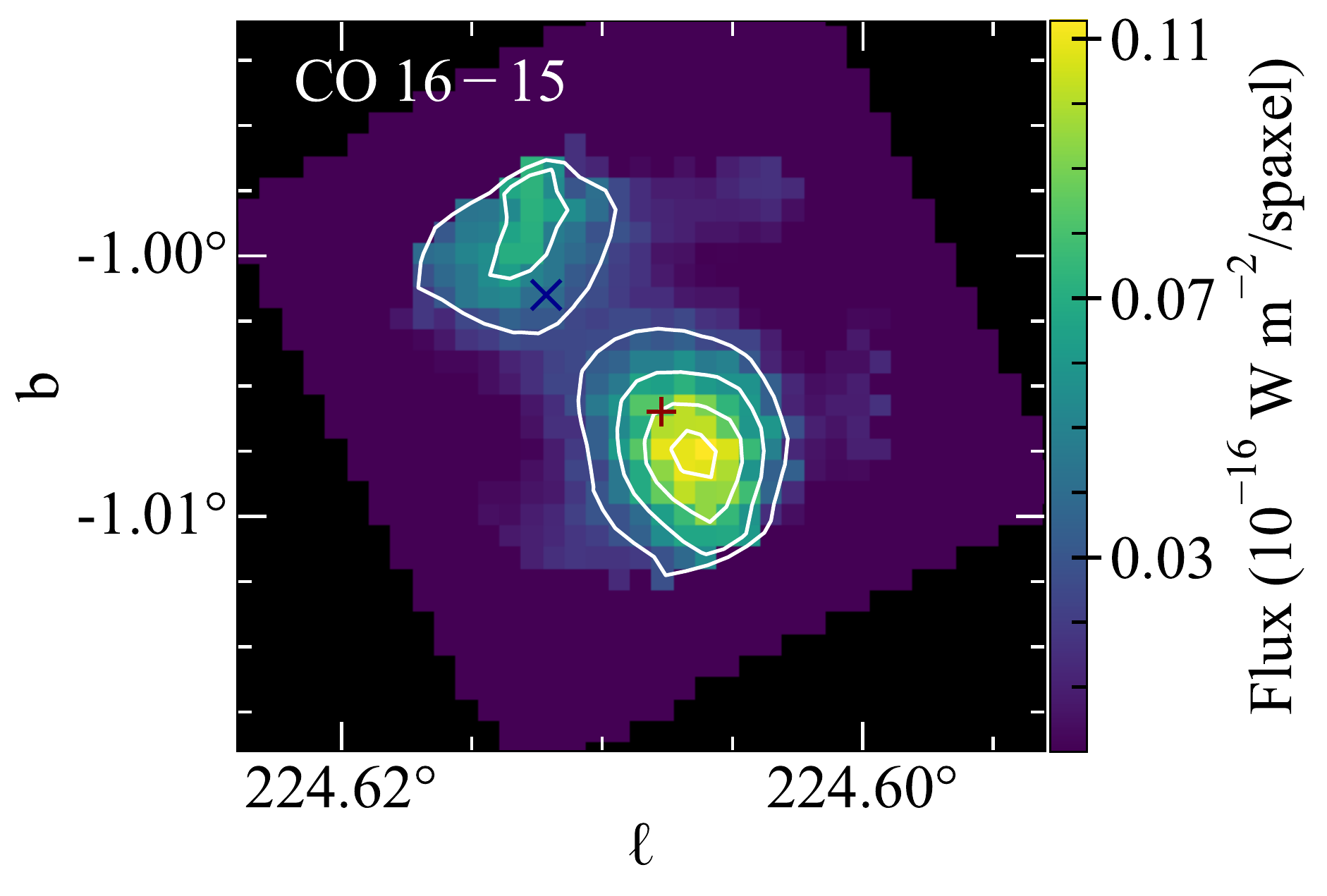}
\includegraphics[height=0.28\linewidth]{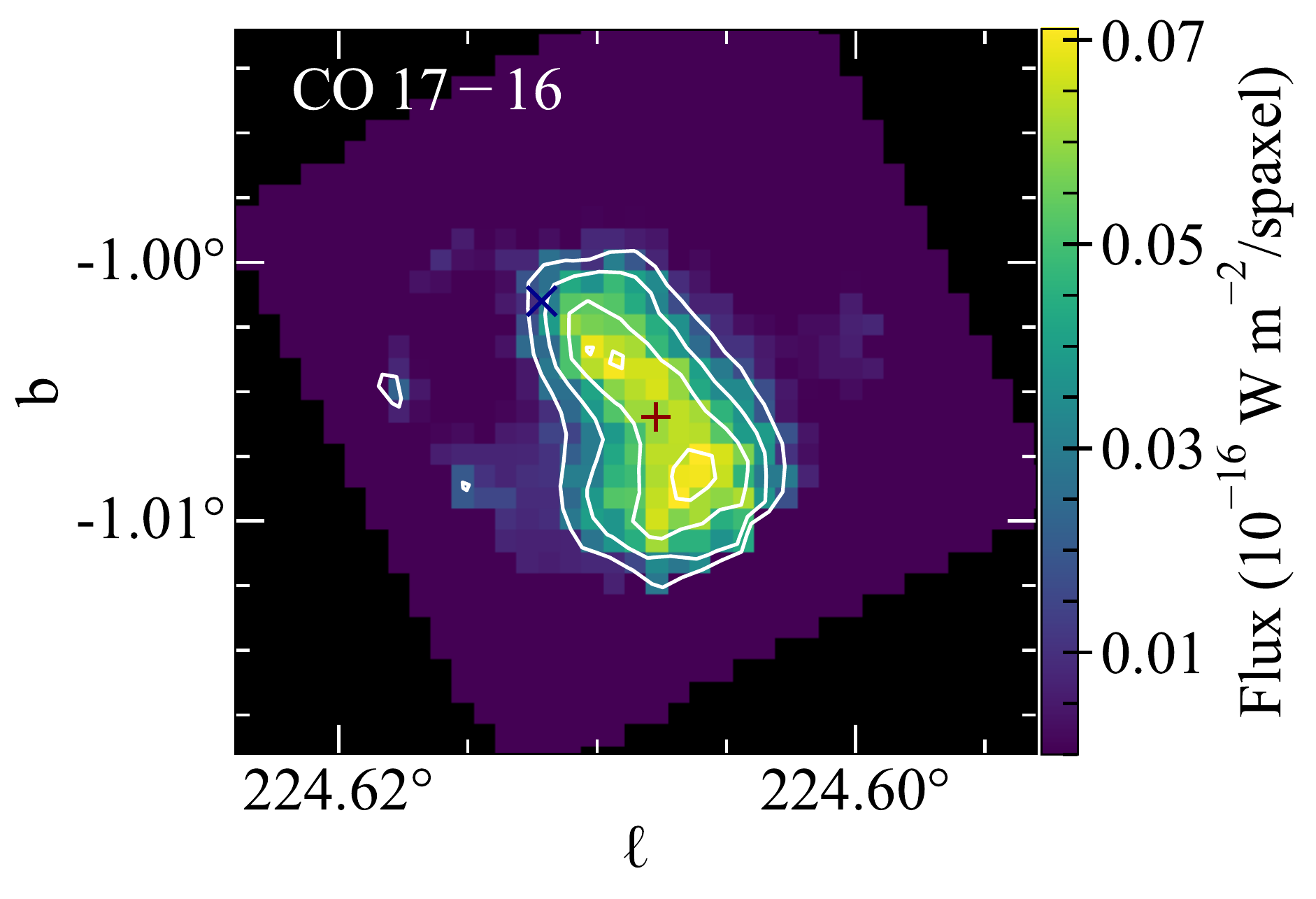}
\caption{FIFI-LS integrated intensity maps of the [\ion{O}{i}] lines at 63.2 and 145.5~$\mu$m, the
[\ion{C}{ii}] line at 157.7~$\mu$m, the CO lines $J=14-13$, $16-15$, $17-16$ at 186, 162.8,
153.3~$\mu$m, respectively. The white contours show line emission at 25$\%$, 50$\%$, 75$\%$, and
95\% of the corresponding line emission peak. 
The "+" and "x" signs show the positions of the dense cores A and B, respectively.}
\label{fig:sofia_emission_map}
\end{figure*}

\begin{figure*}[tb]
\centering
\includegraphics[height=0.28\linewidth]{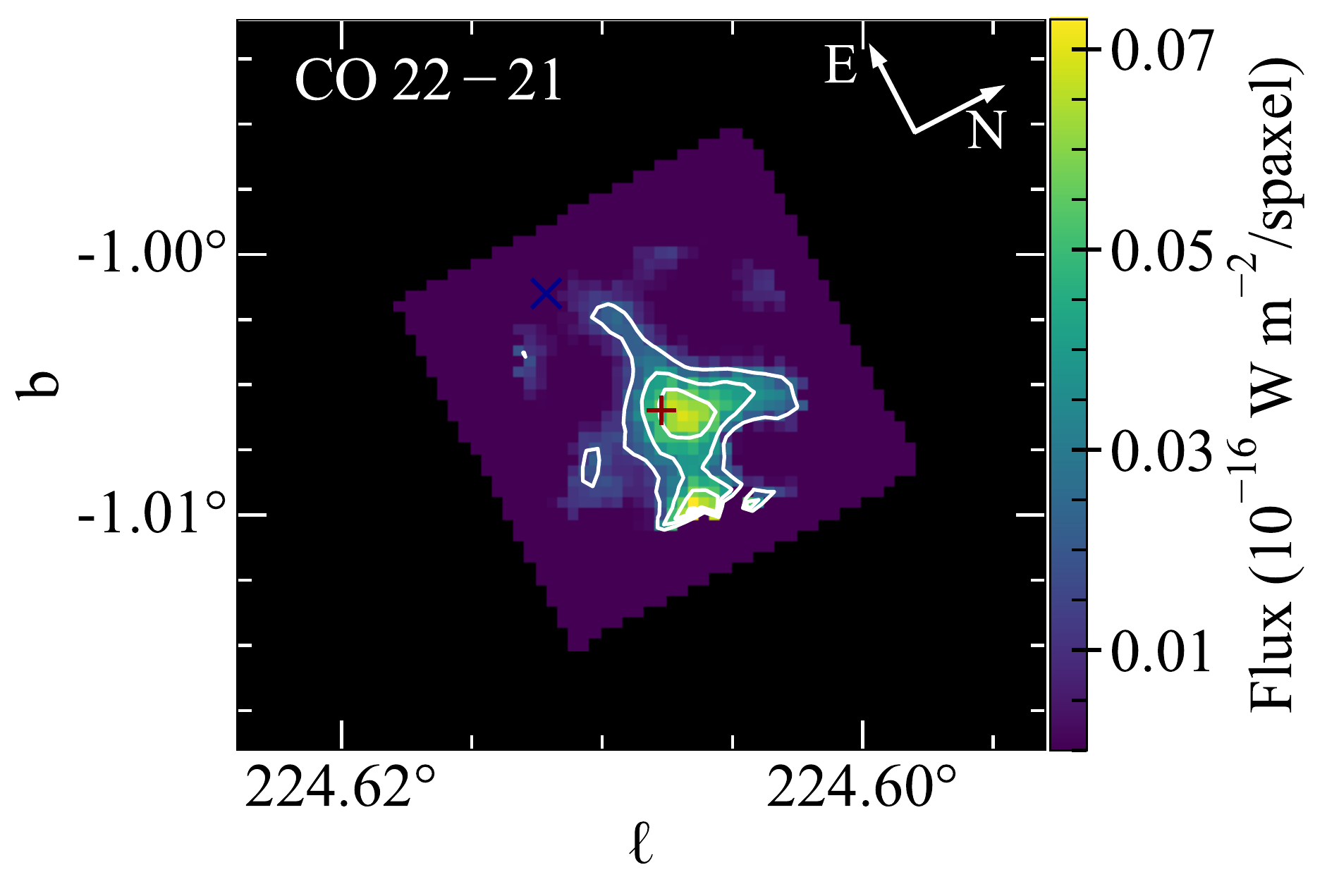}
\includegraphics[height=0.28\linewidth]{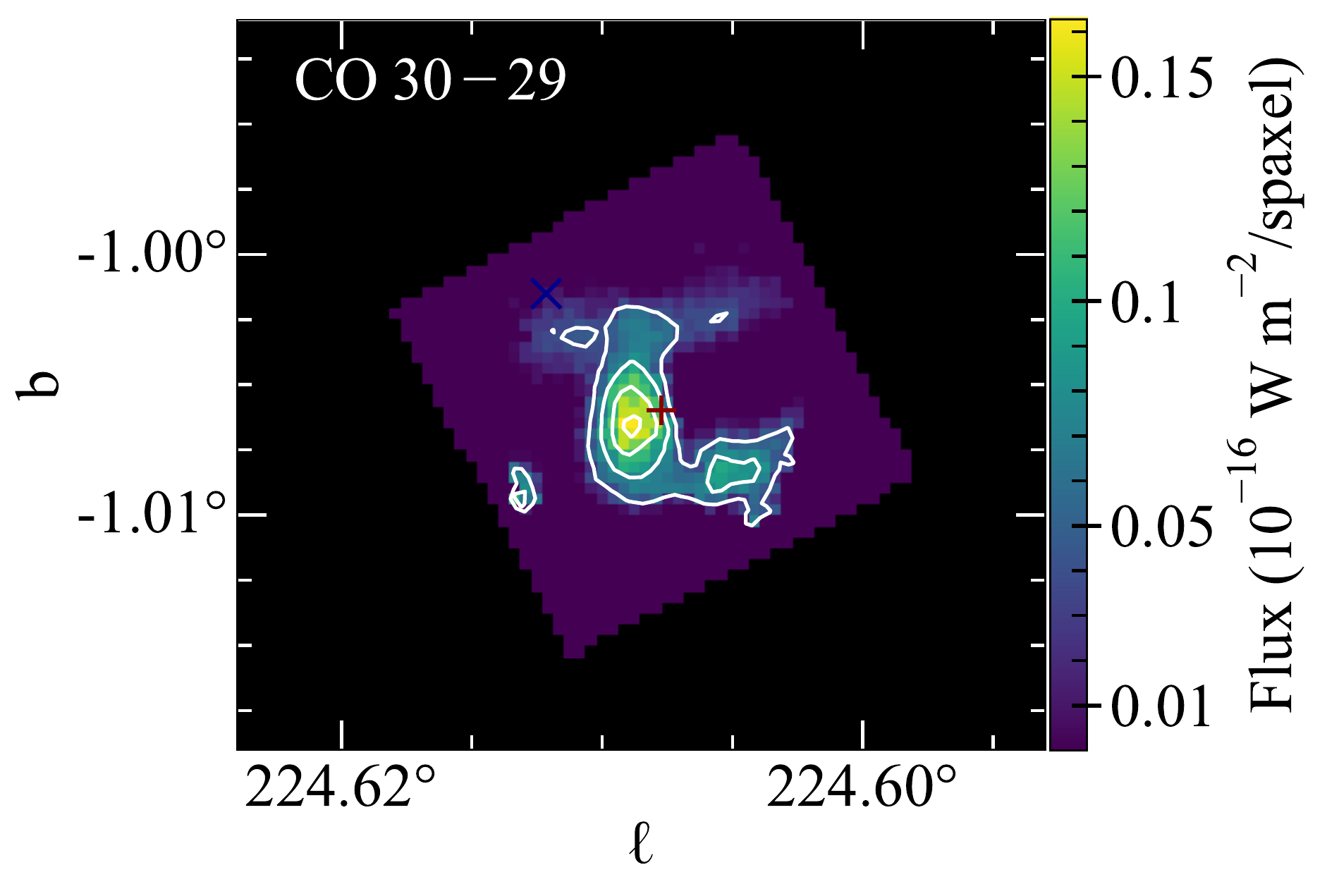}
\includegraphics[height=0.28\linewidth]{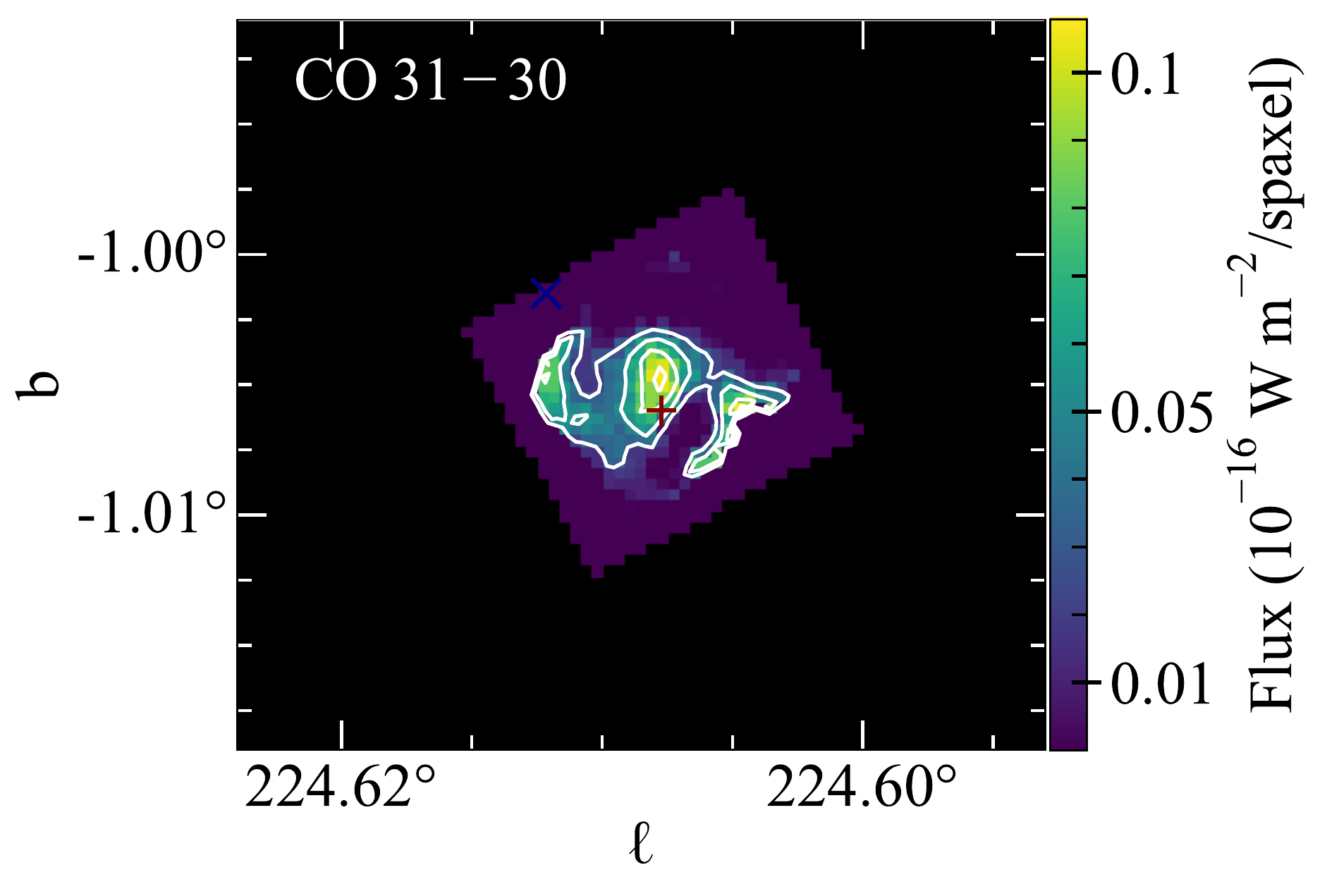}
\includegraphics[height=0.28\linewidth]{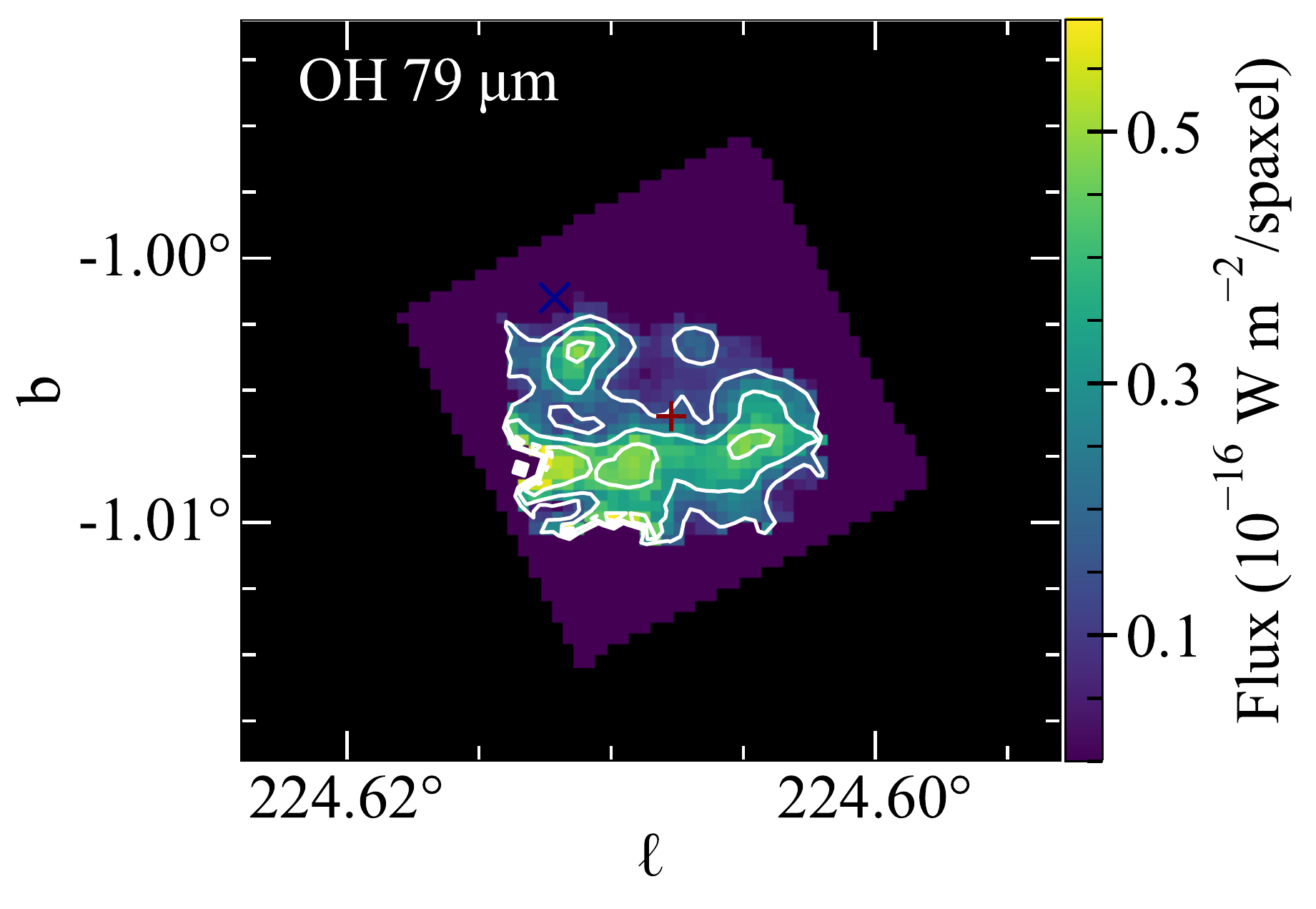}
\caption{FIFI-LS integrated intensity maps of the CO~$22-21$, CO~$30-29$ , and
CO~$31-30$ transitions at 118, 84, and 87~$\mu$m, respectively, and the OH line at 79.2~$\mu$m.
The white contours show line emission at 25$\%$, 50$\%$, 75$\%$, and
95\% of the corresponding line emission peak, see also Figure~\ref{fig:sofia_emission_map}.}
\label{fig:sofia_emission_map2}
\end{figure*}

Figures~\ref{fig:sofia_emission_map} and \ref{fig:sofia_emission_map2} show the spatial extent of
the [\ion{O}{i}] lines at 63 and 145~$\mu$m, the [\ion{C}{ii}] line at 157~$\mu$m, the CO lines with
$J_\mathrm{up}=14-31$, and the OH line at 79.2~$\mu$m. We calculated the flux of these emission line
toward the two dense cores A and B, within a beam size of 20$\arcsec$ (Table~\ref{tab:flux_2cores}). 
Since all lines are spectrally-unresolved, the line fluxes are obtained using Gaussian fits and their
 uncertainties are estimated as 1$\sigma$ of the distribution of 10~000 Gaussian profiles generated
  based on the parameters of the Gaussian fit and their uncertainty.


\begin{table}
\caption{Flux SOFIA FIFI-LS toward the two dense cores within a beam size of
20$\arcsec$.\label{tab:flux_2cores} }
\centering 
\begin{tabular}{r r r }
\hline\hline 
$\lambda$ ($\mu$m) & \multicolumn{2}{c}{Flux (10$^{-16}$ W m$^{-2}$)}\\
~&       \multicolumn{1}{c}{core~A}     &        \multicolumn{1}{c}{core~B}             \\
\hline
63.18 &$ 29.65 \pm 1.60 $&$ 24.37 \pm 1.72$\\
84.41 &$4.93 \pm 1.80 $& \multicolumn{1}{c}{--}\\
87.19 &$ 1.81 \pm 0.53 $& \multicolumn{1}{c}{--}\\
118.58 &$ 2.68 \pm 0.21 $& \multicolumn{1}{c}{--}\\
157.74 &$ 1.15 \pm 0.07 $&$ 1.90 \pm 0.09$\\
145.53 &$ 2.68 \pm 0.09 $&$ 1.97 \pm 0.12$\\
153.27 &$ 2.09 \pm 0.05 $&$ 1.09 \pm 0.07$\\
162.81 &$ 2.84 \pm 0.05 $&$ 1.80 \pm 0.07$\\
186.00 &$ 2.73 \pm 0.08 $&$ 2.56 \pm 0.13$\\
\hline 
\end{tabular}
\end{table} 

\section{Multi-wavelength photometry and SED fitting results}\label{app:sed}
\begin{figure*}
\begin{center}
\includegraphics[scale=0.8, trim=4cm 0cm 4cm 0cm, clip]{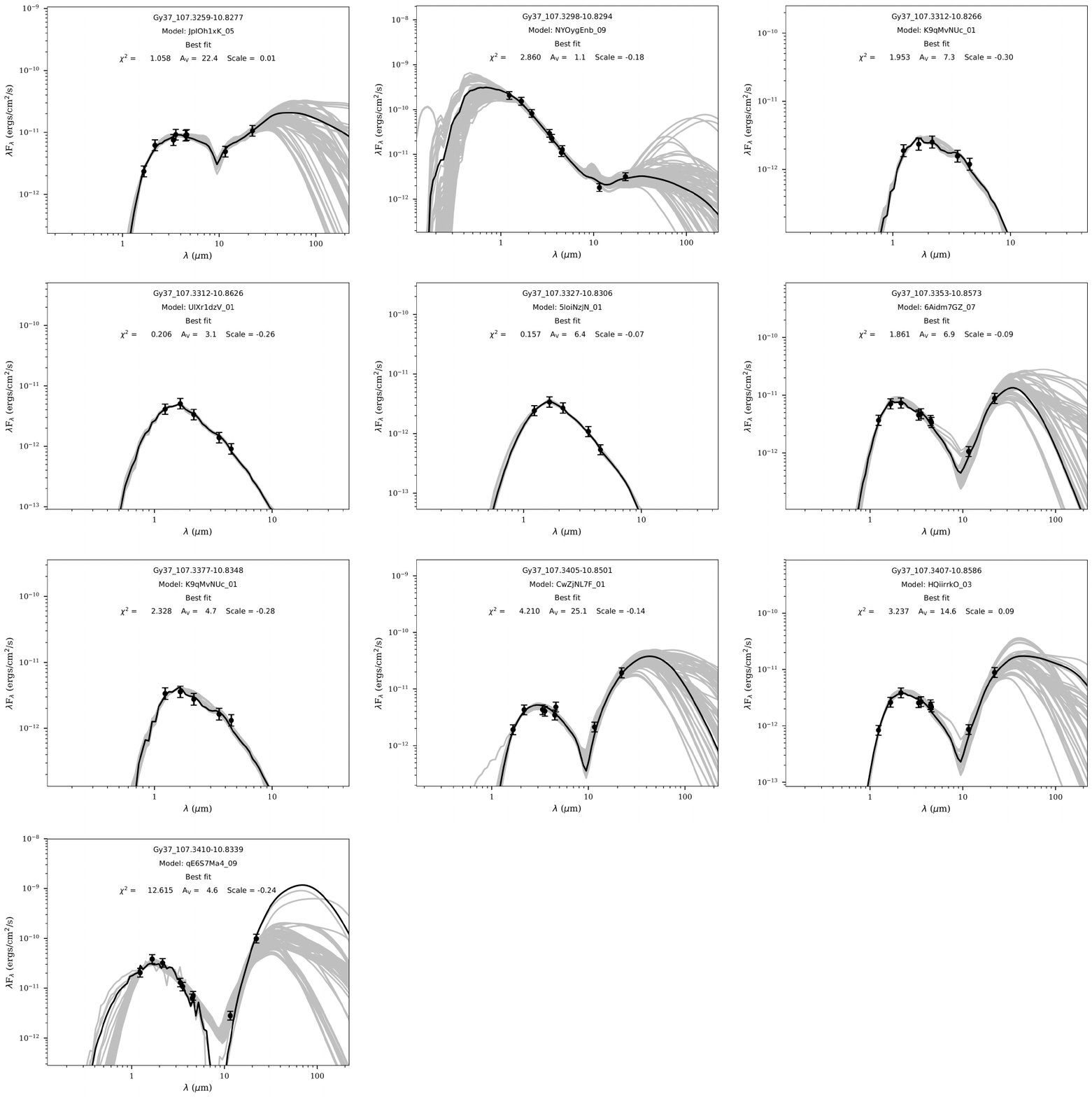} 
\end{center}
\caption{SEDs of YSO candidates with well-fitted \cite{robitaille2017} YSO models. The best-fit
model is indicated with the black solid line. Gray lines show the YSO models with $\chi^{2}$ between
$\chi^{2}_\mathrm{best}$ and $\chi^{2}_\mathrm{best}$+ $F \times n$, where $n$ is the number of data
points and $F$ is a threshold parameter which we set to 3 \citep{sewilo19}. Filled circles and
triangles are valid flux values and flux upper limits, respectively. The values of a reduced
$\chi^{2}$ and interstellar visual extinction for the best-fit model are indicated in the plots.}
\label{fig:sed2}
\end{figure*}
\begin{figure}
\begin{center}
\includegraphics[scale=0.67]{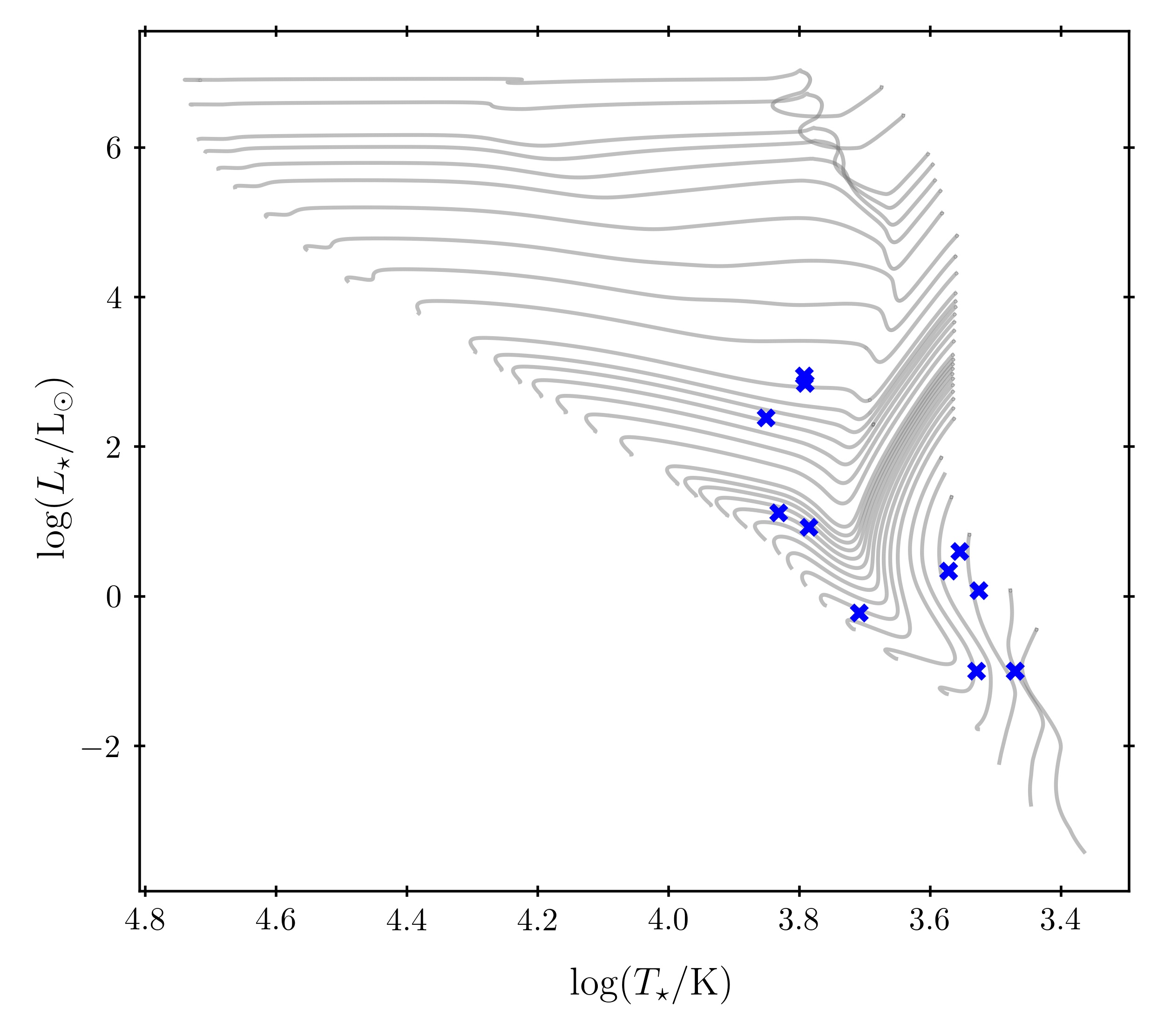} 
\end{center}
\caption{HR diagram with YSOs in Gy~3--7 (blue \lq$\times$' symbols) and the PARSEC evolutionary
tracks \citep{bressan2012,chen2014,chen2015,tang2014}.}
\label{fig:hr}
\end{figure}

Table \ref{tab:phot} shows the multi-wavelength photometry for 15~YSO candidates in Gy~3--7. 
Figure~\ref{fig:sed2} shows the SEDs of YSO candidates in Gy~3--7 with the best-fit
\cite{robitaille2017} models. 
Figure~\ref{fig:hr} shows the Hertzsprung-Russell diagram with
the positions of the YSOs obtained from the SED modeling in line with the PMS tracks. 

We note that sources numbered 3 -- 8 and source 11 lack continuum measurements at $>4.5$ $\mu$m, which
limits the SED modeling. The best-fit models for sources No. 4 -- 6 and 11 are stars (Table
\ref{tab:params}), since the envelope emission could not be traced. Near-IR observations show some
IR excess toward source No. 6 \citep{tapia97}, but the confirmation of its YSO status would require
additional observations, which are outside of the scope of this paper.
\begin{landscape}
\begin{table}
\begin{center}
\tiny \tiny \tiny 
\caption{Multi-wavelength photometry of YSO candidates in the IRAS field. The columns represent the
2MASS $JHK_s$, {\it{Spitzer}} IRAC 3.6 and 4.5~$\mu$m, AllWISE, {\it{Herschel}} PACS, and SPIRE}
\label{tab:phot}
\setlength{\tabcolsep}{3.5 mm}
\begin{normalsize}
\begin{tabular}{r c c c c c c c c c}
\hline
\hline
ID & RA & Dec & $S_J$ & $S_H$ & $S_K$ & $S_{3.6}$ & $ S_{4.5}$ & $S_{\rm W1}$ & $S_{\rm W2}$ \\
 & (deg) & (deg) & (mJy) & (mJy) & (mJy) & (mJy) & (mJy) & (mJy) & (mJy) \\
\hline
1        & 107.32592 & -10.82772 & ...                                                   & $\phantom{1}1.33 \pm 0.06$     &
$\phantom{1}4.56 \pm 0.13$       & $11.11 \pm 0.43$                              & $13.57 \pm 0.30$                              
& $\phantom{1}8.59 \pm 0.18$     & $\phantom{1}14.28 \pm 0.27$  \\
2        & 107.32975 & -10.82944 & $86.55 \pm 2.07$                              & $86.28 \pm 2.07$                              
& $60.13 \pm 1.27$                               & $27.44 \pm 1.05$                              & $16.67 \pm
0.57$                            & $33.60 \pm 0.72$                              & $\phantom{1}20.02 \pm 0.43$   \\
3        & 107.33013 & -10.83094 & ...                                                   & $\phantom{1}2.33 \pm 0.33$     &
$\phantom{1}3.15 \pm 0.26$       & $\phantom{1}2.61 \pm 0.10$    & $\phantom{1}2.63 \pm 0.07$        &
...                                                      & ...                                                  \\
4        & 107.33121 & -10.82658 & $\phantom{1}0.79 \pm 0.07$    & $\phantom{1}1.33 \pm 0.07$        &
$\phantom{1}1.85 \pm 0.12$       & $\phantom{1}1.89 \pm 0.08$    & $\phantom{1}1.82 \pm 0.05$        &
...                                                      & ...                                                  \\
5        & 107.33121 & -10.86256 & $\phantom{1}1.72 \pm 0.06$    & $\phantom{1}2.86 \pm 0.12$        &
$\phantom{1}2.45 \pm 0.13$       & $\phantom{1}1.67 \pm 0.06$    & $\phantom{1}1.38 \pm 0.04$        &
...                                                      & ...                                                  \\
6        & 107.33275 & -10.83064 & $\phantom{1}1.02 \pm 0.06$    & $\phantom{1}1.92 \pm 0.09$        &
$\phantom{1}1.98 \pm 0.12$       & $\phantom{1}1.31 \pm 0.06$    & $\phantom{1}0.82 \pm 0.04$        &
...                                                      & ...                                                  \\
7        & 107.33379 & -10.84222 & $\phantom{1}2.33 \pm 0.24$    & $14.68 \pm 1.47$                                &
$23.61 \pm 2.36$                                 & ...                                                   & ...                                                   
& ...                                                    & ...                                                  \\
8        & 107.33467 & -10.84192 & $\phantom{1}2.59 \pm 0.35$    & $16.75 \pm 1.67$                                &
$32.03 \pm 3.20$                                 & ...                                                   & ...                                                   
& ...                                                    & ...                                                  \\
9        & 107.33529 & -10.85725 & $\phantom{1}1.55 \pm 0.08$    & $\phantom{1}4.01 \pm 0.11$        &
$\phantom{1}5.40 \pm 0.19$       & $\phantom{1}5.79 \pm 0.21$    & $\phantom{1}5.58 \pm 0.16$        &
$\phantom{1}5.16 \pm 0.12$       & $\phantom{11}5.31 \pm 0.11$  \\
10       & 107.33563 & -10.84117 & $\phantom{1}1.17 \pm 0.16$    & $\phantom{1}4.57 \pm 0.65$        &
$21.06 \pm 1.46$                                 & ...                                                   & ...                                                   
& $87.24 \pm 1.79$                               & $403.08 \pm 7.87$                    \\
11       & 107.33771 & -10.83483 & $\phantom{1}1.41 \pm 0.07$    & $\phantom{1}2.02 \pm 0.07$        &
$\phantom{1}2.03 \pm 0.10$       & $\phantom{1}1.98 \pm 0.07$    & $\phantom{1}2.02 \pm 0.07$        &
...                                                      & ...                                                  \\
12       & 107.33879 & -10.84178 & $\phantom{1}1.47 \pm 0.15$    & $\phantom{1}1.16 \pm 0.20$        &
$\phantom{1}3.22 \pm 0.30$       & ...                                                   & ...                                                    &
$24.38 \pm 2.28$                                 & $\phantom{1}52.26 \pm 1.41$   \\
13       & 107.34054 & -10.85011 & ...                                                   & $\phantom{1}1.08 \pm 0.07$     &
$\phantom{1}3.15 \pm 0.10$       & $\phantom{1}4.89 \pm 0.21$    & $\phantom{1}5.30 \pm 0.11$        &
$\phantom{1}4.91 \pm 0.10$       & $\phantom{11}7.53 \pm 0.10$  \\
14       & 107.34071 & -10.85858 & $\phantom{1}0.35 \pm 0.05$    & $\phantom{1}1.48 \pm 0.09$        &
$\phantom{1}2.82 \pm 0.13$       & $\phantom{1}3.25 \pm 0.12$    & $\phantom{1}3.62 \pm 0.14$        &
$\phantom{1}2.93 \pm 0.06$       & $\phantom{11}3.38 \pm 0.07$  \\
15       & 107.34100 & -10.83386 & $\phantom{1}8.58 \pm 0.29$    & $21.79 \pm 0.60$                                &
$23.65 \pm 0.50$                                 & $12.97 \pm 0.61$                              & $\phantom{1}9.56 \pm 0.36$    
& $14.79 \pm 0.27$                               & $\phantom{1}11.00 \pm 0.18$   \\
\hline
\hline
ID & RA & Dec & $S_{\rm W3}$ & $S_{\rm W4}$ & $S_{70}$ & $S_{160}$ & $S_{250}$ & $S_{350}$ &
$S_{500}$ \\
 & (deg) & (deg) & (mJy) & (mJy) & (Jy) & (Jy) & (Jy) & (Jy) & (Jy) \\
\hline
1        & 107.32592 & -10.82772 & $\phantom{11}19.26 \pm 0.41\phantom{1}$       & $\phantom{11}79.72 \pm
2.92\phantom{1}$         & ...                                                   & ...                                                    &
...                              & ...                           & ...                          \\
2        & 107.32975 & -10.82944 & $\phantom{111}7.16
\pm 0.28\phantom{1}$     & $\phantom{11}23.74 \pm 1.56\phantom{1}$       &
...                                                      & ...                                                   & ...                            &
...                              & ...                          \\
3        & 107.33013 & -10.83094 & ...                                                                           &
...                                                                              & ...                                                    &
...                                                      & ...                           & ...                            & ...                          \\
4        & 107.33121 & -10.82658 & ...                                                                           &
...                                                                              & ...                                                    &
...                                                      & ...                           & ...                            & ...                          \\
5        & 107.33121 & -10.86256 & ...                                                                           &
...                                                                              & ...                                                    &
...                                                      & ...                           & ...                            & ...                          \\
6        & 107.33275 & -10.83064 & ...                                                                           &
...                                                                              & ...                                                    &
...                                                      & ...                           & ...                            & ...                          \\
7        & 107.33379 & -10.84222 & ...                                                                           &
...                                                                              & ...                                                    &
...                                                      & ...                           & ...                            & ...                          \\
8        & 107.33467 & -10.84192 & ...                                                                           &
...                                                                              & ...                                                    &
...                                                      & ...                           & ...                            & ...                          \\
9        & 107.33529 & -10.85725 & $\phantom{111}4.18 \pm 0.19\phantom{1}$       & $\phantom{11}66.19 \pm
2.99\phantom{1}$         & ...                                                   & ...                                                    &
...                              & ...                           & ...                          \\
10       & 107.33563 & -10.84117 & $4017.01 \pm 33.44$                                           & $28890.97 \pm
160.10$                                  & $165.8 \pm 0.5$                               & $1030.6 \pm 0.4$                               & $
57.0 \pm 0.5$    & $ 23.5 \pm 0.1$       & $ 10.6 \pm 0.1$      \\
11       & 107.33771 & -10.83483 & ...                                                                           &
...                                                                              & ...                                                    &
...                                                      & ...                           & ...                            & ...                          \\
12       & 107.33879 & -10.84178 & $117.25 \pm 2.62$                                             & $4541.11 \pm
46.24$                                           & $\phantom{1}42.5 \pm 0.2$     & $\phantom{11}60.8 \pm 0.2$     & $
62.0 \pm 0.6$    & $ 20.8 \pm 0.1$       & ...                          \\
13       & 107.34054 & -10.85011 & $\phantom{111}8.38 \pm 0.17\phantom{1}$       & $144.66 \pm
1.34$                                            & ...                                                   & ...                                                    &
...                              & ...                           & ...                          \\
14       & 107.34071 & -10.85858 & $\phantom{111}3.38 \pm 0.20\phantom{1}$       & $\phantom{11}66.43
\pm 2.62\phantom{1}$     & ...                                                   & ...                                                    &
...                              & ...                           & ...                          \\
15       & 107.34100 & -10.83386 & $\phantom{11}11.04 \pm 0.14\phantom{1}$       & $743.98 \pm
1.37$                                            & ...                                                   & ...                                                    &
...                              & ...                           & ...                          \\
\hline
\end{tabular}
\end{normalsize}
\end{center}
\end{table}
\end{landscape}

\section{CO rotational temperature of the intermediate-to high-mass YSOs in the Milky
Way}\label{app:Trot_MW}

We compare the rotational temperatures towards the two dense cores to the results found in the
samples of the IM and HM~YSOs in the Milky Way. There are six IM and ten HM~YSOs presented in
\cite{karska14} and \cite{matuszak2015}, respectively using the data from \textit{Herschel}/PACS. To
be consistent with our SOFIA/FIFI-LS data, we recalculated the rotational temperatures derived by
fitting the rotational diagram using 4 CO transitions with $J_\mathrm{up}$ = (14, 16, 18, 22) and
(14, 16, 17, 22) for the IM and HM~YSOs, respectively.

We checked the new results with the ones presented in the aforementioned studies and find that using
only four CO transitions returns a lower \Trot~than the case of using all the observable CO
transitions. The relative difference is in between 4-27$\%$ with a median of 5$\%$ for the HM~YSOs.
The comparison of these differences is shown in Table~\ref{tab:Trot_compared}. 

Table~\ref{tab:summary_distance_Luminosity_YSOs_MW_MCs} shows heliocentric distances,
Galactocentric radii, metallicities, and luminosities of the sources in the Milky Way and
Magellanic Clouds. We used heliocentric distances together with source coordinates to calculate Galactocentric radii,
assuming a distance from the Sun to the Galactic center of 8.34~kpc. We estimated the metallicity
toward the Milky Way sources using the O/H galactocentric radial gradient based on \ion{H}{ii} 
regions in the Galactic disk \citep{balser11}. 


\begin{table}
\caption{CO rotational temperature and number of emitting molecules for HM~YSOs, derived by fitting
the rotational diagram of CO using all and only four available CO transitions. }
\label{tab:Trot_compared} 
\centering 
\begin{tabular}{l | r l c  } 
\hline \hline 
Source  & \multicolumn{1}{c}{$T\mathrm{_{rot,4}}^{a}$}  &       
\multicolumn{1}{c}{$T\mathrm{_{rot,all}}^{b}$} &$T\mathrm{_{rot}}$ Diff.$^{c}$  \\      
        &       \multicolumn{1}{c}{(K)} &               \multicolumn{1}{c}{(K)} & ($\%$) \\     
\hline
G327-0.6        &$      275     \pm     60      $&$     265     \pm     30      $       &       4       \\
W51N-e1 &$      270     \pm     15      $&$     260     \pm     15      $       &       4       \\
DR21OH  &$      225     \pm     25      $&$     235     \pm     15      $       &       4       \\
W33A    &$      135     \pm     5       $&$     185     \pm     25      $       &       27      \\
G34.26+0.15     &$      330     \pm     60      $&$     320     \pm     25      $       &       3       \\
NGC6334-I       &$      320     \pm     55      $&$     300     \pm     25      $       &       7       \\
NGC7538-I1      &$      175     \pm     25      $&$     185     \pm     15      $       &       5       \\
AFGL2591        &$      175     \pm     10      $&$     195     \pm     15      $       &       10      \\
W3IRS5  &$      335     \pm     40      $&$     320     \pm     15      $       &       5       \\
G5.89-0.39      &$      345     \pm     40      $&$     285     \pm     15      $       &       21      \\
\hline 
\end{tabular} 
\begin{flushleft}
\tablefoot{ $^{(a)}$ Rotational temperatures calculated using 4 CO transitions with $J_{\rm{up}}=
(14, 16, 17, 22)$.
 $^{(b)}$ Rotational temperatures calculated using all detected CO transitions, data is adopted from
\cite{karska14}. 
 $^{(c)}$ The relative difference of \Trot~is taken as $|1-T_{\rm{rot,4}}/
T_{\rm{rot,all}}|\times100$. }
\end{flushleft}
\end{table}
\begin{table*}[hpt]
\caption{FIR line cooling luminosity of the sample in the Milky Way (MW), LMC, and SMC used to
compare with results of this study}\label{tab:summary_distance_Luminosity_YSOs_MW_MCs}
\centering 
\tiny\tiny
\begin{tabular}{lrrrcrcrrrc} 
\hline \hline 
\multicolumn{1}{c}{Source}      &       \multicolumn{1}{c}{$D$} &       \multicolumn{1}{c}{$D$ Ref. } &
        \multicolumn{1}{c}{$R{\rm _{GC}} ^a$}& \multicolumn{1}{c}{$Z^b$} &
\multicolumn{1}{c}{log($L\mathrm{_{bol}}$/$L\mathrm{_\odot}$)} &
\multicolumn{1}{c}{$L\mathrm{_{bol}}$ Ref.}& \multicolumn{1}{c}{$L\mathrm{_{CO}}^c$}&
\multicolumn{1}{c}{ $L\mathrm{_{[\ion{O}{i}]}}^d$} &
\multicolumn{1}{c}{$L\mathrm{_{CO}}$/$L\mathrm{_{[\ion{O}{i}]}}$}&
\multicolumn{1}{c}{$L\mathrm{_{lines}}$ Ref.}\\
 ~&\multicolumn{1}{c}{(kpc)}    & &\multicolumn{1}{c}{(kpc)} &
\multicolumn{1}{c}{($Z\mathrm{_\odot}$)}&~&~& \multicolumn{1}{c}{(10$^{-2}L\mathrm{_\odot}$)}&
\multicolumn{1}{c}{(10$^{-2}L\mathrm{_\odot}$)}& &\\
\hline 
\multicolumn{10}{c}{LMC YSOs} \\
\hline

IRAS04514$-$6931        &       50.0    $\pm$   1.1     &       1       &       ---     &       0.4     &       4.8     &
        16      &       5.3                     &       8.8                     &       0.605                   &       16      \\
N113
YSO3    &       50.0    $\pm$   1.1     &       1       &       ---     &       0.4     &       5.4     &       16      &
        29.5                    &       57.8                    &       0.511                   &       16      \\
SAGE045400.9$-$691151.6 &       50.0    $\pm$   1.1     &       1       &       ---     &       0.4     &       5.1
        &       16      &       14.6                    &       29.2                    &       0.5                     &       16      \\
SAGE051351.5$-$672721.9 &       50.0    $\pm$   1.1     &       1       &       ---     &       0.4     &       5.1
        &       16      &       5.5                     &       15.4                    &       0.357                   &       16      \\
SAGE052202.7$-$674702.1 &       50.0    $\pm$   1.1     &       1       &       ---     &       0.4     &       4.5
        &       16      &       4.3                     &       9.4                     &       0.459                   &       16      \\
SAGE052212.6$-$675832.4 &       50.0    $\pm$   1.1     &       1       &       ---     &       0.4     &       5.5
        &       16      &       14.9                    &       25.4                    &       0.585                   &       16      \\
SAGE053054.2$-$683428.3 &       50.0    $\pm$   1.1     &       1       &       ---     &       0.4     &       4.9
        &       16      &       6.8                     &       5.4                     &       1.24                    &       16      \\
ST01    &       50.0    $\pm$   1.1     &       1       &       ---     &       0.4     &       4.6     &       16      &       7
.1                      &       8.5                     &       0.833                   &       16      \\
N113
YSO1    &       50.0    $\pm$   1.1     &       1       &       ---     &       0.4     &       5.4     &       16      &
        47.5                    &       47.5                    &       1                       &       16      \\
\hline 
\multicolumn{10}{c}{SMC YSOs} \\
\hline
IRAS00464$-$7322        &       62.1    $\pm$   2.0     &       2       &       ---     &       0.2     &       4.1     &
        16      &       3.0                     &       2.8                     &       1.077                   &       16      \\
IRAS00430$-$7326        &       62.1    $\pm$   2.0     &       2       &       ---     &       0.2     &       4.9     &
        16      &       1.9                     &       10.9                    &       0.176                   &       16      \\
N81     &       62.1    $\pm$   2.0     &       2       &       ---     &       0.2     &       4.7     &       16      &       2.
4                       &       36.4                    &       0.067                   &       16      \\
SMC012407-730904        &       62.1    $\pm$   2.0     &       2       &       ---     &       0.2     &       5.3     &
        16      &       6.8                     &       71.2                    &       0.096                   &       16      \\
\hline 
\multicolumn{10}{c}{ MW IM YSOs} \\
\hline

AFGL
490     &       0.97    $\pm$   0.40    &       3       &       9.12    &       0.7     &       3.7     &       17      &
        24.7    $\pm$   2.5     &       36.6    $\pm$   0.2     &       0.7     $\pm$   0.1     &       32      \\
L
1641    &       0.428   $\pm$   0.010   &       4       &       8.69    &       0.8     &       1.8     &       18      &
        2.7     $\pm$   0.4     &       1.6     $\pm$   0.1     &       1.7     $\pm$   0.3     &       32      \\
NGC
2071    &       0.422   $\pm$   0.050   &       4       &       8.71    &       0.8     &       2.7     &       19      &
        44.8    $\pm$   4.3     &       33.4    $\pm$   0.1     &       1.3     $\pm$   0.1     &       32      \\
Vela
17      &       0.7                     &       5       &       8.44    &       0.8     &       2.9     &       20      &       13.2    $\
pm$     2.0     &       44.8    $\pm$   0.1     &       0.3     $\pm$   0.1     &       32      \\
Vela
19      &       0.7                     &       5       &       8.45    &       0.8     &       2.9     &       5       &       9.7     $\pm$
        2.0     &       14.4    $\pm$   0.1     &       0.7     $\pm$   0.1     &       32      \\
NGC
7129    &       1.25                    &       6       &       8.75    &       0.8     &       2.6     &       21      &       20.9    
$\pm$   3.0     &       6.9     $\pm$   0.1     &       3.0     $\pm$   0.4     &       32      \\
\hline 
\multicolumn{10}{c}{MW HM~YSOs} \\
\hline

G327-0.6        &       3.3                     &       7       &       5.84    &       1.0     &       4.7     &       22      &       180
        $\pm$   60      &       30      $\pm$   10      &       6.0     $\pm$   2.8     &       33      \\
W51N-e1 &       5.1     $\pm$   2.2     &       8       &       12.42   &       0.5     &       5.0     &       23      &
        2530    $\pm$   680     &       760     $\pm$   180     &       3.3     $\pm$   1.2     &       33      \\
DR21(OH)        &       1.50    $\pm$   0.08    &       9       &       8.26    &       0.8     &       4.1     &       24      
&       120     $\pm$   30      &       9       $\pm$   3       &       13.3    $\pm$   5.6     &       33      \\
W33A    &       2.40    $\pm$   0.16    &       10      &       6.02    &       1.0     &       4.6     &       25      &
        60      $\pm$   20      &       5       $\pm$   2       &       12.0    $\pm$   6.2     &       33      \\
G34.26+0.15     &       3.30                    &       11      &       5.91    &       1.0     &       5.5     &       26      &
        770     $\pm$   240     &       190     $\pm$   50      &       4.1     $\pm$   1.7     &       33      \\
NGC6334I        &       1.7     $\pm$   0.3     &       12      &       6.66    &       0.9     &       5.4     &       27      &
        340     $\pm$   100     &       4       $\pm$   3       &       85.0    $\pm$   68.5    &       33      \\
NGC7538-IRS1    &       2.7     $\pm$   0.1     &       13      &       9.66    &       0.7     &       5.1     &       28
        &       280     $\pm$   50      &       1100    $\pm$   250     &       0.2     $\pm$   0.1     &       33      \\
AFGL2591        &       3.3     $\pm$   0.08    &       9       &       8.36    &       0.9     &       5.3     &       29      &
        160     $\pm$   50      &       340     $\pm$   90      &       0.5     $\pm$   0.2     &       33      \\
W3-IRS5 &       2.00    $\pm$   0.07    &       14      &       9.83    &       0.7     &       5.2     &       30      
&       1050    $\pm$   250     &       420     $\pm$   10      &       2.5     $\pm$   0.8     &       33      \\
G5.89-0.39      &       1.30    $\pm$   0.09    &       15      &       7.05    &       0.9     &       4.7     &       31
        &       390     $\pm$   90      &       370     $\pm$   90      &       1.1     $\pm$   0.4     &       33      \\

\hline 
\end{tabular} 
\begin{flushleft}
 \textbf{References:} (1) \cite{Pietrzynski13}, (2) \cite{Graczyk14}), (3) \cite{Purser21}, (4)
\cite{Wilson05}, (5) \cite{Liseau92},(6) \cite{Shevchenko_Yakubov89}, (7) \cite{Minier09}, (8)
\cite{Xu09}, (9) \cite{Rygl12} , (10) \cite{Immer13}, (11) \cite{Kuchar_Bania94}, (12)
\cite{Neckel78}, (13) \cite{Moscadelli09}, (14) \cite{Hachisuka06}, (15) \cite{Motogi11}, (16)
\cite{Oliveira19},(17) \cite{Navarete15}, (18) \cite{Stanke2000} , (19) \cite{Butner90}, (20)
\cite{Giannini05}, (21)\cite{Johnstone10}, (22) \cite{Urquhart12}, (23) \cite{Dishoeck11},(24)
\cite{Jakob07}, (25) \cite{Faundez04}, (26) \cite{Hatchell03}, (27) \cite{Sandell2000}, (28)
\cite{Sandell04}, (29) \cite{vandertak99}, (30) \cite{Ladd93}, (31) \cite{vandertak12}, (32)
\cite{matuszak2015}, (33) \cite{karska14}. \\

$^a$ Galactocentric radius toward the sources in the Milky Way is calculated using the information
from their coordinate and distance to the Sun, assuming the distance from the Sun to the Galactic
center is 8.34~kpc. \\
$^b$ Metallicity towards the sources in the Milky Way is calculated using the relation of O/H
gradients inferred from the \ion{H}{ii} regions: $12 + \log\rm{(O/H)} = -0.0446 R_{\rm GC} + 8.962$
\citep{balser11}.\\
$^c$ Total FIR luminosity of CO derived from the rotational diagram fitting. \\
$^d$ Total FIR luminosity of [\ion{O}{i}] calculated by summing the luminosity of [\ion{O}{i}]
lines at 63 and 145~$\mu$m. 
\end{flushleft}
\end{table*}

\end{appendix}

\end{document}